\documentclass[12pt]{iopart}   %% Institute of Physics
\usepackage{iopams}
\usepackage[english]{babel}
\usepackage{graphicx}% Include figure files
\usepackage{dcolumn}% Align table columns on decimal point
\usepackage{bm}% bold math
\usepackage{verbatim}
\usepackage{amsfonts}
\usepackage{amssymb}
\usepackage{color}
\def\be#1{\begin{equation}\label{#1}}
\def\ee{\end{equation}}
\def\bea#1{\begin{eqnarray}\label{#1}}
\def\eea{\end{eqnarray}}
\def\Bea{\begin{eqnarray*}}
\def\Eea{\end{eqnarray*}}
\def\Eq#1{Eq.(\ref{#1})}  % AIP eqn reference
\def\sp{\hspace{.5em}}

\def\spp{\sp\sp}
\def\no{\nonumber \\}
\def\Fig#1{Fig.(\ref{#1})}

\def\ket#1{|#1\rangle}
\def\bra#1{\langle #1|}

\def\lstateDefn#1 #2{|n_{po} - #1 - #2\rangle_{p}|n_{so} + #1\rangle_s|#1\rangle_{\bar{i}}}
\def\myover#1{\myoverDefn#1}
\def\myoverDefn#1#2{\hbox{\space \raise-2mm\hbox{$\textstyle{#1} \atop \scriptstyle{#2}$} }}
\def\ns0{n_{s0}}
\def\nsbar0{n_{\bar{s}0}}
\def\np0{n_{p0}}
\def\nc0{n_{c0}}
\def\nb0{n_{bh0}}
\def\sech{\textrm{sech}}
\newcommand{\binom}[2]{\left(\begin{array}{c} #1 \\ #2 \end{array}\right)}
\begin{document}
%=============================================================
% APS
%=============================================================
%%\draft{DRAFT:version 2: 13Aug2014}
%\title{Parametric down conversion with a depleted pump \\ as a heuristic model for black hole particle production}
%%\title{Parametric down conversion with a depleted pump \\ and the black hole information problem}
%%Lines break automatically or can be forced with \\
%\author{Paul M. Alsing}
% \altaffiliation[]{Corresponding author}  %  optional
% \affiliation{Air Force Research Laboratory, Information Directorate, Rome, NY 13441}
% \email{paul.alsing@us.af.mil}   %optional
%\date{\today}

%=============================================================
%IOP
%=============================================================
\title[Parametric down conversion with depleted a pump and quantum black holes]{Parametric down conversion with a depleted pump as a model for classical information transmission
capacity of quantum black holes}

\author{Paul M. Alsing}
\address{Air Force Research Laboratory, Information Directorate, Rome, N.Y., USA}
%=============================================================

\begin{abstract}
In this paper we extend the investigation of Adami and Ver Steeg [Class. Quantum Grav. \textbf{31}, 075015 (2014)]  to treat the process of black hole particle emission effectively as the analogous quantum optical process of parametric down conversion (PDC) with a dynamical (depleted vs. non-depleted) `pump' source mode which models the
evaporating black hole (BH) energy degree of freedom.
We investigate both the short time (non-depleted pump) and long time (depleted pump) regimes of the quantum state and its impact on the Holevo channel capacity for communicating information from the far past to the far future
in the presence of Hawking radiation.
The new feature introduced in this work is the coupling of the emitted Hawking radiation modes through the common black hole `source pump' mode which phenomenologically represents a quantized energy degree of freedom of the gravitational field.
This (zero-dimensional) model serves as a simplified arena to explore BH particle production/evaporation and back-action effects under an explicitly unitary evolution which enforces quantized energy/particle conservation.
Within our analogous quantum optical model we examine the entanglement between two emitted particle/anti-particle and anti-particle/particle pairs coupled via the black hole (BH) evaporating `pump' source. We also analytically and dynamically verify the `Page information time' for our model which refers to the conventionally held belief that the information in the BH radiation becomes significant after the black hole has evaporated half its initial energy into the outgoing radiation. Lastly, we investigate the effect of BH particle production/evaporation on two modes in the exterior region of the BH event horizon that are initially maximally entangled, when one mode falls inward and interacts with the black hole, and the other remains forever outside and non-interacting.

%=======================
% testing equations
%=======================
% $\asbardag \lstate{n m}$
%=======================
%%\begin{subequations}
%\begin{align}
%(i)   & \int_{-s_{max}}^{s_{max}} ds < \infty, \\
%\label{DirBCs} (ii)  & \lim_{s\rightarrow\pm s_{max}} \psi(s) = 0, \\
%\label{NeumBCs}(iii) & \lim_{s\rightarrow\pm s_{max}} \psi_s(s) = \mp 1, \\
%\label{HigherDerivBCs} (iii) & \lim_{s\rightarrow\pm s_{max}} \frac{d^{2k}\psi(s)}{ds^{2k}} =  0, \quad k > 1.
%\end{align}
%%\end{subequations}
%=======================

\end{abstract}
%\maketitle

%=============================================================
% APS
%=============================================================
%\pacs{PACS numbers:  03.67.-a, 03.67.Bg, 04.70.Dy}
%
%\maketitle
%=============================================================
% APS
%=============================================================
%===============================================================================================================

\section{Introduction}
Recently, Adami and Ver Steeg \cite{Adami:2014} (see also Br\'{a}dler and Adami \cite{Bradler_Adami:2014}) have proposed a solution to the black hole information problem (BHIP) by essentially modeling
the black hole (BH) particle emission
%as with
from a
constant mass black hole (BH)
%as parametric down conversion (PDC) with a non-depleted pump,
and calculating the channel capacity of the non-thermal stimulated emission around the black hole in the presence of the Hawking radiation treated as spontaneous emission. By demonstrating that the channel capacity is non-zero, even for infinite black hole surface gravity with as little as one initial particle present before particle creation, they conclude that a classical copying of infalling information is performed by the out going unitary stimulated emission process around the black hole, and therefore classical information is preserved.
In this paper, we extend the approach of Adami, Ver Steeg and Br\'{a}dler
to treat the process of black hole particle emission effectively as the analogous quantum optical process of parametric down conversion (PDC) with a dynamical (a depleted vs. a non-depleted) `pump' source which provides
a simplified arena to phenomenologically  model the
evaporating black hole (BH) energy degree of freedom, and which unitarily incorporates back-action effects.

As discussed more fully below, parametric down conversion is the process by which a `pump'  source quanta (boson) of higher energy creates a pair of lower energy particles such that energy (and momentum) are conserved.
%This is the process typically considered in standard treatments of Hawking radiation production.
Under energy conservation the  reverse process (sum frequency generation) simultaneously occurs in which two lower energy  (bosons) combine to produce a particle of higher energy.
In standard treatments of Hawking radiation production
(see \cite{Hawking:1975,Unruh:1976,Gerlach:1976,Giddings:1992,Mathur:2009}
and references therein), the BH `pump' source is (implicitly) taken to have such a large number of quanta that it is reasonable to approximate the source to be a constant (mass/energy) c-number, and pair production proceeds from the vacuum state as spontaneous emission  (typically referred to as spontaneous parametric down conversion, SPDC). If there is already population present in either of the emitted modes, as often used to model the formation of the BH by infalling matter, then stimulated emission can occur simultaneously with spontaneous emission.
While stimulated emission in general requires
a pump to create an inversion, it is certainly valid to model the process by a constant (non-depleted) `pump source,' as performed by Adami and Ver Steeg \cite{Adami:2014} (and Br\'{a}dler and Adami \cite{Bradler_Adami:2014}) in their
channel capacity calculations.
In this work we maintain the quantized nature of the  BH 'pump' source mode in addition to the conventionally considered quantized Hawking radiation pairs.
The trilinear Hamiltonian model  utilized in this work to investigate BH particle production/evaporation is the simplest generalization of the purely pair production Hamiltonian utilized in standard treatments of Hawking radiation that also enforces quantized energy and particle conservation during the unitary process. As such, the newly introduced BH `pump source' mode  phenomenologically models a quantized energy degree of freedom of the BH gravitation field. Its chief feature is a `natural' incorporation of BH evaporation, expressed as energy/particle number conservation of the BH `pump source' with the emitted Hawking radiation pairs.

Another justification for the inclusion of a quantized BH `pump' source mode is the following. In the spirit of a quantum information approach to understanding entanglement issues in BH (as advocated e.g. in Br\'{a}dler and Adami \cite{Bradler_Adami:2014} and references therein), the evolution of pure states into mixed states in \textit{open systems dynamics} (quantum channels) is in accord with standard quantum mechanics when one traces over the inaccessible degrees of freedom, typically referred to as the \textit{environment}. This produces a system evolution with loss (to the environment, modeled as an effective bath) via a quantum master equation \cite{Walls:1994}.
The evolution on the system (sans environment) is a
\textit{completely positive map}, mapping density matrices to density matrices (for details see e.g. Chapter 8 of \cite{Nielsen:2000}). In a \textit{purified} system that includes both system and environment, the evolution in the higher Hilbert space is unitary on a composite pure state. The inclusion of the quantized BH `pump source' mode in our trilinear Hamiltonian model then represents the simplest such purification of the Hawking radiation system
with its quantized source.

One of the primary motivations of this work is to re-investigate the channel capacity calculation of Adami and Ver Steeg in light of the non-thermal nature of the late-time joint quantum state of the evaporating BH  and the emitted Hawking radiation pairs.
At early stages of evolution when the `population' in the BH `pump source' is large (and effectively constant)
results should agree with standard treatments of Hawking radiation which focus primarily on the pair production process.
The joint quantum state of such a system factorizes into a product of the BH `pump source' state (usually taken to be a classical c-number source) and the entangled (across the horizon) correlated state of the emitted Hawking pairs.
However, at late-times, when the population in the emitted pairs becomes on the order of the population of the BH `pump source' the quantum state of the joint system no longer factorizes and new effects are anticipated.
The motivation of this work is to explore the consequences of these effects in a simple, zero-dimensional unitary model that inherently incorporates BH evaporation and hence, back-action effects.
\footnote{Previous work by
Nation and Blencowe \cite{Nation:2010} has also explored a model of an evaporating BH as PDC with depleted pump source. Those authors use a related but different analytic approach utilized here, and their work has less emphasis on entanglement and channel capacity issues than this present work.}
In this work we investigate the short time (non-evaporating) and long time (evaporating) regimes (analogous to a the non-depleted and depleted laser regimes, respectively in PDC) and its impact on the channel capacity, entanglement entropy across the event horizon, and the emergence of information from the BH at late-times. The new feature introduced here is the coupling of the emitted Hawking radiation modes through the common black hole `source pump' mode (which can be depleted), which introduces new entanglement effects that we explore.
%modeling coupled black hole particle production/evaporation.
%

Recently, there has intense renewed interest in entanglement issues across the event horizon as the BH evaporates. The central issue is most succinctly described in the introduction to the recent paper by Lloyd and Preskill \cite{Lloyd:2014,Mathur:2009} which we quote \textit{The crux of the puzzle is this: if a pure
quantum state collapses to form a black hole, the geometry of the evaporating black hole contains
spacelike surfaces crossed by both the collapsing body inside the event horizon and nearly all of the
emitted Hawking radiation outside the event horizon. If this process is unitary, then the quantum
information encoded in the collapsing matter must also be encoded (perhaps in a highly scrambled
form) in the outgoing radiation; hence the infalling quantum state is cloned in the radiation,
violating the linearity of quantum mechanics}. The authors go on to note that this puzzle has spawned many recent highly innovative ideas to rescue unitarity including BH complementarity \cite{Susskind:1993,Susskind:1994}, firewalls \cite{Almheiri:2013,Braunstein:2009,Braunstein:2013} and even wormholes \cite{Maldacena:2013}. In the majority of these approaches the Hawking radiation is canonically taken to be of the form $\sum_n \sqrt{p_n}\,\ket{n}_{int}\ket{n}_{ext}$ where Hilbert space of the BH is taken to be of the tensor product form ${\mathcal{H}} ={\mathcal{H}_{ext}}\otimes{\mathcal{H}_{int}}$ for the interior (int) and exterior (ext) of the BH. The action of evaporation is to move some subsystem from the black hole interior to the exterior \cite{Braunstein:2009,Braunstein:2013}
${\mathcal{H}_{int}} \rightarrow {\mathcal{H}_{bh}}\otimes{\mathcal{H}_{r}}$ via
$\ket{n}_{int}\rightarrow (U\ket{n}_{bh,r})$ where U denotes the unitary process that might be thought
of as "selecting" the subsystem to``eject." Here $\ket{n}_{int}$ is the initial
state of the black hole interior, $bh$ denotes the reduced size subsystem corresponding to the remaining interior after evaporation, and $r$ denotes the subsystem that escapes as radiation.

The central theme of this present work is to consider a model for BH particle production/evaporation as ${\mathcal{H}} ={\mathcal{H}_{bh}}\otimes {\mathcal{H}_{ext}} \otimes {\mathcal{H}_{int}}$ with the pure state wave function
$\ket{\psi} = \sum_{n=0}^{\nb0}\, c_n\, \ket{\nb0-n}_{bh}\ket{n_{e0}+n}_{ext}\ket{n}_{int}$ evolving unitarily under the trilinear interaction Hamiltonian
$H_{bh,ext,int} = r( a_{bh}\, a^\dagger_{ext}\,a^\dagger_{int}
+ a^\dagger_{bh}\, a_{ext}\,a_{int})$. Here $\ket{n}_{int}$ represents a state of $n$ emitted anti-particles in the interior of the BH, and $\ket{n_{e0}+n}_{ext}$ represents the state of $n$ emitted particles in the exterior of the BH when there were initially $n_{e0}$ particles present in the mode. The new feature introduced is $\ket{\nb0-n}_{bh}$ which phenomenologically models the quantized `state' (energy degree of freedom) of the BH containing $\nb0-n$ particles, when initially the BH contains $\nb0$ particles in this mode.
In the short-time regime when $\nb0 \gg n,n_{e0}$ one can approximate
$\ket{\nb0-n}_{bh}\sim\ket{\nb0}_{bh}$ and factor the wave function into the biseparable state
$\ket{\psi}\approx\ket{\nb0}_{bh}\otimes\ket{\psi}_{ext,int}$ where $\ket{\psi}_{ext,int}=\sum_{n=0}^{\nb0}\, c_n\,\ket{n_{e0}+n}_{ext}\ket{n}_{int}$ is the usual highly entangled state considered in the literature for the Hawking radiation. However, in the late-time regime when $n\sim\nb0 \gg n_{e0}$ and the BH is undergoing evaporation this factorization cannot be performed and the full state $\ket{\psi}$ is non-separable and exhibits additional intermodal entanglement. This (zero-dimensional) model then serves as a simplified arena to explore BH particle production/evaporation and back-action effects under an explicitly unitary evolution.

The model discussed above is well known in the quantum optics literature \cite{Nation:2010,Walls:1970,Bonifacio:1970,Walls:1994,Gerry:2004,Agarwal:2013} as the trilinear Hamiltonian describing parametric down conversion (PDC) via a nonlinear crystal where the pump is now quantized and can be depleted. In PDC a pump of frequency $\omega_p$ and wave vector $\vec{k}_p$ is converted into lower frequency particles, the signal of frequency/wavevector  $\omega_s, \vec{k}_s$ and the idler of frequency/wavevector $\omega_i, \vec{k}_i$ such that energy and momentum are conserved $\omega_p=\omega_s + \omega_i$, $\vec{k}_p = \vec{k}_s + \vec{k}_i$. If there are no initial particles in the signal and idler modes in the crystal (which is typically the case in laboratory experiments) the process is called spontaneous parametric down conversion (SPDC) and is the workhorse for generating entangled photons for
quantum optical information processing applications \cite{Kwiat:1994,Kwiat:1999,O'Brien:2007,Pan:2012,Fanto:2011,Peters:2012}.
The state $\ket{\psi}_{ext,int}$ is the well known and studied two-mode squeezed state vacuum.

The above trilinear PDC model is described and explored in more detail below. For now the correspondence between PDC and BH particle production that we exploit is $(pump, signal, idler) \leftrightarrow (bh, ext, int)$.
The work of Adami and Ver Steeg \cite{Adami:2014} (upon which this present work expands, see also Br\'{a}dler and Adami \cite{Bradler_Adami:2014}) treats $\nb0$ as constant and ignores the $bh$ modes, which is valid in the regime $n_{bh0}\gg n, n_{e0}$, and sufficient for the computations presented in that work. In quantum optics this is the "non-depleted pump" regime where the pump is treated as a classical c-number and absorbed into the Hamiltonian coupling constant.   In laboratory quantum optics experiments, such an approximation is eminently justifiable due to the small,  finite length  of the non-linear crystal used to generate the SPDC. The main focus of this paper is to include the effects of the quantized pump, which can then transfer particles unitarily into the signal and idler, and treat this as a model for unitary BH particle production/evaporation.
%
% $H_{bh,ext,int} = r( a_{bh}\, a^\dagger_{ext}\,a^\dagger_{int}
%+ a^\dagger_{bh}\, a_{ext}\,a_{int})$. Here $\ket{n}_{int}$
The first term $a_{bh}\, a^\dagger_{ext}\,a^\dagger_{int}$  of $H_{bh,ext,int}$ above represents the parametric down conversion process in which the BH mode $bh$ of frequency $\omega_{bh}$ produces correlated Hawking radiation pairs $ext$ and $int$ of lower energy
such that $\omega_{bh} = \omega_{int} + \omega_{ext}$. This term (with $a_{bh}$ treated as a constant c-number source) has been predominantly used in the literature (see \cite{Hawking:1975,Unruh:1976,Gerlach:1976,Giddings:1992,Mathur:2009}
and references therein) to investigate Hawking radiation production.
The (Hermitian conjugate) term $a^\dagger_{bh}\, a_{ext}\,a_{int}$ represents sum frequency generation by which the modes $int$ and $ext$ combine to yield a higher frequency $\omega_{int} + \omega_{ext}=\omega_{bh}$ in mode $bh$. $H_{bh,ext,int}$ represents the most general Hermitian form of energy/particle number conservation.
Though both process occur simultaneously, during the early stages of evolution, when $\nb0\gg n$, energy flows from the BH to the Hawking radiation modes such that $dn_{bh}/dt<0$, where $n_{bh} = \nb0 - n$ in our model. The model considered here is only valid in the regime $dn_{bh}/dt\le0$, thus prohibiting the unphysical reverse flow of Hawking radiation back into the BH. We will see below that this always occurs after the time for which
$n_{bh} = n_{ext}$ and so there exists a \textit{late-time} regime beyond the two-mode squeezed state vacuum commonly used to model the Hawking radiation.

%======================================
The outline of this paper is as follows.
In Section \ref{AvS:review} we review the work of Adami and Ver Steeg \cite{Adami:2014} and their main results for the classical cloning of infalling information in the stimulated emission of the BH radiation in the presence of the Hawking spontaneous emission.
In Section \ref{BHPRwDP} we define our first model and explore its short and long time behaviors.
Our first trilinear Hamiltonian models the BH with particles exterior to the BH event horizon and anti-particles inside the event horizon. We explore entanglement between various bipartite subdivisions of the global pure state.
In Section \ref{fullH} we include a second trilinear Hamiltonian which models anti-particles exterior to the event horizon and particles inside the event horizon. The full Hamiltonian is required in order to redo the (Holevo) channel capacity calculation of Adami and Ver Steeg. Derivation details are
relegated to \ref{appendix:fullH} for purpose of exposition.
In Section \ref{EntHpsibarHpsbari} we explore a new  entanglement feature that arises between an emitted particle/anti-particle (ext/int) pair with an emitted anti-particle/particle pair (ext/int) due to the common
BH `pump' source mode.
In Section \ref{HolevoCapacity} we redo the Adami and Ver Steeg's channel capacity calculation, now with a dynamically evaporating BH (the `depleted pump' regime).
In Section \ref{GrayBody} we follow Sorkin \cite{Sorkin:1987}, Adami and Ver Steeg \cite{Adami:2014} and
Br\'{a}dler and Adami \cite{Bradler_Adami:2014}, and compute gray body effects which model classical scattering of late infalling particles off the emitted outgoing BH radiation.
In Section \ref{Page} we re-examine, using our model, the conventionally held notion developed by Page \cite{Page:1993a,Page:1993b} that the information in the BH radiation essentially emerges when the BH has emitted half its particles, the \textit{Page time}. Our simplified unitary evaporating BH model re-produces Pages anticipated result dynamically (see also Nation and Blencowe \cite{Nation:2010}). We show that one can also interpret the Page time as that time in which the variances of the probability distributions for the evaporating BH and Hawking radiation become equal.
In Section~\ref{EntMonog} we examine entanglement between two modes $s$ and $c$ outside the BH event horizon that are initially maximally entangled, where one mode falls inward and interacts with the BH and the other remains forever outside and non-interacting. We show that the entanglement in the initially maximally entangled (outside) state
is degraded (though not completely destroyed) by the interaction with the BH with finite absorption.
(Related work for a perfectly absorbing and perfectly reflecting BH was performed recently by Br\'{a}dler and Adami \cite{Bradler_Adami:2014}). These results suggest the necessity for the concept of a BH firewall
to preclude quantum cloning as dispensable.

Finally, in Section \ref{Summary} we summarize our results and present numerical evidence that suggests the trilinear Hamiltonian model utilized in this work suggests the transference (at least partially) of information initially in the probability distribution of the BH `pump' source mode to the outgoing Hawking radiation modes at late-times.
We also indicate future directions that can be explored with this heuristic model for BH particle production/evaporation.

\section{Review of work by Adami and Ver Steeg (A\&VS)}
\label{AvS:review}
\begin{figure}[h]
\begin{center}
\includegraphics[width=2.5in,height=3.0in]{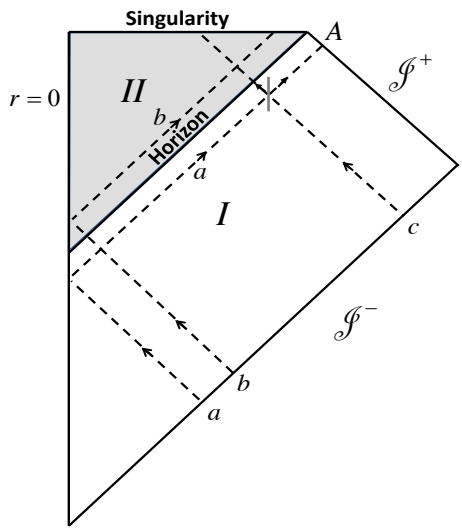}
\caption{\label{AvS:fig1} Penrose diagram for pair production in neighborhood of a black hole (adapted from Fig. 1 of Adami and Ver Steeg \cite{Adami:2014}). Modes $a$ and $b$ are early time infalling modes emanating from the far past impinging on matter collapsing into a black hole (BH). Near the BH event horizon mode $a$ lies just outside the BH exterior (region $I$, white), while mode $b$ lies just inside the BH interior (region $II$, gray). Mode $c$ is an infalling late time mode (after the BH formation) which can scatter off the region $I$ outgoing radiation $a$. This scattering is phenomenologically modeled as a 'beam splitter' (gray vertical line) interaction between modes $a$ and $c$.
}
\end{center}
\end{figure}

In the work of Adami and Ver Steeg (A\&VS) \cite{Adami:2014} the authors consider the regions $I$ (outside), and $II$ (inside) a Schwarzschild black hole as illustrated in \Fig{AvS:fig1}. Here the operators that annihilate the curved spacetime vacuum at late time ${\mathcal{J_+}}$ are given by the Bogoliubov transformation
%
%\begin{subequations}
\bea{AvS:eq1_2}
A_k            & = & e^{-i H} a_k e^{i H}       =   \hspace{0.75em} \cosh r_k \, a_k           - \sinh r_k \, b^{\dagger}_{-k}, \\
B_{k}  & = &  e^{-i H} b^{\dagger}_{-k} e^{i H}  =  -\sinh r_k \, a_{k} + \cosh r_k \, b^{\dagger}_{-k},
\eea
%\end{subequations}
%
where $a^{\dagger}_k$ creates a particle, and $a_{-k}$ annihilates an antiparticle of mode $k$ in region $I$,  while $b^{\dagger}_{-k}$ creates a antiparticle, and $b_{k}$ annihilates a particle of mode $k$ in region $II$.
Here $r_k$ is the rapidity of the local Lorentz transformation about which curved spacetime metric is linearized for a stationary observer with $\tanh r_k = \exp[-\omega_k/(2T)]$ with $\omega_k=|k|c$ and $T = \hbar\,(\kappa/c)/(2\pi)$ is the temperature of the black hole. The hyperbolic functions ensure the `out' mode commutation relations are preserved: $[A_k, A^{\dagger}_{k'}] = \delta_{k,k'}$,
$[B_k, B^{\dagger}_{k'}] = \delta_{k,k'}$, and $[A_k, B_{k'}] = 0$. The quantity $\kappa=G M /R_s^2 = c^4/(4 G M)$ is the surface gravity  of the black hole of Schwarzschild radius $R_s=2G M_{bh}/c^2$ and hence a measure of the local acceleration of the stationary observer. For the related case of the observer undergoing constant acceleration $a$ in the flat Minksowski spacetime, the Unruh temperature is given by analogous expression $T_U =\hbar (a/c)/(2\pi)$, (for a recent review, see \cite{Alsing:2012} and references therein).

The Bogoliubov transformation \Eq{AvS:eq1_2} provides a mapping from the \textit{Boulware} `in'-vacuum $\ket{0}_{in} = \ket{0}_a \ket{0}_b$ to the Unruh `out'-vacuum $\ket{0}_{out}$ using the $S$-matrix $U=e^{-i H}$ with
\be{AvS:eq3}
\ket{0}_{out} = e^{-i H} \,\ket{0}_{in}
\ee
with the two-mode `squeezing' Hamiltonian \cite{Walls:1994,Gerry:2004,Agarwal:2013}
\be{AvS:eq4}
H = \sum_{k=-\infty}^{\infty} H_k = \sum_{k=-\infty}^{\infty} i r_k ( a^{\dagger}_{k}\, b^{\dagger}_{-k} - a_{k} \, b_{-k} ).
\ee
The vacuum state is given by
\bea{AvS:eq8_12}
\fl \ket{0}_{out} &=& \prod_{k=-\infty}^{\infty} \, e^{r_k ( a^{\dagger}_{k}\, b^{\dagger}_{-k} - a_{k} \, b_{-k} )} \, \ket{0}_{out} \no
\fl &=& \prod_{k} \left[ \frac{1}{\cosh r_k} \sum_{n_k} \, \tanh^{n_k} r_k   \ket{n_{k}}_a \ket{n_{-k}}_b  \right] \,
              \left[ \frac{1}{\cosh r_k} \sum_{n'_k} \, \tanh^{n'_k} r_k  \ket{n'_{-k}}_a \ket{n'_{k}}_b \right],
\eea
where in the left square bracket
$\ket{n_k}_a = \displaystyle{\frac{(a^\dagger_k)^{n_k}}{\sqrt{n_k !}}} \, \ket{0}_a$ represents an $n$-particle state of mode $k$  just outside ($a$, region $I$) the BH, and $\ket{n_{-k}}_b$ an $n$-antiparticle state of mode $k$ just inside ($b$, region $II$) the BH, and the right square brackets swaps the particle-antiparticle labels.
The above state is most readily computed from the (normally ordered form) disentangling theorem \cite{Perelomov:1986} for the two boson mode Schwinger representation of $SU(1,1)$ given by \cite{Perelomov:1986,Agarwal:2013}
%\begin{subequations}
\bea{su11DisEntThm}
S_{su(1,1)}(\xi) &\equiv e^{\xi K_- - \bar{\xi} K_+} \qquad & = e^{\zeta K_+} \, e^{-\eta K_0} \, e^{\zeta' K_-}, \\
\zeta &= e^{i \phi} \, \tanh |\xi|, \quad \eta              & = \ln\cosh^2|\xi| = -\ln(1-|\zeta|^2), \quad \zeta' = -\zeta^*, \,\quad\\
K_+   & = a^{\dagger}_{k}\, b^{\dagger}_{-k},
 \quad\qquad K_- & = a_{k}\, b_{-k},\,\, K_0 = \frac{1}{2} ( a^{\dagger}_{k}\, a_{k}  + b^{\dagger}_{-k}\, b_{-k} + 1),\\
 \left[ K_0, K_\pm \right]  & =  \pm K_\pm,  \quad  \left[ K_-, K_+ \right] &= 2 K_0.
\eea
%\end{subequations}

The density matrix for the outgoing radiation appropriate for region $I$ is given by
\bea{AvS:eq13}
\fl \rho^{I}_a &=& Tr_{II}[\,\ket{0}_{out}\bra{0}\,] \no
\fl &=& \prod_{k}\, \left[ \frac{1}{\cosh^2 r_k} \sum_{n_k} \, \tanh^{2 n_k} r_k \, \ket{n_{k}}_a\bra{n_{k}}  \right] \, \left[ \frac{1}{\cosh^2 r_k} \sum_{n'_k} \, \tanh^{2 n'_k} r_k \, \ket{n'_{-k}}_a\bra{n'_{-k}}  \right], \\
\fl &\equiv & \prod_{k} \rho_{k}\otimes\rho_{-k},
\eea
where it is useful to note that $1/\cosh^2 r_k = (1-\tanh^2 r_k)$.
Here $\rho_{k}$ is the region $I$ density matrix for outgoing particles and
and $\rho_{-k}$ is the region $I$ density matrix for outgoing anti-particles.
The mean number of particles emitted in region I (mode $a$) is calculated as
\be{AvS:15}
\fl \langle N_a \rangle = Tr[ a^\dagger_k a_k \rho_a^I ] = \sum_{k=-\infty}^{\infty} {}_{out}\bra{0} a^\dagger_k a_k \ket{0}_{out}
= \sum_{k=-\infty}^{\infty} \sinh^2 r_k = \sum_{k=-\infty}^{\infty} \frac{ e^{-\omega_k/T} }{1 - e^{-\omega_k/T}},
\ee
where the last term contains the famous Planck distribution of the Hawking radiation.

The von Neumann entropy of the frequency mode $k$ of the outside or inside region can be computed as
\bea{AvS:17}
S(\rho_k) &=& -Tr[\rho_k \log\rho_k] = \frac{\omega/T}{e^{\omega/T}-1} + \ln(1 - e^{-\omega/T}) \no
 &=& - \sinh^2 r_k \, \ln(\tanh^2 r_k)\,  - \ln(1-\tanh^2 r_k), \\
 &\equiv& -\frac{z_k}{1-z_k} \ln z_k -\ln(1-z_k), \nonumber
\eea
where it is conventional to define the parameter $z_k = \tanh^2 r_k$.
Stimulated emission in frequency mode $k$ from the BH can be treated by considering $m$ initial particles in (say) mode $a$, and no antiparticles in mode $b$
\be{AvS:18}
\fl \ket{\psi}_{out} = e^{-i H_k}\,\ket{m}_a \ket{0}_b
= \frac{1}{(\cosh r_k)^{m+1}} \, \sum_{n=0}^{\infty} \tanh^{n} r_k
\binom{m+n}{n}^{1/2}\, \ket{m+n}_a\ket{n}_b
\ee
which is readily computed with the disentangling theorem \Eq{su11DisEntThm}, since $e^{\zeta' K_-} = e^{\zeta' a_k b_{-k}}$ acts trivially (as the identity) on $\ket{\psi}_{in} = \ket{m}_a \ket{0}_b$.
Again the region $I$ density matrix for the outgoing radiation can be easily computed to be
\be{AvS:eq19}
\rho^{I}_a = Tr_{II}[\,\ket{0}_{out}\bra{0}\,] = \rho_{k|m} \otimes \rho_{-k|0} ,
\ee
where $\rho_{-k|0}$ is the density matrix of anti-particles given that zero anti-particles were present in mode $k$ initially, and
$\rho_{k|m}$ is the density matrix of particles given that $m$ particles were initially present:
\be{AvS:20}
\rho_{k|m} = \frac{1}{(\cosh^2 r_k)^{m+1}} \, \sum_{n=0}^{\infty} \tanh^{2n} r_k
\binom{m+n}{n} \,\ket{m+n}\bra{m+n}.
\ee
It is useful to note the binomial identity \cite{Sorkin:1987,Abramowitz:1972}
$\sum_{n=0}^\infty z^n \binom{m+n}{n} = (1-z)^{-(m+1)}$.
The mean number of particles in region $I$ is given by
\bea{AvS:24}
\fl \langle N_I \rangle = Tr[ a^\dagger_k a_k \rho_a^I ] =  {}_{out}\bra{\psi} a^\dagger_k a_k \ket{\psi}_{out}
&=& {}_{in}\bra{\psi} \cosh^2 r_k a^\dagger_k a_k + \sinh^2 r_k \ket{\psi}_{in} \no
&=&  \cosh^2 r_k \, m + \sinh^2 r_k,
\eea
which is interpreted as $m$ incident particles having stimulated emission of $\cosh^2 r_k \, m$ particles outside the horizon in region $I$, in addition to the $\sinh^2 r_k$ spontaneously emitted particles (Hawking radiation) due to the vacuum $\ket{0}_a\ket{0}_b$. Since particle number is conserved, $\sinh^2 r_k \, m$ anti-particles are simultaneously stimulated behind the horizon in region $II$.
Adami and Ver Steeg \cite{Adami:2014} note that because the incident particles carry energy and momentum, the BH does not have to donate mass in order to allow the emission of stimulated pairs, as it does for virtual pairs. We will return to this point in the next section.

One of the main questions addressed by Adami and Ver Steeg \cite{Adami:2014} is whether or not the particles stimulated in region $I$ carry the information that was inherent in the particles that were absorbed during the formation of the black hole. To answer this question the authors compute the Holevo capacity \cite{Schumacher:1997,Nielsen:2000} of the stimulated emission in the presence of the noisy background of the spontaneous emission. The authors imagine a dual rail encoding of classical information at early times  on $\mathcal{J_-}$ in which a logical `0' is represented by a particle launched towards the forming black hole, and a logical `1' by an anti-particle. The relevant question at hand is whether or not an observer measuring particles and anti-particles at late time on $\mathcal{J_+}$ can determine the classical message encoded by the preparer. The answer is affirmative if the  Holevo capacity is non-zero, which they find is the case, even for infinite rapidity $r_k$.

The relevant quantity to compute is the mutual information \cite{Nielsen:2000} between the preparer $X$ and the radiation field outside the black hole (in region $I$):
\be{AvS:25}
H(X:I) = S(\rho_I) + S(\rho_X) - S(\rho_{I,X}).
\ee
Here $S(\rho_I)$ is the von Neumann entropy of the reduced density matrix $\rho_I$ for the radiation field in region $I$, $S(\rho_X)$ is the von Neumann entropy of the preparer, and $S(\rho_{I,X})$ is the joint von Neumann entropy between the preparer and the region $I$ radiation field. The capacity of the channel $\chi$ is the maximum of the shared information over the probability distribution of the signal states \cite{Schumacher:1997},
\be{AvS:26}
\chi = \myover{{\textrm{max}} p}H(X:I).
\ee
Below we outline the calculation of $\chi$.

If the preparer sends the state `0' with probability $p$ and state `1' with probability $1-p$ the preparer's entropy is given by Shannon entropy (in bits)
\be{AvS:27}
S(\rho_X) = -p \log_2 p - (1-p) \log_2(1-p) \equiv H(p),
\ee
which is maximized at one bit for $p=1/2$.
Here $H(p) = -\sum_n \, p_n\,\log_2 p_n$ is the Shannon entropy (in bits) of the probability distribution $\{p_n\}$.

The density matrix $\rho_I$ in region $I$ is given by the probabilistic mixture of the density matrices $\rho_I(0)$ that describes  `0' being sent and $\rho_I(1)$ for a  `1' being sent:
\be{AvS:30}
\rho_I(p) = p \rho_I(0) + (1-p) \rho_I(1).
\ee
Here $\rho_I(0) = \rho_{k|1} \otimes \rho_{-k|0}$
describes one ($m=1$) particle sent, with zero ($m=0$) anti-particles sent, while
$\rho_I(1) = \rho_{k|0} \otimes \rho_{-k|1}$
describes one ($m=1$) anti-particle sent, with zero ($m=0$) particles sent.

\begin{figure}[h]
\begin{center}
\includegraphics[width=3.5in,height=2.0in]{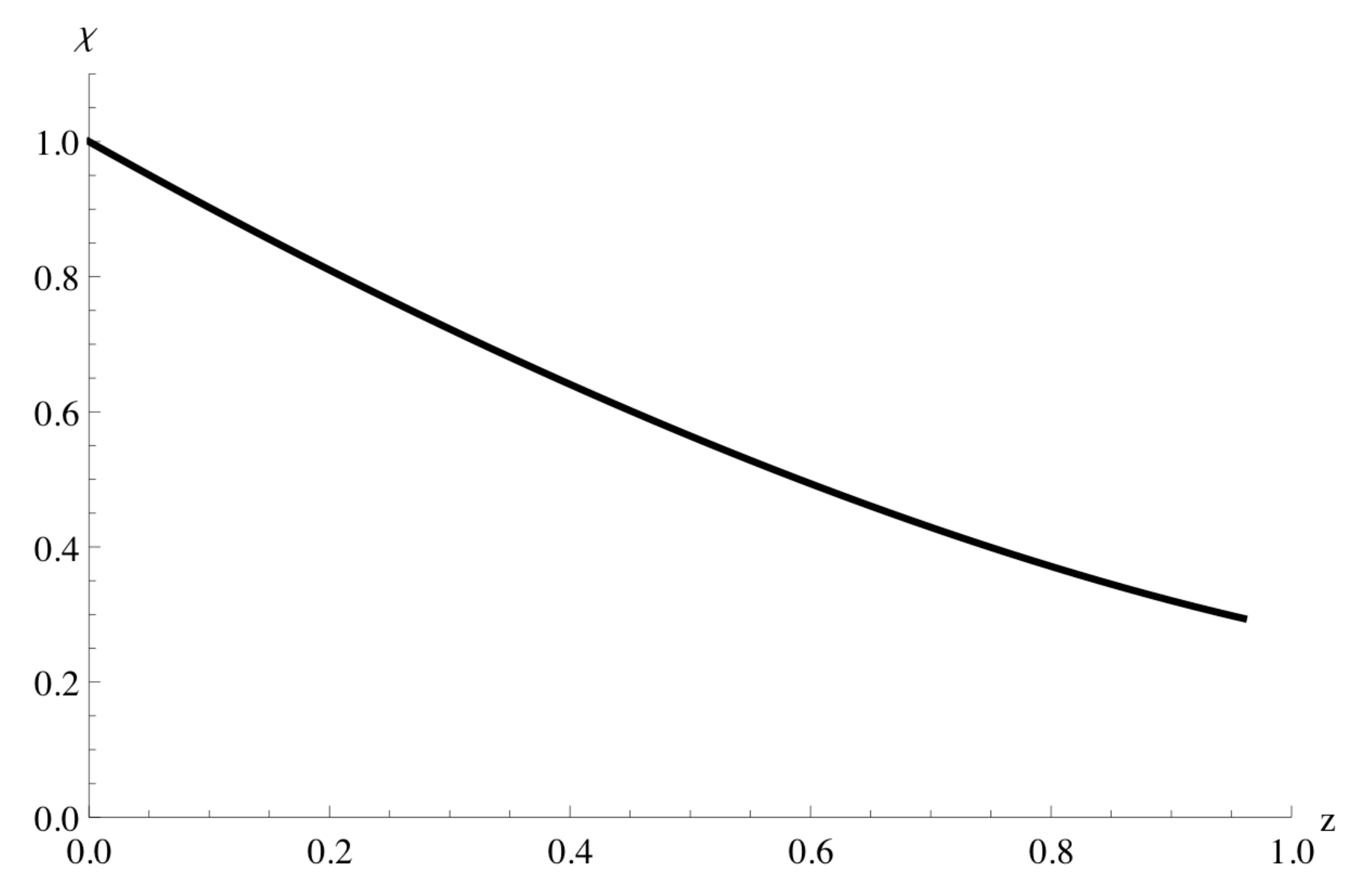}
\caption{\label{AvS:fig2} Channel capacity $\chi$ \Eq{AvS:40} as a function of $z=\tanh^2 r_k$
(Fig. 2 of Adami and Ver Steeg \cite{Adami:2014}).}
\end{center}
\end{figure}

Lastly,
\be{AvS:31}
\rho_{I,X} = \sum_{i=0}^1 p(i) \rho_I(i) \otimes \ket{i}\bra{i},
\ee
is the joint, block-diagonal density matrix
between the region $I$ sent states $\rho_I(i)$ and the preparer's basis states $\ket{i}$.
From the definition of the von Neumann entropy and the block-diagonal structure of \Eq{AvS:31} one can easily compute
\be{AvS:32}
S(\rho_{I,X}) = H(p) +  \sum_{i=0}^1 p(i) S(\rho_I(i)).
\ee
%where $H(p) = -\sum_n \, p_n\,\log_2 p_n$ is the Shannon entropy (in bits) of the probability distribution $\{p_n\}$.
This expression is further reduced by using the factorization of the particle and anti-particle density matrices
(which are pure tensor products) yielding
\bea{AvS:33}
S(\rho_I(0)) &=& S(\rho_{k|1}) + S(\rho_{-k|0}), \no
S(\rho_I(1)) &=& S(\rho_{k|0}) + S(\rho_{-k|1}),
\eea
where the individual terms on the right hand side of \Eq{AvS:33} can be readily computing using the
probabilities from the diagonal density matrices for $m=0$ and $m=1$ using \Eq{AvS:20},
%\begin{subequations}
\bea{AvS:34:35}
\fl S(\rho_{k|0}) &=& S(\rho_{-k|0}) =  -\log_2(1-z) - \frac{z}{1-z}\,\log_2 z,\\
\fl S(\rho_{k|1}) &=& S(\rho_{-k|1}) = -2 \log_2 (1-z) - \frac{2z}{1-z}\,\log_2 z - (1-z)^2 \sum_{n=0}^\infty \, z^n\,(n+1)\log_2 (n+1). \quad
\eea
%\end{subequations}
Since the density matrix \cite{Adami:2014}
\be{AvS:38}
\rho_I(p) = (1-z)^3 \sum_{n,n'=0}^{\infty} z^{n+n'+1} \, (p n' + (1-p) n) \, \ket{n,n'}\bra{n,n'},
\ee
is symmetric under the interchange of $p\rightarrow 1-p$ the mutual information $H(X:I)$ is maximized at $p=1/2$, at which $S(\rho_I(p))$ is evaluated.
The resulting channel capacity has the form
\bea{chi:formula}
\chi &=& S \big( p\,\rho_{k|1}\otimes\rho_{-k|0} + (1-p)\,\rho_{k|0}\otimes\rho_{-k|1}\big) \no
&-& p\,\left( S(\rho_{k|1}) + S(\rho_{-k|0}) \right)
- (1-p)\,\left( S(\rho_{k|0}) + S(\rho_{-k|1}) \right),
\eea
which is maximized over $p$ at the value $p=1/2$.
An expression for $\chi$ is readily computed in closed form and given by \cite{Adami:2014} (with $z\equiv\tanh^2 r_k$ for mode $k$)
\be{AvS:40}
\fl \chi(z) = 1 -\frac{1}{2} (1-z)^3 \sum_{n=0}^{\infty} z^n (n+1) (n+2) \log(n+1)
+ (1-z)^2 \sum_{n=0}^{\infty} (n+1) \log(n+1),
\ee
which is plotted in \Fig{AvS:fig2}.
The important point to note in \Fig{AvS:fig2} is that the channel capacity $\chi$ is non-zero for $z=1$, or infinite rapidity (acceleration) $r_k\rightarrow\infty$, or in the case of a black hole, infinite surface gravity.

The work discussed above dealt with `early-time' modes and indicates that classical information is not lost just before the formation of the BH. To determine the fate of information after the formation of the BH, Adami and Ver Steeg utilized the approach of `late-time' modes introduced by Sorkin \cite{Sorkin:1987}. In \Fig{AvS:fig1} mode $c$ is the strongly blue-shifted mode relative to the early-time modes $a$ and $b$, the particles and anti-particles  introduced above that propagate just outside and inside the BH horizon respectively. Because of this blue-shift of mode $c$, modes $a$ and $b$ are unexcited by mode $c$ and as a consequence commute with it. The particle mode $c$ can scatter off of mode $a$ and introduce so called 'gray-body' factors. The mode $A_k$ that escapes to infinity on ${\mathcal{J_+}}$ in \Fig{AvS:fig1} is given by
\bea{AvS:41}
A_k = e^{-i {\mathcal{H}_k}} a_k e^{i {\mathcal{H}_k}} &=& \alpha_k \, a_k - \beta_k \,b^\dagger_{-k} + \gamma_k \, c_k \no
&=& \cosh r_k \, \cos\theta_k \, a_k - \sinh r_k \,b^\dagger_{-k} + \cosh r_k \, \sin\theta_k \, c_k,
\eea
such that $\alpha^2_k - \beta^2_k + \gamma^2_k = 1$.
 The transformation in \Eq{AvS:41} can be produced by \cite{Adami:2014} combining the previous squeezing Hamiltonian $H_k$ \Eq{AvS:eq4} with the `beam splitter' Hamiltonian  $H'_k$ \cite{Walls:1994,Gerry:2004,Agarwal:2013}
\bea{AvS:42}
{\mathcal{H}_k} &=& H_k + H'_k, \no          %\sum_{k=-\infty}^{\infty}
  &=&  i r_k ( a^{\dagger}_{k}\, b^{\dagger}_{-k} - a_{k} \, b_{-k} ) + i \theta_k (a^{\dagger}_{k}\, c_k - a_{k} \, c^{\dagger}_k).
\eea
Note that $H_k$ and $H'_k$ in \Eq{AvS:42} do not commute and also imply that the scattering between modes $a$ and $c$ at late times is occurring commensurately with the particle anti-particle production via $H_k$.
Alternatively, one could also effect the transformation in  \Eq{AvS:42} by the successive (in time) transformations
%$H'_k = i \theta_k (a^{\dagger}_{k}\, c_k - a_{k} \, c^{\dagger}_k)$
\be{AvS:eq42_v2}
A_k = e^{-i H'_k} \left( e^{-i H_k} a_k e^{i H_k} \right) e^{i H'_k}
\ee
where the `beam-splitter' Hamiltonian $H'_k$ effects the rotation via \cite{Gerry:2004,Agarwal:2013}
%\begin{subequations}
\bea{BStransf}
e^{-i H'_k} a_k e^{ i H'_k} &=& \cos\theta_k \, a_k + \sin\theta_k c_k, \\
e^{-i H'_k} c_k e^{ i H'_k} &=& \cos\theta_k \, c_k + \sin\theta_k a_k.
\eea
This successive evolution is what occurs naturally in laboratory quantum optical experiments.
%\end{subequations}
State evolution can then be readily computed by successively applying the $SU(1,1)$ disentangling theorem of \Eq{su11DisEntThm}
and the corresponding disentangling theorem for $SU(2)$ \cite{Perelomov:1986}
%\begin{subequations}
\bea{su2DisEntThm}
S_{su(2)}(\xi) &\equiv e^{\xi J_- - \bar{\xi} J_+} \qquad & = e^{\zeta J_+} \, e^{-\eta J_0} \, e^{\zeta' J_-}, \\
\zeta &= -e^{i \phi} \, \tan |\xi|, \quad \eta &= \ln\cos^2|\xi| = -\ln(1+|\zeta|^2), \quad \zeta' = -\zeta^*, \\
J_+ &= a^{\dagger}_{k}\, c_{k}, \quad\qquad J_- &= a_{k}\,  c^{\dagger}_{k},
\quad J_0 = \frac{1}{2} ( a^{\dagger}_{k}\, a_{k}  + c^{\dagger}_{k}\, c_{k} + 1), \\
 \left[ J_0, J_\pm \right]  &=  \pm J_\pm,  \quad  \left[ J_-, J_+ \right] &= -2 J_0.
\eea
%\end{subequations}
Under $H'_k$ Fock states $\ket{n}_a\ket{n'}_c$ are transformed into states of the form $\sum_{p=0}^{n+n'}\, f_p \,\ket{p}_a\ket{n+n'-p}_c$ since the total particle number $a^\dagger a_k + c^\dagger c_k$ is preserved.
Here $f_p$ is given by \cite{Agarwal:2013} $f_p = \sum_{q=0}^{n}\,\sum_{q'=0}^{n'} \, \delta_{p,q+q'}\binom{n}{q}\binom{n'}{q'}\big((q+q')! (n+n'-q-q')!/(n! n'!)\big)^{1/2} (\cos\theta)^{n'+q-q'} \, (\sin\theta)^{n-q+q'}(-1)^{n-q}$.

Considering the case of $m$ incident particles in the late-time mode $c$, $\ket{\psi}_{in}=\ket{0}_a \ket{0}_b \ket{m}_c$, the
evolution  $\ket{\psi}_{out} = e^{-i H'_k} e^{-i H_k}\ket{\psi}_{in}$ produces the number of particles emitted into the outgoing mode $a$ as
\bea{AvS:52}
\langle N_I \rangle =  {}_{out}\bra{\psi} a^\dagger_k a_k \ket{\psi}_{out}
&=& \gamma_k^2 m + \beta^2_k, \no
&=&  \cosh^2 r_k \sin^2\theta_k\, m + \sinh^2 r_k.
\eea
Thus, in addition to the spontaneously emitted particles $\beta^2_k$ that will be detected at future null infinity ${\mathcal{J_+}}$, $\gamma^2_k m \equiv (1-\alpha^2_k) m + \beta^2_k m$ particles will also be detected. This result can be interpreted  \cite{Adami:2014} as $(1-\alpha^2_k) m$ particles arising  due to the elastic scattering of mode $a$ with mode $c$ with an absorption probability $\alpha^2_k$ along with the usual $\beta^2_k m$ particles due to spontaneous emission, as before. A calculation of the channel capacity in the presence of this additional late-time scattering off the BH (gray-body factors) produces \cite{Adami:2014,Bradler_Adami:2014}, for the case of $r_k = \theta_k$, the value $\chi\big(z/(1+z)\big)$, i.e. the same functional form for the channel capacity in \Eq{AvS:40}, with $z\rightarrow z/(1+z)$, which at a given value of $z$, increases the value of $\chi$ over the previous case.

%===========================================================
\section{Black hole particle production with the gravitational field modeled as a `depleted pump'}\label{BHPRwDP}

For the rest of this work, we wish to make the analogy of particle emission in the presence of the BH with the quantum optical process of parametric down conversion with a depleted (vs the usual non-depleted) pump.
In this analogy, the gravitational field (mass) of the BH plays the role of the `laser pump' while the Hawking radiation plays the role of the spontaneously emitted signal and idler bosons (`photons') such that energy is conserved
$\omega_p = \omega_s + \omega_{\bar{i}}$. Here we have introduced the quantum optical notation $p$ for pump to represent an idealized quantized gravitational energy mode of the BH (to enforce energy conservation), $s$ for the region $I$ signal boson, which we will take as the particle (mode $a$ of wave vector $k$ in the above discussion), and $\bar{i}$ as the (`anti'-)idler boson, which we will take as the region $II$ anti-particle (mode $b$ of wavevector $-k$). We will also use the notation of $\bar{s}$ and $i$ for the (region $I$, anti-particle) signal and (region $II$ particle) idler bosons of wavevector $k$ and $-k$, respectively. Thus, the modes $(s,\bar{s})$ propagate just outside the BH, while $(i,\bar{i})$ propagate just inside the BH, and the `pump' $p$ models the BH source. The overbar indicates anti-particles, though in the usual laboratory case of PDC with photons, no such anti-particles are present. Henceforth we continue to use the `pump, signal, idler' nomenclature of the photon-based laboratory PDC case in order to reinforce the analogy with the BH particle/anti-particle emission.
The Hamiltonian (for the single mode $k$) is now of the form
\bea{fullHamil}
\mathcal{H} &=& H_{p,s,\bar{i}} + H_{p,\bar{s},i}, \no
&=& r (a_p \, a_s^\dagger \, a_{\bar{i}}^\dagger + a_p^\dagger \, a_s \, a_{\bar{i}} ) + r' (a_p \, a_{\bar{s}}^\dagger \, a_i^\dagger + a_p^\dagger \, a_{\bar{s}} \, a_i ).
\eea
We will investigate the two cases when (i) for early times the pump occupation number $n_p$ is very much greater than the emitted particle/anti-particle occupation numbers $n_s, n_{\bar{i}}$ and $n_{\bar{s}}, n_i$, and (ii) late-times when these later boson occupation numbers are on the order of $n_p$. While all boson modes commute $[a_i,a^\dagger_j] = \delta_{i,j}$, for $i,j \in \{p,s,\bar{i},\bar{s},i\}$, the region $I//II$ particle/anti-particle $H_{p,s,\bar{i}}$ and anti-particle/particle $H_{p,\bar{s},i}$ Hamiltonians do not commute $[H_{p,s,\bar{i}},H_{p,\bar{s},i}]\ne 0$ due to the coupling of the emitted modes $\{s,\bar{i},\bar{s},i\}$ with the pump mode.

We analyze this model in stages. We will first look at the trilinear Hamiltonian
$H_{p,s,\bar{i}}$ for particle/anti-particle creation in region $I$/$II$ just outside/inside the BH respectively,
and the influence of the pump on early and late times before we  turn to analysis of the full Hamiltonian in \Eq{fullHamil}.
We will explore the entanglement created between various bipartite division of our system in two subsystems before turning our attention to the exploration of the influence of the pump (BH) mode on the channel capacity. Lastly we explore the role of gray-body factors on the channel capacity, and entanglement issues with a maximally entangled initial state.

\subsection{The trilinear Hamiltonian $H_{p,s,\bar{i}}$}
The investigation of the trilinear Hamiltonian
\be{trilinearHamil}
H_{p,s,\bar{i}} = r (a_p \, a_s^\dagger \, a_{\bar{i}}^\dagger + a_p^\dagger \, a_s \, a_{\bar{i}} )
\ee
for spontaneous parametric down conversion (SPDC) (as well as emission from superradiant Dicke-states) has a long history dating back to works of Walls and Barakat \cite{Walls:1970} and Bonifacio and Preparata \cite{Bonifacio:1970} in 1970 and continuing today (for a recent review see \cite{Bandilla:2000} and references therein). There are many approaches to handling the trilinear Hamiltonian \Eq{trilinearHamil}. Walls and Barakat took a numerical approach by investigating the eigenvalues of the  tridiagonal matrix representation of $H_{p,s,\bar{i}}$ in the computational basis states $\ket{n}_L \equiv\ket{n_{p0}-n}_p\ket{n}_s\ket{n}_{\bar{i}}$, where $L$ indicates the `logical basis' state and $n_{p0}$ is the initial number of particles (photons) in the pump mode. Bonifacio and Preparata used the two-boson mode Schwinger representation of $SU(2)$ given by \Eq{su2DisEntThm} with $J^{(p\bar{i})}_+ = a_p^\dagger \, a_{\bar{i}}$ to convert \Eq{trilinearHamil} to the spin-boson Hamiltonian
$H_{p,s,\bar{i}} = r (J^{(p\bar{i})}_+ \, a_s + J^{(p\bar{i})}_- \, a_s^\dagger)$. Note that there is no closed form disentangling theorem for this Hamiltonian since the individual terms $J^{(p\bar{i})}_+ \, a_s$ and $J^{(p\bar{i})}_- \, a_s^\dagger$ do not form the raising or lower operators of a Lie group with a finite number of generators (or commutators). The authors develop differential-difference equations for the quantum amplitudes
$c_n(t) = {}_L\bra{n} e^{-i H_{p,s,\bar{i}} t}\ket{\psi}_{in}$  (with $\ket{\psi}_{out} = \sum_{n=0}^\infty \, c_n \ket{n}_L$) and examine the behavior in the short time (non-depleted pump) and long-time (depleted pump) regimes.
We develop and adapt the methodology of Bonifacio and Preparata first for the trilinear Hamiltonian \Eq{trilinearHamil}, and subsequently for the full Hamiltonian \Eq{fullHamil}.

The trilinear Hamiltonian in \Eq{trilinearHamil} can also be considered as
\be{trilinearapK}
H_{p,s,\bar{i}} = r (a_p^\dagger \, K^{(s\bar{i})}_-  + a_p \, K^{(s\bar{i})}_+),
\ee
with $K^{(s\bar{i})}_+ = a_s^\dagger \, a_{\bar{i}}^\dagger$
(this is also the form considered by Nation and Blencowe \cite{Nation:2010}).
Here we consider the logical states
\be{logicalstates}
\ket{n}_L \equiv\ket{n_{p0}-n}_p\ket{n_{s0}+n}_s\ket{n}_{\bar{i}}, \quad \ket{\psi}_{in} = \ket{0}_L = \ket{n_{p0}}_p\ket{n_{s0}}_s\ket{0}_{\bar{i}},
\ee
where the initial state $\ket{\psi}_{in} = \ket{0}_L$ contains $n_{p0}$ pump particles and $n_{s0}$ signal particles with $n_{p0}\gg n_{s0}$.

Here a few words are in order to justify the form of our initial state in \Eq{logicalstates}. Since we consider the initial state of the outgoing radiation field modes $(s,\bar{i})$ to have very small occupation number it is reasonable to model them starting in the Fock state $\ket{n_{s0}}_s\ket{0}_{\bar{i}}$. For the case of the BH 'pump' source with a very high initial occupation number $n_{p0}\gg n_{s0}$ it might be more appropriate to model its initial state as the
classical-like coherent state \cite{Walls:1994,Loudon:1983} (appropriate for a laser)
$\ket{\alpha}_p = e^{-|\alpha|^2/2}\,\sum_{m=0}^\infty \alpha^m/\sqrt{(m!)}\,|m>_p$
vs. the highly non-classical Fock number state $\ket{n_{p0}}_p$. The later has the property that
${}_p\bra{\np0} (a^\dagger_p + a_p)/2  \ket{\np0}_p = 0$, while the former has the property
$a_p\,\ket{\alpha}_p = \alpha \,\ket{\alpha}_p$ with $\alpha = \sqrt{n_{p0}}$ such that
${}_p\bra{\alpha} (a^\dagger_p + a_p)/2  \ket{\alpha}_p = \textrm{Re}(\alpha)$.
However, we choose the initial BH 'pump' source to be the Fock state $\ket{n_{p0}}_p$ both for computational convenience and because it more clearly elucidates the essential features of the BH - outgoing radiation modes $(p) - (s,\bar{i})$ interactions. Later in Section~\ref{Page} we present results and numerical simulations in which the initial BH state is modeled as the coherent state $\ket{\alpha}_p$ (with the $(s,\bar{i})$ in the vacuum state $\ket{0}_s\ket{0}_{\bar{i}}$), which we then interpret in terms of the collection of Fock states under the coherent state (Poisson) probability distribution.

To develop a differential difference equation for the quantum amplitudes we compute  $i \dot{c}_n(t) = {}_L\bra{n} H_{p,s,\bar{i}} e^{-i H_{p,s,\bar{i}} t}\ket{\psi}_{in}$.
We note that a discrete series representation of $SU(1,1)$ is defined \cite{Perelomov:1986} by the set of states $\ket{\kappa; \kappa + n}$, $n=\{0,1,\ldots,\infty\}$. Here $\kappa$
is the constant defined by the $SU(1,1)$ Casimir operator $-1/4 + 1/4 ( \ldots ) = \kappa(\kappa-1) I$ where $I$ is the identity operator.
The $SU(1,1)$ generators $K^{(s\bar{i})}_\pm$ and $K^{(s\bar{i})}_0$ act on these basis states as follows (in analogy with the usual $SU(2)$ angular momentum operators $J_\pm$ and $J_0$) as
%\begin{subequations}
\bea{KRaisingLoweringOps}
K_- \, \ket{\kappa; \kappa} &=& 0, \\
K_+ \, \ket{\kappa; \kappa+n} &=& \sqrt{(n+1) (2k+n)}   \, \ket{\kappa; \kappa+n+1},  \\
K_- \, \ket{\kappa; \kappa+n} &=& \sqrt{ n\, (2k+n-1)} \, \ket{\kappa; \kappa+n-1}.
\eea
%\end{subequations}
In the basis of two-boson Fock states $\{\ket{m}\ket{n}\}$ we have $\ket{\kappa; \kappa+n} = \ket{n_0+n}\ket{n}$ with $\kappa=1/2(1+n_0)$ reflecting the fact that $SU(1,1)$ preserves the difference of the number of bosons in the two modes.
For our case of interest $\ket{\kappa; \kappa+n} = \ket{n_{s0}+n}_s\ket{n}_{\bar{i}}$ with $n_0 = n_{s0}$, the initial number of signal particles present. With the standard Heisenberg algebra (simple harmonic oscillator) raising and lowering operations $a \ket{n} = \sqrt{n}\,\ket{n-1}_p$ and $a^\dagger \ket{n} = \sqrt{n+1}\,\ket{n+1}$ one obtains the equation for the amplitude $c_n(t)$ of the logical state $\ket{n}_L$
\bea{cn:eqn}
\fl i \frac{d c_n (t)}{d t} &=& r \sqrt{n_{p0} - n} \, \sqrt{(n+1) (2\kappa + n)} \, c_{n+1}(t) \no
\fl                        &+& r \sqrt{(n_{p0} - n + 1)} \, \sqrt{n (2\kappa + n-1)} \, c_{n-1}(t),  \quad c_n(0) = \delta_{n,0}, \quad 2\kappa = n_{s0}+1.
\eea

\subsection{Early times: the non-depleted pump regime}
For early times the condition $n_{p0} \gg n_{s0},\, n$ holds, and the simplest approximation is to approximate the terms $\sqrt{n_{p0} -n}$ and $\sqrt{n_{p0} -n+1}$ by $\sqrt{n_{p0}}$, which leads to
\be{short-time:eqn}
i\,\frac{d c_n (t)}{d \tau} =   \sqrt{n\,(n+n_{s0})} \, c_{n+1}(t) + \sqrt{(n+1) (n+1+n_{s0})} \, c_{n-1}(t).
\ee
This has solution
%\begin{subequations}
\bea{short-time:soln}
\label{short-time:soln:ent}
\ket{\psi_{<}(\tau)}_{out} &=& \sum_{n=0}^{\np0} \, c_n(\tau) \, \ket{n}_L =
\sum_{n=0}^{\np0} \, c_n(\tau) \, \ket{n_{p0}-n}_p\ket{n_{s0}+n}_s\ket{n}_{\bar{i}}  \\
%
%\label{short-time:soln:sep}
\label{psiout:shorttime}
&\approx&  \ket{n_{p0}}_p\otimes\,\sum_{n=0}^{\infty} \, c_n(\tau) \,\ket{n_{s0}+n}_s\ket{n}_{\bar{i}}
\equiv \ket{n_{p0}}_p\otimes\,\ket{\psi_{t_<}(\tau)}_{s,\bar{i}}, \\
\label{cn:shorttime}
c^<_n(\tau) &=& \frac{(-i\,\tanh\tau)^n}{(\cosh\tau)^{n_{s0}+1}} \, \sqrt{\binom{n_{s0} + n}{n}},  \qquad \tau = \sqrt{n_{p0}}\,r\, t,
\eea
%\end{subequations}
where the subscript `$<$' denotes the short-time solution, and we have used $\np0\gg n, 1$ to factorize the pump mode from the signal and idler modes. For all practical calculations, we can effectively take the upper limit $\np0$ in sums over $n$ to be infinite, $\np0\to\infty$.
The solution \Eq{psiout:shorttime} is essentially the same solution as that of \Eq{AvS:18} (with $m\rightarrow n_{s0}$), since the pump occupation number is effectively constant. What determines the validity and extent of the short-time solution is the condition $n_{p0} \gg n,\, n_{s0}, 1$.  The reduced density matrix $\rho_s^I = Tr_{p,\bar{i}}[\,\ket{\psi(\tau)}_{out}\bra{\psi(\tau)}\,]$ for the signal particles in region $I$ is the same as \Eq{AvS:eq19}.
\Eq{short-time:soln} yields the generalized thermal probability distribution
\be{pn:thermal}
\fl p_<(n,\tau) = |c^<_n(\tau)|^2 =
\frac{\tanh^{2 n}\tau}{(\cosh^2\tau)^{n_{s0}+1}} \, \binom{n_{s0} + n}{n} \equiv (1-z)^{\ns0+1} z^n \binom{\ns0+n}{n}, \, z=\tanh^2\tau.
\ee
One has $\sum_{n=0}^\infty \, p_<(n,\tau) = 1$ upon noting the identity \cite{Sorkin:1987,Abramowitz:1972}
$\sum_{n=0}^\infty \, z^n \binom{\ns0+n}{n} = (1-z)^{-(\ns0+1)}$.
The average number of particles in region $I$ is given by  $\bar{n}_<(\tau)$ where (taking $\np0\to\infty$)
\be{nbar:shorttime}
\bar{n}_<(\tau) = \sum_{n=0}^\infty \, n \, p(n,\tau) = (\ns0+1) \, \frac{z}{1-z} = (\ns0+1)\sinh^2\tau,
\ee
 which allows one to write \Eq{pn:thermal} as
\be{pn:shortime}
p_<(n,\tau) = (\ns0+1)^{\ns0+1} \,\binom{\ns0+n}{n} \frac{\bar{n}_<^n(\tau)}{(\bar{n}_<(\tau)+\ns0+1)^{n+\ns0+1}},
\ee
which reduces to the standard thermal probability distribution
$p_{thermal}(n,\tau) = \bar{n}^n /(\bar{n}+1)^{n+1}$ with $\bar{n}_{thermal} = \sinh^2\tau$
when $\ns0=0$.

The same solution could have also been obtained by a slightly more accurate approximation by invoking a Holstein-Primakoff approximation (HPA) \cite{Holstein:1940}.
In the formulation of $H_{p,s,\bar{i}} = r (J^{(p\bar{i})}_+ \, a_s + J^{(p\bar{i})}_- \, a_s^\dagger)$ the HPA would entail invoking the group contraction of $SU(2)$ to the Heisenberg group $\{a, a^\dagger\}$ under the condition of large total angular momentum $j$ via \cite{Sanders:2000}, $J_0\rightarrow jI -  a^\dagger\, a$, $J_+ \rightarrow \sqrt{j+\bar{m}} \,a$ and $J_- \rightarrow \sqrt{j+\bar{m}} \,a^\dagger$. Here, angular momentum state $\ket{j; m}$ for $m\approx\bar{m}\gg -j$ is replaced by the Fock state $\ket{\mu} = \ket{j-m}$. The translation between the states is made via the association \cite{Bonifacio:1970} $j=(n_p + n_s)/2 = (n_{p0} + n_{s0})/2=\textrm{constant}$ and $m=(n_p - n_s)/2 = j - n$, so that $\mu = n$ is close to zero. (A similar group contraction of $SU(1,1)$ to the Heisenberg group could also be invoked with analogous results \cite{Perelomov:1972}). Another possible approximation would be to assume the decoupling of the pump from the emitted signal and (anti)idler modes allows one to factorize as follows:
$e^{-i r (a_p^\dagger \, K^{(s\bar{i})}_-  + a_p \, K^{(s\bar{i})}_+) t} \approx e^{-i (\alpha a_p^\dagger + \alpha^* a_p) t} \, e^{-i r ( \alpha\, K^{(s\bar{i})}_-  + \alpha^* K^{(s\bar{i})}_+) t}$ which would create the state
$\ket{\psi}_{out} = \ket{\alpha}_p \, \ket{\psi_{s,\bar{i}}}_{out}$ where $\ket{\alpha}_p$ is a coherent state \cite{Walls:1994} of the pump such that $a_p\,\ket{\alpha}_p = \alpha \,\ket{\alpha}_p$ with $\alpha = \sqrt{n_{p0}}$ and
$\ket{\psi_{s,\bar{i}}}_{out}$ is the two mode squeezed state \Eq{AvS:18} with $m=n_{s0}$ signal particles initially in the field.

\subsection{Late times: the depleted pump regime}\label{Hpsibar_latetime_derivation}
For late times we wish to examine the conditions under which $n \sim \mathcal{O}(n_{p0}) \gg n_{s0}, 1$ holds. Here, we follow the methodology of Bonifacio and Preparata \cite{Bonifacio:1970}
(who used the Hamiltonian form $r (J^{(p\bar{i})}_+ \, a_s + J^{(p\bar{i})}_- \, a_s^\dagger)$ instead of our form in \Eq{trilinearHamil}) and develop a partial differential equation (pde) approximation to \Eq{cn:eqn} for continuous $n$, which is then evaluated at discrete values of $n$. Connection to the short time solution given by \Eq{short-time:soln} developed above is made by treating it as the initial condition for the pde solution.

We first put \Eq{cn:eqn} in a more amenable form by defining the functions
%======================================
%\cite{Bonifacio:1970:note}
\footnote{This formula corrects a minor, but important typographical error in the definition of $g(n)$ in \cite{Bonifacio:1970}.}
%======================================
\be{g:eqn}
g(n) = \sqrt{2} \, \frac{\Gamma(1 + n/2)}{\Gamma(1/2 + n/2)}, \qquad g(n-1)\, g(n) = n,
\ee
and
\be{G:eqn}
G(n) = g(n_{p0}-n) \, g(n) \, g(2\kappa + n -1).
\ee
We further define
\be{Cn:defn}
\tilde{C}_n(t) = \sqrt{G(n)} \,\tilde{c}_n(t), \qquad \tilde{c}_n(t) = (-i)^n \, c_n(t),
\ee
to obtain the exact equation
\be{Cn:eqn}
\frac{d\tilde{C}_n(t')}{dt'} + G(n) \, \big(\,\tilde{C}_{n+1}(t') - \tilde{C}_{n-1}(t') \, \big) = 0, \qquad t' = r\,t.
\ee
We now treat $n$ as a continuous variable and define $\theta$ by
\be{n:eqn}
n = n_{p0} \, \sin^2\theta, \qquad \theta_n = \sin^{-1}(\sqrt{n/n_{p0}}), \qquad \frac{d\theta_n}{d n} =  \frac{1}{2 \sqrt{n\,(n_{p0}-n)}}.
\ee
Using the definitions in \Eq{n:eqn} we develop the approximate pde
\bea{Cn:pde}
0&=&\frac{d\tilde{C}_n(t')}{dt'} + 2 \, G(n) \, \left( \,\frac{\tilde{C}_{n+1}(t') - \tilde{C}_{n-1}(t')}{(n+1) - (n-1)} \,\right) \no
&\approx& \frac{\partial\tilde{C}_n(\theta,t')}{dt'} + 2 \, G(n)\, \frac{d\theta}{d n} \, \frac{\partial\tilde{C}(\theta,t')}{\partial\theta}, \no
&\equiv&  \frac{\partial\tilde{C}(\theta,t')}{\partial t'} + v(\theta)\, \frac{\partial\tilde{C}(\theta,t')}{\partial\theta},
\eea
where second and higher order partial derivatives in $\theta$ have been dropped, and we define $v(\theta)$ by
\bea{v:eqn}
\fl v(\theta) =  2 \, G(n)\, \frac{d\theta}{d n} &=& \frac{g(n_{p0}-n)}{\sqrt{g(n_{p0}-n-1)\,g(n_{p0}-n)}} \, \frac{g(n)}{\sqrt{g(n-1)\,g(n)}} \, g(2\kappa + n -1), \no
\fl & \approx & g(2\kappa + n -1) \approx \sqrt{n + 2\kappa}, \no
\fl &=& \sqrt{n_{p0} \sin^2\theta + n_{s0} + 1}.
\eea
In \Eq{v:eqn} we have made the long-time approximation that $n_{p0},\, n \gg \ns0, 1$ and additionally for large argument $g(x)\approx \sqrt{x+1} \approx \sqrt{x}$.
This long-time approximation becomes increasingly more accurate for larger pump depletion as $\theta \gg 0$ is closer to its maximum value at $\theta_{max}=\pi/2$.

\Eq{Cn:pde} can be rewritten in the form
\be{Cnu:pde}
\frac{\partial\tilde{C}(u, t')}{\partial t'} +  \frac{\partial\tilde{C}(u, t')}{\partial u} = 0,
\ee
where
\be{u:eqn}
u(\theta) = \int_0^{\theta} \, \frac{d\theta'}{v(\theta')}
\equiv \int_0^{\theta} \, \frac{d\theta'}{\sqrt{1 - \left(-\displaystyle{\frac{\np0}{(\ns0+1)}}\right)\sin^2\theta'}}
\ee
is an elliptic integral of the first kind with imaginary parameter.
%
%\be{u:eqn2}
%u(\theta) = \int_0^{\theta} \, \frac{d\theta'}{v(\theta')}
%\equiv \int_0^{\theta} \, \frac{d\theta'}{\sqrt{1 - \left(-\frac{\alpha}{\beta}\right)\sin^2\theta'}},
%\quad \alpha = n_{s0}+1, \quad \beta = n_{p0}
%\ee
The solution of \Eq{u:eqn} is given by the Jacobi elliptic functions \cite{Abramowitz:1972}
%\begin{subequations}
\bea{u:soln}
\label{u:soln:sn}
\fl \sin\theta &=& sn\left( \sqrt{\ns0+1}\,u\,| \sqrt{-\np0/(\ns0+1)} \right), \\
\label{u:soln:sd}
\fl &=& \sqrt{\frac{\ns0+1}{\np0+\ns0+1}} \,\,
%\frac{sn\left(\sqrt{\np0+\ns0+1})\, u\,| \sqrt{\np0/(\np0+\ns0+1)}\,\right)}
%     {dn\left(\sqrt{\np0+\ns0+1})\,u \,| \sqrt{\np0/(\np0+\ns0+1)}\,\right)}, \\
\frac{sn\left(\sqrt{\np0+\ns0+1})\,u\,|\,k_e\,\right)}
     {dn\left(\sqrt{\np0+\ns0+1})\,u \,|\,k_e\,\right)}, \quad k_e \equiv \frac{\np0}{(\np0+\ns0+1)} < 1,  \\
\label{u:soln:cn}
%&=& cn\left(\sqrt{\np0+\ns0+1})\,u - T_q \,| \sqrt{\np0/(\np0+\ns0+1)}\right).
\fl &=& cn\left(\sqrt{\np0+\ns0+1})\,u - T_q \,| \, k_e\right).
\eea
%\end{subequations}
\Eq{u:soln:sn} is the defining Jacobi elliptic function solution $\sin\theta$ in terms of $u$, with an imaginary argument. By standard elliptic function \cite{Abramowitz:1972}  transformations this  can be written in terms of the function $sd(x) = sn(x)/dn(x)$ of real parameter
 $k_e\equiv\np0/(\np0+\ns0+1)$ in \Eq{u:soln:sd}, which is extremely close to, though strictly less than unity. Lastly, \Eq{u:soln:cn} writes the solution in terms of the quarter period $T_q$ such that
\be{Tq:eqn}
\fl T_q =  \textsf{a}(k_e) + \frac{1}{2} \ln\left(\frac{\np0+\ns0+1}{\ns0+1}\right)
= \textsf{a}(k_e) - \frac{1}{2} \ln(1-k_e),
\,\, \myover{{\ln 4} {k_e\rightarrow 0}} \le \textsf{a}(k_e) \le \myover{{\pi/2} {k_e\rightarrow 1}}.
\ee

The general, stationary solution of \Eq{Cnu:pde} is given by
\be{Cn:gensoln}
\tilde{C}(u, t') = \tilde{\mathcal{C}}(u-t'),
\ee
where $\tilde{\mathcal{C}}$ is an arbitrary function that is determined by the initial condition.
At $t'=0$, we formally have, $\tilde{\mathcal{C}}(u-t') = (-i)^n\,\sqrt{G(n)}\,c_n(0)=\sqrt{G(0)}\,\delta_{n,0}$, which is sharply about $\theta=0$. Such an initial condition would violate neglecting the higher order derivatives in $\theta$ in the derivation of \Eq{Cn:pde}. As such we take the initial condition to be at a time $t=t^* > 0$ such that the short-time solution \Eq{short-time:soln} is valid, i.e. $\tilde{\mathcal{C}}(u-t^*) = (-i)^n \, \sqrt{G\big(n(u)\big)} \,\, c^<(n(u),t^*)$. Working directly with the quantum amplitude $c_n$, this initial condition will allow us construct $c^>_n$ by first writing $c^<_n$ in terms of the most probable value of $n$, namely the mean number $\bar{n}_<$ \Eq{nbar:shorttime}, and then replacing $\bar{n}_<$ by $\bar{n}_>$, which will be derived below in \Eq{nbar:longtime}.

One can consider $\tilde{\mathcal{C}}(u-t')$ as the probability distribution in $u$-space, and conservation of probability can be written as
\be{normlztn}
\sum_{n=0}^{n_{p0}} |c_n(t')|^2 \approx \int\,du\, |\tilde{\mathcal{C}}(u-t')|^2 = 1.
\ee
For all practical computations we can take the upper limit as $\np0\to\infty$, and treat $u$ as a continuous variable.
Moments can be calculated as \cite{Bonifacio:1970}
\be{moments}
\langle n^k \rangle = \sum_{n=0}^{n_{p0}} n^k \, |c_n(t')|^2 \approx \int\,du\, n^k(u) |\tilde{\mathcal{C}}(u-t')|^2.
\ee
Using \Eq{n:eqn} and \Eq{u:soln:cn} we can now write $n$ as
\be{n:soln}
%n = \np0 \, cn^2\left(\sqrt{\np0+\ns0+1}\,u_n - T_q \,| \sqrt{\np0/(\np0+\ns0+1)}\right),
n(u) = \np0 \, cn^2\left(\sqrt{\np0+\ns0+1}\,u_n - T_q \,| \, k_e \right),
\ee
which defines and relates the discrete set of values $u_n$ to the discrete values of $n$ from $0$ to $\np0$.
The long-time state $\ket{\psi}_{out}$ can be written as
\be{psiout:longtime}
\fl \ket{\psi(t')}_{out} = \sum_{n=0}^{\np0} c_n(t')\,\ket{n}_L = \sum_{n=0}^{\np0} c_n(t')\,\ket{\np0-n}_p\ket{\ns0+n}_s\ket{n}_{\bar{i}}, \,\, c_n(t') = i^n \frac{\tilde{\mathcal{C}}(u_n-t')}{G(n)}.
\ee
Here $G(n)$ acts as the `metric' to connect $u$-space to $n$-space.

The stationary solution \Eq{Cn:gensoln} is to be matched to the short-time solution \Eq{nbar:shorttime}.
We first note that $\tilde{\mathcal{C}}(u-t')$ in \Eq{Cn:gensoln} is a very sharply peaked function about $u=t'$ \cite{Bonifacio:1970} so that we can
approximate the mean occupation number  $\bar{n}_>(t')$ in the long time limit by its most probable value  obtained by setting $u\rightarrow t$ in \Eq{n:soln} \cite{Bonifacio:1970},
%\begin{subequations}
\bea{nbar:longtime}
\bar{n}_>(t') &=& \np0 \, cn^2\left(\sqrt{\np0+\ns0+1}\,t' - T_q) \,| \, k_e \right), \\
\label{nbar:longtime:sd}
&=& \frac{\np0\,(\ns0+1)}{\np0+\ns0+1}\,sd^2\left(\sqrt{\np0+\ns0+1}\,t'\,|\,k_e \right).
\eea
%\end{subequations}
To first order in $x$, $sd(x|k_e)\approx x$ so that $\bar{n}_>(t')\approx \np0\,(\ns0+1) r^2 t^2$ recalling that $t'= r t$.
This agrees with $\mathcal{O}(t)$ approximation of \Eq{nbar:shorttime}
$\bar{n}_<(\tau)=(\ns0+1)\,\sinh^2\tau\approx \np0\,(\ns0+1) r^2 t^2$ recalling that $\tau = \sqrt{\np0}\,r\,t$. Therefore, it is
reasonably appropriate to define $t^*$ as the overlap time when these two approximations agree.
The matching of the mean number of particles allows us to
replace $\bar{n}_<$ in the expression for $p_<(n)$ in \Eq{pn:shortime}
by $\bar{n}_>$  in \Eq{nbar:longtime} to construct $p_>(n)$ as
%\begin{subequations}
\bea{pn:longtime}
\fl p_>(n,\tau) &=& |c^>_n(\tau)|^2 =(\ns0+1)^{\ns0+1} \,\binom{\ns0+n}{n} \frac{\bar{n}_>^n(\tau)}{(\bar{n}_>(\tau)+\ns0+1)^{n+\ns0+1}},\\
\label{tau:new}
\fl \tau &\equiv& \sqrt{\np0+\ns0+1}\, r\, t \; (\sim \sqrt{\np0}\, r\, t),
\eea
%\end{subequations}
where within our approximations that $\np0\gg\ns0, 1$ we have redefined  $\tau$ henceforward for both short-time and long-time quantities by \Eq{tau:new}.

As discussed in \cite{Bonifacio:1970}
the quarter period $T_q$ in \Eq{nbar:longtime} is the `build up time' of the laser pulse, i.e. time in which the mean number of photons (in the case of the laser) increases from zero to its maximum value $\np0$. In the case of the BH, it would represent the idealized evaporation time of the BH, in which the total energy of the gravitation field was emitted into correlated particle/anti-particle pairs. From the expression for $\bar{n}_>(\tau)$ in \Eq{nbar:longtime} the quarter period $T_q$ increases very slowly and reaches a maximum value at
\be{Tq:ke1}
T_q \myover{\approx {k_e\to 1}} \frac{\pi}{2} + \frac{1}{2}\,\ln\left(\frac{\np0+\ns0+1}{\ns0+1}\right).
\ee

Since the elliptic parameter $k_e$ as defined in \Eq{u:soln:sd} is very nearly unity
(depending only on the initial populations $\np0$ and $\ns0$),
the Jacobi elliptic function $cn$ can be well approximated
by the hyperbolic function $\sech$
\footnote{Analogous results were obtained by Nation and Blencowe \cite{Nation:2010} using a different analytical approach.}
\Eq{nbar:longtime}
\be{nbar:longtime:sech}
\bar{n}_>(\tau) \approx \np0 \, \sech^2\left(\tau - (2 m + 1)\,T_q \right), \quad m = 0, 1, \ldots,
\ee
In the case of the laser this is interpreted as periodic train of pulses separated by a distance $2 T_q$. For the case of the BH, the radiation escaping to infinity $\mathcal{J}_+$ would not feed back into the evaporated (`depleted') BH and the whole evaporation process could be considered as analogous to one long  `laser pulse' emission.

The expressions $\bar{n}_<(\tau)$ and $\bar{n}_>(\tau)$ in \Eq{nbar:shorttime} and \Eq{nbar:longtime:sech} (with $m=0$, i.e. only one period of the output pulse) and the expression for $T_q$ in \Eq{Tq:ke1} allow us to more precisely define the crossover time $\tau^*$ through the equation
\be{zstar:eqn}
\fl \bar{n}_<(\tau^*) = \bar{n}_>(\tau^*) \Leftrightarrow
(\ns0+1)\,\frac{\zeta^2}{1-\zeta^2} = \np0\,\left[1 - \frac{(\zeta-\zeta_{T_q})^2}{(1-\zeta\zeta_{T_q})^2}\right],
\,\, \zeta=\tanh\tau^*, \,\, \zeta_{T_q}=\tanh(T_q),
\ee
where we have used $\sech^2 x=1-\tanh^2 x$ have expanded out $\tanh(\tau^*-T_q)$ in terms of $\zeta$ and $\zeta_{T_q}$.
Writing $\zeta_{T_q}\approx1-\epsilon_{T_q}$ to first order in the small parameter $\epsilon_{T_q} \ll 1$ and using
\Eq{Tq:ke1} yields $\epsilon_{T_q} = 2 e^{-2{T_q}} = 4 e^{-\pi} (\ns0+1)/(\np0+\ns0+1)$. Substituting this expression into
\Eq{zstar:eqn} and solving to first order in $\epsilon_{T_q}$ and $\epsilon\equiv (\ns0+1)/\np0 \ll 1$ yields the cross over time
\be{zstar}
\fl z^* \equiv \zeta^{*2}=\tanh^2\tau^*
= \left(\displaystyle \frac{2\,\epsilon_{T_q} + \sqrt{2\,\epsilon_{T_q} \epsilon}}{\epsilon -2\,\epsilon_{T_q}}\right)^2
\myover{{\longrightarrow}{\np0\to\infty}} \frac{1}{(e^{\pi/2}/2-1)^2}\approx 0.506407,
\ee
where $0\le z \equiv \tanh^2\tau\le 1$.

\begin{figure}[ht]
\begin{center}
%\includegraphics[width=5in,height=3.5in]{AvS:fig1}
%\begin{tabular}{cc}
  %\includegraphics[width=3in,height=2.25in]{Fidelity_vs_tau_analytic_27Dec2014} %&
  \includegraphics[width=3in,height=2.25in]{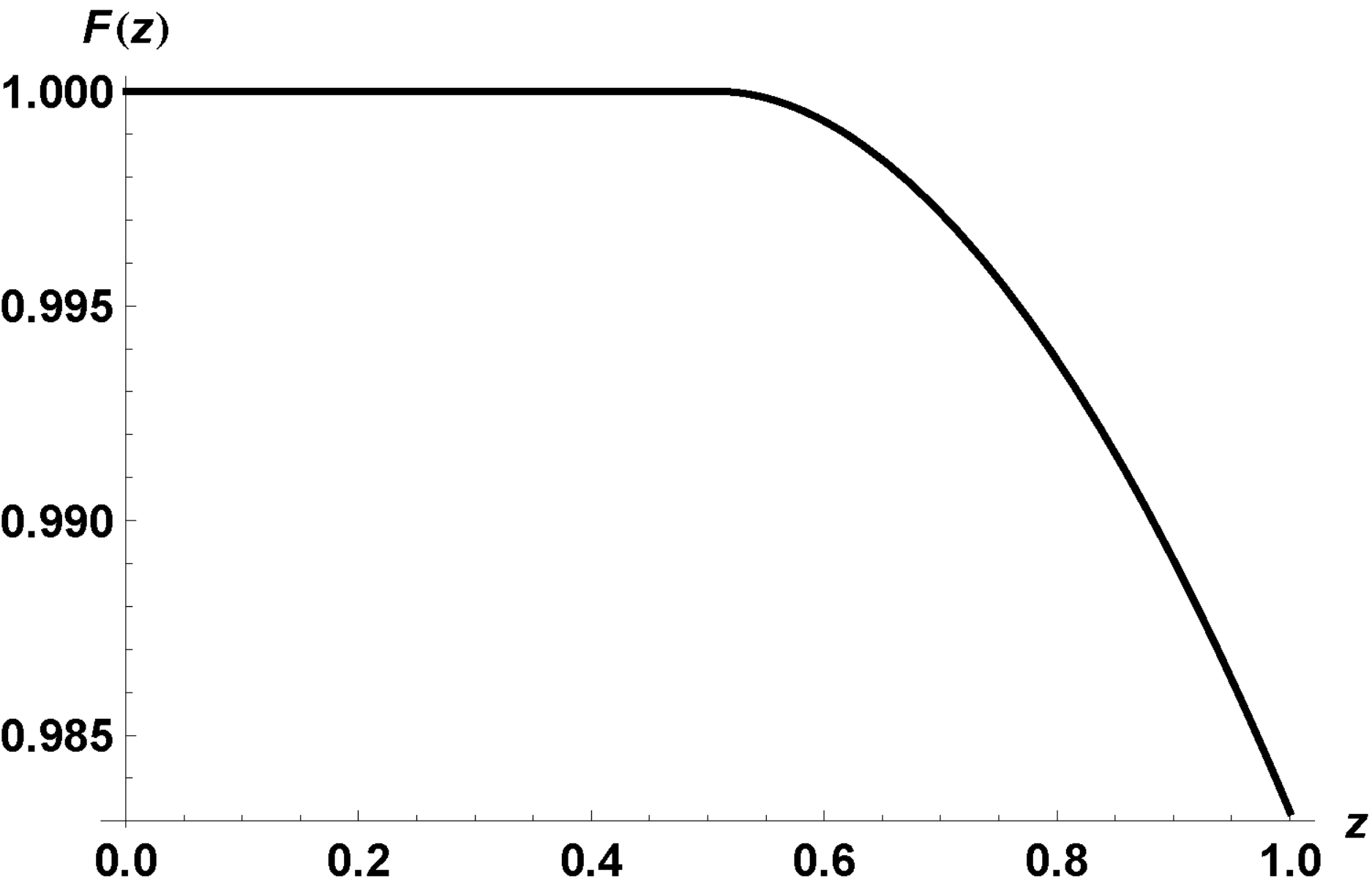} %&
%  \includegraphics[width=3in,height=1.75in]{fig_3_right}
%  \includegraphics[width=3in,height=1.875in]{LogNeg_pibar_s_ns0_0_shorttime-blue_longtime-red} &
%  \includegraphics[width=3in,height=1.875in]{LogNeg_pibar_s_ns0_1_shorttime-blue_longtime-red}
%\end{tabular}
\caption{\label{fig:fidelity:analytic} Fidelity \Eq{fidelity:analytic} of  early \Eq{pn:shortime} and late-time \Eq{pn:longtime} analytic solutions with the standard thermal Hawking radiation state ($n_p(\tau)\equiv\np0$).
}
\end{center}
\end{figure}

As a measure of the difference between the early \Eq{pn:shortime} and late-time \Eq{pn:longtime} probability distributions
we plot in \Fig{fig:fidelity:analytic} the
fidelity \cite{Nation:2010,Nielsen:2000} between $\rho_<(\tau)$ and $\rho_>(\tau)$ defined as
\be{fidelity:analytic}
\hspace{-1.75cm}
F(\tau) = Tr\, \sqrt{\rho^{1/2}(\tau) \, \rho_{thermal}(\tau) \,  \rho^{1/2}(\tau)} =
\left\{
  \begin{array}{cc}
    1 & z  \le z^* \\
      &  \\
    \sqrt{\sum_{n=0}^\infty \,p_{<}(n,\tau)\,p_>(n,\tau)} \spp &  z > z^*. \\
  \end{array}
\right.
\ee
%=============================================
%===========================================================
Here we treat $\rho_{<}(\tau)$ as the usually considered ($n_p(\tau)\equiv\np0$) Hawking thermal density matrix
$\rho_{thermal}(\tau)$ for all time $\tau$,  and  $\rho=\rho_{<}$ for $z\le z^*$ and
$\rho=\rho_{>}$ for $z > z^*$ ($z=\tanh^2\tau$). The long-time solution $\rho_>$ exhibits deviations from
standard thermal distribution $\rho_<$, due to the coupled nature of the late-time BH particle production/evaporation
state $\sum_n\,c^>_n(\tau)\,\ket{\np0-n}_p\ket{\ns0+n}_s\ket{n}_{\bar{i}}$ which can no longer be factorized.
We explore the consequences of this behavior in the subsequent sections.

\subsection{Entanglement produced by $H_{p,s,\bar{i}}$}
\label{section:Ent:Hpsibar}
We now investigate the entanglement of the short-time \Eq{psiout:shorttime} and long-time state \Eq{psiout:longtime}.
We use the log-negativity \cite{Vidal:2002,Plenio:2005,Agarwal:2013} as the useful bipartite entanglement measure given by
\be{logneg}
E_{\mathcal{N}}(\rho) = \log_2[1 + 2 {\mathcal{N}}(\rho)],
\ee
where ${\mathcal{N}}(\rho)$ is the sum of the absolute values of the negative eigenvalues
of the partial transpose $\rho^{PT}$ on one subsystem of a bipartite density matrix $\rho$.
%$\rho = \sum_{n,m} c_n\, c^*_m \ket{n,n}\bra{m,m}$.
For short-times the density matrix given by \Eq{psiout:shorttime} has the form
$\rho^{(s,\bar{i})}_< = \sum_{n,m} c_n\, c^*_m \ket{n+\ns0,n}_{s,\bar{i}}\bra{m+\ns0,m}$, with partial transpose (on the idler mode)
$\rho_<^{(s,PT_{\bar{i}})} = \sum_{n,m} c_n\, c^*_m \ket{n+\ns0,m}_{s,\bar{i}}\bra{m+\ns0,n}$.
Following Agarwal \cite{Agarwal:2013}, for a given $n \ne m$
the Hermitian combination $c_n\, c^*_m \ket{n+\ns0,m}_{s,\bar{i}}\bra{m+\ns0,n}
+ c^*_n\, c_m \ket{m+\ns0,n}_{s,\bar{i}}\bra{n+\ns0,m}$ can be written in diagonal form as
$|c_n\,c_m|\,(\ket{\phi_+}\bra{\phi_+}-\ket{\phi_-}\bra{\phi_-})$ with negative eigenvalue $-|c_n\,c_m|$,
where $c_n\,c^*_m = |c_n\,c_m|\,e^{i\theta{n m}}$
and $\ket{\phi_\pm}=\left(\ket{n+\ns0,m}_{s,\bar{i}} \pm e^{-i\theta{n m}}\ket{m+\ns0,n}_{s,\bar{i}} \right)/\sqrt{2}$.
This yields $E_{\mathcal{N}}(\rho) =  \log_2[\,1 + \sum_{n,m} |c_n\,c_m|\,]$. Using the fact that $\sum_n |c_n|^2=1$ the argument
of the logarithm can be identically written as the square of the sum of the absolute values of the quantum amplitudes, namely
$E_{\mathcal{N}}(\rho) =  \log_2\Big[\sum_n |c_n|\,\Big]^2$.
For short-times, using $c_n(\tau)$ from \Eq{cn:shorttime} (and recalling our definition of $\tau$ in \Eq{tau:new}) we have
\bea{logneg:shorttime}
\fl E^{s,\bar{i}}_{\mathcal{N}}(\rho_<) &=&  2\,\log_2\Big[\sum_n |c^<_n(\tau)|\,\Big]
=  2\,\log_2\left[ \sum_n\,\frac{\tanh^n\tau}{(\cosh\tau)^{n_{s0}+1}} \, \sqrt{\binom{n_{s0} + n}{n}}\right],\no
\fl &=& 2\,\log_2\left[\sum_n \left((1-z)^{n_{s0}+1}\, z^n \,\binom{n_{s0} + n}{n}\right)^{1/2}\right],
\,\, 0 \le z\equiv \tanh^2(\tau) \le z^*. \qquad
\eea
The presence of the square root of the binomial term in \Eq{logneg:shorttime} prevents the closed form solution of the summation for $\ns0>0$. For $\ns0=0$ (the spontaneous emission, Hawking radiation case) the binomial term is unity and one readily computes
the well known result for the two-mode squeezed vacuum state \cite{Agarwal:2013}
$E^{s,\bar{i}}_{\mathcal{N}}(\rho_<^{\ns0=0})= 2\log_2\left[e^\tau\right] = 2\tau/\ln 2$ which grows linearly in time.

\begin{figure}[ht]
\begin{center}
\begin{tabular}{cc}
  \includegraphics[width=3in,height=1.75in]{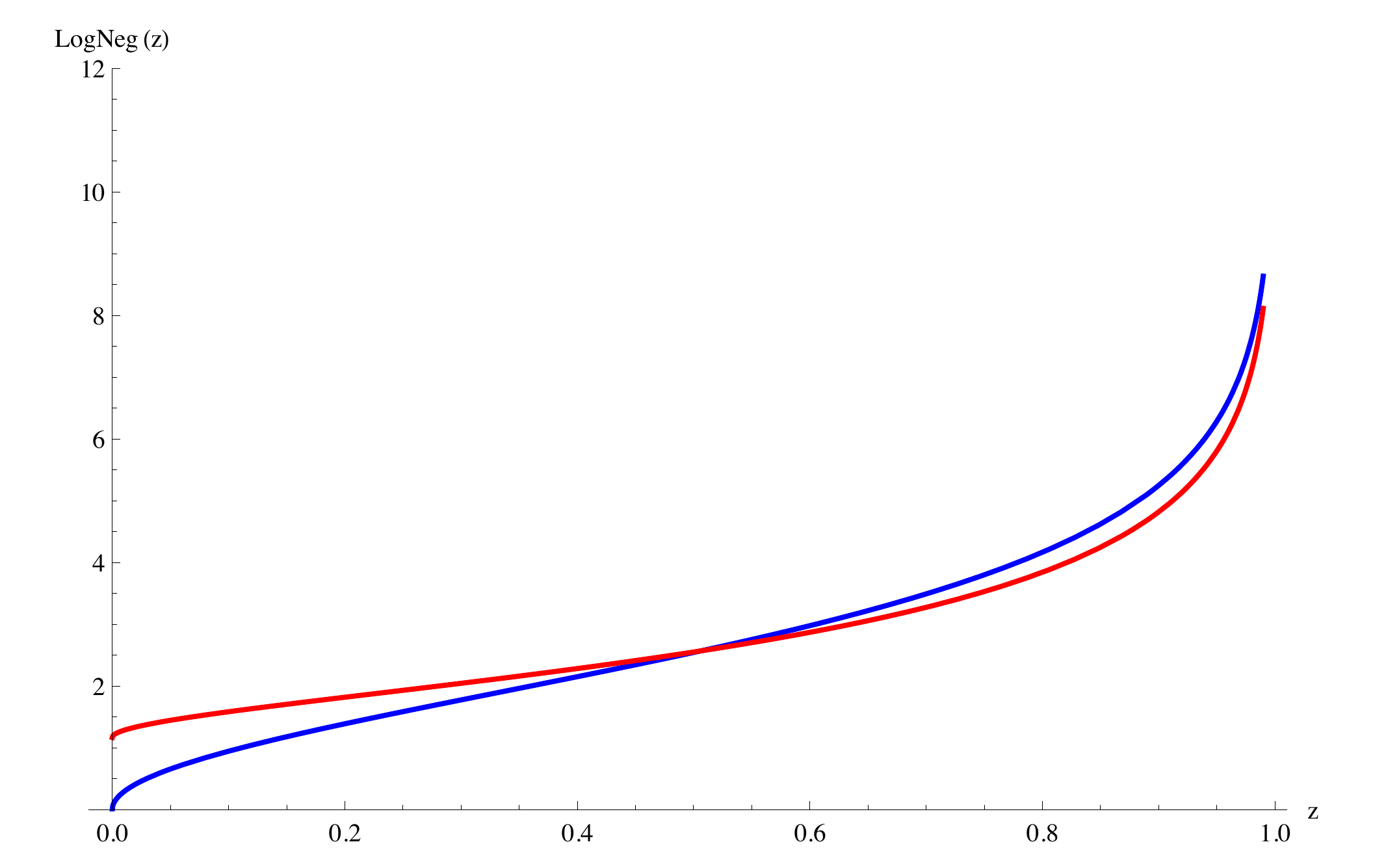} &
  \includegraphics[width=3in,height=1.75in]{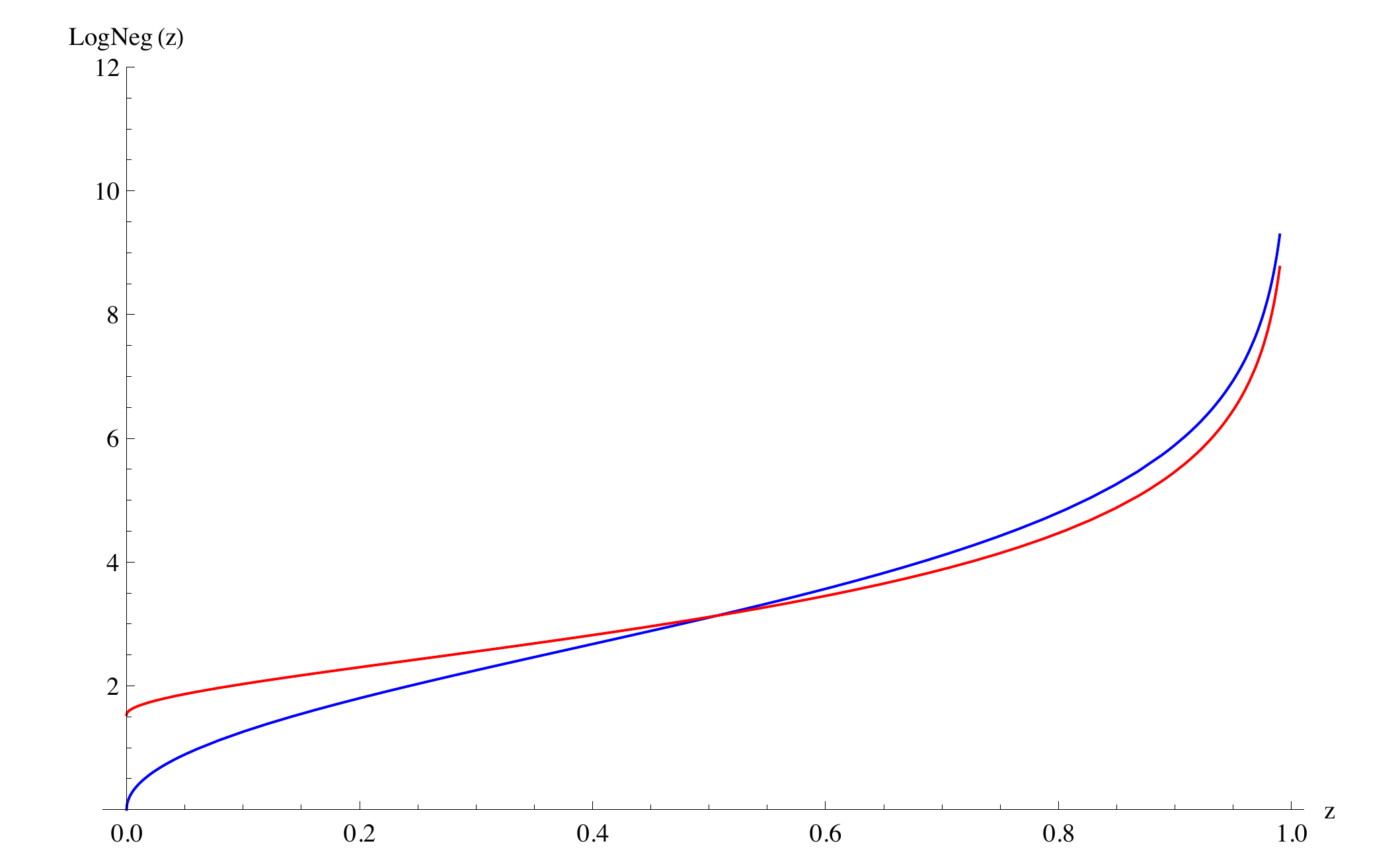}
\end{tabular}
\caption{\label{fig_3} Log-Negativity: $E^{(p,\bar{i}),s}_{\mathcal{N}}(\rho_<)$ short-times (blue curve) and $E^{(p,\bar{i}),s}_{\mathcal{N}}(\rho_>)$ long-times (red curve) plotted for $0\le z=\tanh^2\tau \lesssim 1$ for $\ns0=0$ (left) and $\ns0=1$ (right). (color online)
}
\end{center}
\end{figure}

\begin{figure}[ht]
\begin{center}
\includegraphics[width=3.5in,height=1.75in]{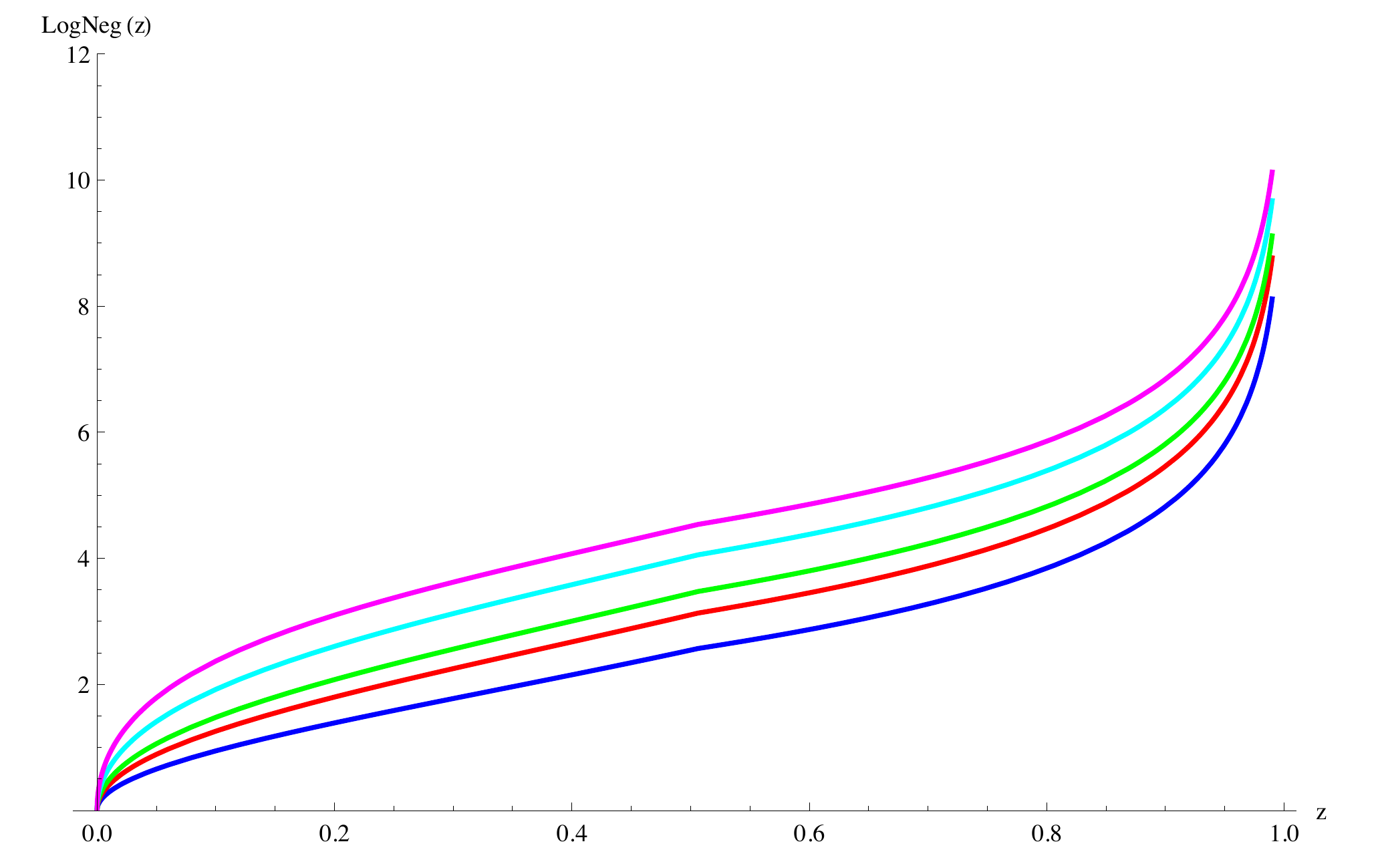}
\caption{\label{fig_4} Log-Negativity: combined  $E^{(p,\bar{i}),s}_{\mathcal{N}}(\rho_<)$ short-time and $E^{(p,\bar{i}),s}_{\mathcal{N}}(\rho_>)$ long-time formulas with crossover time $z^*=0.506407$ \Eq{zstar} in the limit $\np0\to\infty$ for $\ns0=(0,1,2,5,10)$ (lowest to upper curves). (color online)}
\end{center}
\end{figure}

Due to the specific correlated nature of the out-state
$\ket{\psi}_{out} = \sum_n c_n \ket{n}_L = \sum_n c_n \ket{\np0-n}_p\ket{\ns0+n}_s\ket{n}_{\bar{i}}$ which depends on the single index $n$, we would obtain the same value for the log-negativity in \Eq{logneg:shorttime}, namely $E^{(p,\bar{i}),s}_{\mathcal{N}}(\rho_<)=E^{s,\bar{i}}_{\mathcal{N}}(\rho_<)$,
for the bipartite division $(p,\bar{i}),s$ where we partition the states as
$\ket{\psi}_{out} = \sum_n c_n \ket{\ns0+n}_s\ket{n}_{p,\bar{i}}$ with the definition
$\ket{n}_{p,\bar{i}} \equiv \ket{\np0-n}_p\ket{n}_{\bar{i}}$. For short-times this is academic since we are in the regime where
$\np0\gg n,\ns0, 1$, and the state $\ket{\np0-n}_p\approx\ket{\np0}_p$ effectively factorizes with the remaining signal/idler modes.
For the case of long-times where $n\sim\np0 \gg \ns0$, all modes $\{p,s,\bar{i}\}$ are correlated and
we must to chose a bipartite division of the system in order to utilize the log-negativity. To make connection to the short-time discussion above we consider the bipartite division $(p,\bar{i}),s$ to compute the log-negativity for long-times.
Using the expression for $\bar{n}_>(\tau)$ \Eq{zstar:eqn} in terms of $\zeta=\sqrt{z}=\tanh\tau$, $\zeta_{T_q}=\tanh T_q$ and the approximations that led to the cross over time in \Eq{zstar} in the limit $k_e\approx 1$ given
by \Eq{u:soln:sd}, we can approximate
%=========================================
%Using \Eq{pn:longtime}, \Eq{nbar:longtime} and \Eq{nbar:longtime:sech}
%%\begin{subequations}
%\bea{logneg:longtime}
%E^{(p,\bar{i}),s}_{\mathcal{N}}(\rho_>) &=&  2\,\log_2\Big[\sum_n |c^>_n(\tau)|\,\Big], \no
%%\displaystyle{\frac{\Gamma(n+\ns0+1)}{\Gamma(n+1)\,\Gamma(\ns0+1)}} \,
%|c^>_n(\tau)|^2 &=& (\ns0+1)^{\ns0+1} \,\binom{\ns0+n}{n} \, \frac{\bar{n}_>^n(\tau)}{(\bar{n}_>(\tau)+\ns0+1)^{n+\ns0+1}}, \\
%\bar{n}_>(\tau) &=& \np0 \, cn^2\left(\tau - T_q) \,| \, k_e \right) \approx \np0 \, sech^2\left(\tau - \,T_q \right)
%\quad \textrm{for} \,\, k_e \approx 1.
%\eea
%%\end{subequations}
%=========================================
%\begin{subequations}
\bea{logneg:longtime:approx}
\fl E^{(p,\bar{i}),s}_{\mathcal{N}}(\rho_>) &=&  2\log_2\,\Big[\sum_{n=0}^\infty |c^>_n(\tau)|\,\Big], \no
\label{cngt:longtime:approx}
\fl &\approx& 2\log_2\left[\sum_{n=0}^\infty  \left(\frac{f(z)^n}{\big(1+f(z)\big)^{n+\ns0+1}}\, \binom{\ns0+n}{n}\,\right)^{1/2}\right],
\,\, k_e\approx 1, \,\, z^* \le z \le 1, \\
\label{f:formula}
\fl f(z) &=& 4\,e^{-\pi} \,\left( \frac{1+\sqrt{z}}{1-\sqrt{z}} \right) = 4\,e^{-\pi}\,e^{2\tau},
\eea
%\end{subequations}
where the expression for $|c^>_n(\tau)|$ in \Eq{cngt:longtime:approx} still satisfies $\sum_{n=0}^\infty |c^>_n(\tau)|^2 = 1$.
Again, we can only perform the sum analytically for the case $\ns0=0$ which yields
$E^{(p,\bar{i}),s}_{\mathcal{N}}(\rho_>^{\ns0=0})= 2\log_2\left[\sqrt{1+f(z)}+\sqrt{f(z)}\right]
\myover{{\rightarrow} {\tau\gg1}} 2\,\tau/\ln2 + (4-\pi/\ln2) = E^{s,\bar{i}}_{\mathcal{N}}(\rho_<^{\ns0=0}) -0.53236$.
In
%\Fig{LogNeg:short:long:ns0:0-1}
\Fig{fig_3} we plot the log-negativity for short-times, long-times, and in
%\Fig{LogNeg:combined:ns0:0-10}
\Fig{fig_4} plot the combined formula with crossover time $z^*$ in \Eq{zstar} for various values of $\ns0$.
%=============================================
%===========================================================
%===========================================================
%\vspace{-.25in}
%===========================================================

A further trace over the BH pump mode yields a diagonal form for the density matrix
$\rho_{s,i}=\sum_n |c_n(\tau)|^2 \,\ket{\ns0+n,n}_{s,\bar{i}}\bra{\ns0+n,n}$ so that its partial transpose has no negative eigenvalues, leading to  $E^{s,\bar{i}}_{\mathcal{N}}(\rho_{s,\bar{i}})=0$. Thus, in our model, there is formally no direct bipartite entanglement of the particles of mode $s$ in the exterior (region $I$) and the anti-particles of mode $\bar{i}$ in the interior (region $II$) of the BH. Instead, there is bipartite entanglement of the subsystems $A=(p,\bar{i})$ and $B=s$. In the usual BH literature where the BH 'pump' mode is not modeled (i.e. the short-time 'un-depleted pump' regime $\np0 \gg n,\ns0$ in which the pump is implicitly absorbed into the Hamiltonian coupling constant) the pair of modes $A=(p,\bar{i})\rightarrow \bar{i}$ with the interpretation that the bipartite entanglement occurs between these two effective modes, $s$ and $\bar{i}$.
%

%=============================================
%% Tripartite entanglement discussion
%=============================================
It is worth noting again that the out-state with the three physical (computational) modes $\{p,s,\bar{i}\}$ is indexed by a single integer $n$ via
$\ket{\psi}_{out} = \sum_n c_n \ket{n}_L = \sum_n c_n \ket{\np0-n}_p\ket{\ns0+n}_s\ket{n}_{\bar{i}}$.
Hence each subsystem density matrix obtained by tracing out one or two of the physical modes has the diagonal form
$\rho_{red} =  \sum_n |c_n|^2 \ket{n}_{L^*}\bra{n}$
with identical probability distribution $p_n = |c_n|^2$ (modulo possible reordering)
where
$\ket{n}_{L^*} =
\{
\ket{\ns0+n}_{s}\ket{n}_{\bar{i}},
\ket{\np0-n}_{p}\ket{n}_{\bar{i}},
\ket{\np0-n}_{p}\ket{\ns0+n}_{s},
\ket{\ns0+n}_{s},
\ket{n}_{\bar{i}},
\ket{\np0-n}_{p}
\}$
when we trace out over
$\{
p,
s,
\bar{i},
(p,\bar{i}),
(p,s),
(s,\bar{i})
\}$, respectively.
This implies the entropies of reduced density matrices are all identical, i.e.
$S \equiv S(\rho_{s,\bar{i}})=S(\rho_{p,\bar{i}})=S(\rho_{p,s})=S(\rho_{s})=S(\rho_{\bar{i}})=S(\rho_{p})$.
Hence, this state has mutual information \cite{Nielsen:2000}
\bea{MutualInfo}
I(A : B) &\equiv& S(A) + S(B) - S(AB), \no
&=& S \nonumber
\eea
for $(A,B) \in (p,s,\bar{i})$,
and
\textit{tripartite information} \cite{Hayden:2013} (or in classical information-theory context, the I-measure)
\bea{I3}
\hspace{-2cm} I_3(A : B : C) &\equiv& S(A) + S(B) + S(C) - S(AB) - S(BC) - S(AC) + S(ABC),\\
&=& I(A : B) + I(A : C) - I(A : BC), \no
&=& 0,
\eea
with $(A,B,C) = (p,s,\bar{i})$
where we have used $S(\rho_{p,s,\bar{i}}) = 0$ for the pure state $\rho_{p,s,\bar{i}} = \ket{\psi}_{out}\bra{\psi}$.

\section{The full Hamiltonian $\mathcal{H} = H_{p,s,\bar{i}} + H_{p,\bar{s},i},$}\label{fullH}
We now consider the full Hamiltonian
%\begin{subequations}
\bea{fullHamil2}
\mathcal{H} &=& H_{p,s,\bar{i}} + H_{p,\bar{s},i}, \no
&=& r (a_p \, a_s^\dagger \, a_{\bar{i}}^\dagger + a_p^\dagger \, a_s \, a_{\bar{i}} ) + r' (a_p \, a_{\bar{s}}^\dagger \, a_i^\dagger + a_p^\dagger \, a_{\bar{s}} \, a_i ), \\
\label{fullHamil2:Ks}
 &=& r \,(a_p^\dagger \, K^{(s\bar{i})}_-  + a_p \, K^{(s\bar{i})}_+)
  +  r \,(a_p^\dagger \, K^{(\bar{s} i)}_-  + a_p \, K^{(\bar{s} i)}_+), \quad r'=r,
\eea
%\end{subequations}
where the first term in \Eq{fullHamil2:Ks} is the trilinear Hamiltonian
investigated in the previous section \ref{section:Ent:Hpsibar}
with signal particles $s$ in region $I$ outside the BH and anti-particle idler modes $\bar{i}$ in region $II$ just inside the BH. The second term in \Eq{fullHamil2:Ks} is the mode-reversed situation with anti-particle signal modes $\bar{s}$ in region $I$ and particle idler modes $i$ in region $II$. In the case examined by Adami and Ver Steeg \cite{Adami:2014} these two modes pairs $(s,\bar{i})$ and $(\bar{s},i)$ were uncoupled due to the constancy of the BH `pump' mode $p$ occupation number. In our case, these mode pairs are coupled through the pump mode, in particular at long-times.

The logical states of all modes involved are given by
\be{state:nmL}
\ket{n,m}_L = \ket{\np0-n-m}_p\ket{\ns0+n}_{s}\ket{n}_{\bar{i}}\ket{\nsbar0+m}_{\bar{s}}\ket{m}_{i},
\ee
where $\ns0$ and $\nsbar0$ are the initial number of particles/anti-particles in the $s$ and $\bar{s}$ modes in region $I$.
The output state is given by
\be{psi:out:full}
\fl \ket{\psi}_{out} = \sum_{n=0}\,\sum_{m=0} c_{n,m}(t) \, \ket{n,m}_L, \quad 0\le n+m \le \np0, \quad \ket{\psi}_{in} = \ket{0,0}_L,
\ee
where $c_{n,m}(t) = {}_L\bra{n,m}\,e^{-i\,\mathcal{H}\,t}\ket{\psi}_{in}$.
Note that the sums in \Eq{psi:out:full} can also be written as the ordered sums $\sum_{n=0}^{\np0}\,\sum_{m=0}^{\np0-n}$ or
$\sum_{m=0}^{\np0}\,\sum_{n=0}^{\np0-m}$.

The derivation proceeds in a similar fashion to the previous section, with derivation details germane to the inclusion of two pairs of modes relegated to \ref{appendix:fullH}. A summary of those results are the following.
The cross over time $z^*= (e^{\pi/2}/2-1)^{-2} = 0.506407$ (for $\np0\gg 1$) is identical to that given before in \Eq{zstar}.

For short-times $z\le z*$, we can again factor out $\sqrt{\np0}$, define
$\tau = r\,\sqrt{\np0}\,t$ and obtain the factorized (separable) amplitudes
\bea{cnm:shorttime}
\fl c_{n,m}(\tau) &=& c^{<\,(s,\bar{i})}_n(\tau) \,\, c^{<\,(\bar{s},i)}_m(\tau), \\
\label{cncm:shorttime}
\fl c^{<\,(s,\bar{i})}_n(\tau) &=& \frac{(-i\,\tanh\tau)^n}{(\cosh\tau)^{n_{s0}+1}} \, \sqrt{\binom{n_{s0} + n}{n}},
\quad
c^{<\,(\bar{s},i)}_m(\tau) = \frac{(-i\,\tanh\tau)^m}{(\cosh\tau)^{n_{\bar{s}0}+1}} \, \sqrt{\binom{n_{\bar{s}0} + m}{m}},
%\quad \tau = \sqrt{n_{p0}}\,r\, t,
\eea

For long times $z>z*$, we find similarly to the results of Section \ref{Hpsibar_latetime_derivation}
%\begin{subequations}
\bea{nmbar:longtime}
%\bar{n}_>(\tau) + \bar{m}_>(\tau) = \np0 \, cn^2\left(\tau - T \,| \, k_e \right),
\bar{n}_>(\tau) &=& \frac{\ns0+1}{\ns0+\nsbar0+2}\,\np0\,cn^2\left(\tau - T_q \,| \, k_e \right), \\
\bar{m}_>(\tau) &=& \frac{\nsbar0+1}{\ns0+\nsbar0+2}\,\np0\,cn^2\left(\tau - T_q \,| \, k_e \right),
\eea
%\end{subequations}
where, with a redefinition of $\tau$ within our long-time approximation, we have
\be{taunm:kenm}
\fl \tau = \sqrt{\np0 + \ns0 + \nsbar0 + 2}, \quad k_e = \frac{\np0}{\ns0 + \nsbar0 + 2},
\quad cn^2\left(\tau - T \,| \, k_e \right) \myover{{\rightarrow} {k_e\to 1}} \sech^2\left(\tau - T\right),
\ee
and
\bea{Tqnm}
\fl T_q =  \textsf{a}(k_e) + \frac{1}{2} \ln\left(\frac{\np0+\ns0+\nsbar0+2}{\ns0+\nsbar0+2}\right)
= \textsf{a}(k_e) - \frac{1}{2} \ln(1-k_e),
\,\, \myover{{\ln 4} {k_e\rightarrow 0}} \le \textsf{a}(k_e) \le \myover{{\pi/2} {k_e\rightarrow 1}}.\,\qquad
\eea
This allows us to construct $p_>(n,m,\tau)$ from $p_<(n,m,\tau)$ analogously to \Eq{pn:longtime}.

\section{Entanglement produced by $\mathcal{H} = H_{p,s,\bar{i}} + H_{p,\bar{s},i}$}\label{EntHpsibarHpsbari}
We now consider the entanglement (as measured by the log-negativity) produced by $\mathcal{H} = H_{p,s,\bar{i}} + H_{p,\bar{s},i}$
for the bipartite partition $(s,\bar{i})$ and $(\bar{s},i)$, i.e. the entanglement between the particle/anti-particle pairs produced in region $I$/$II$ and the anti-particle/particles region $I$/$II$ pairs for the density matrix
\be{}
\rho_{(s,\bar{i}),(\bar{s},i)} = Tr_p\left[\,\ket{\psi}_{out}\bra{\psi}\,\right].
\ee
In general, the full output state for both short-time and long-times takes the form of \Eq{state:nmL} and \Eq{psi:out:full} repeated here
\bea{psi:out:full2}
\ket{\psi}_{out} &=& \sum_{n=0}\,\sum_{m=0} c_{n,m}(t) \, \ket{n,m}_L, \quad 0\le n+m \le \np0, \quad \ket{\psi}_{in} = \ket{0,0}_L,\no
\ket{n,m}_L &=&  \ket{\np0-n-m}_p\ket{\ns0+n}_{s}\ket{n}_{\bar{i}}\ket{\nsbar0+m}_{\bar{s}}\ket{m}_{i},
\eea
and the two pairs of emitted modes $(s,\bar{i})$ and $(\bar{s},i)$ are coupled through the common pump mode $p$ via their separate production (`squeezing') Hamiltonians $H_{p,s,\bar{i}}$ and $H_{p,\bar{s},i}$.

For short-times $\np0\gg n,m,1$ we have $\ket{\np0-n-m}_p\approx\ket{\np0}_p$ and the state becomes separable
$\ket{\psi_{<}(\tau)}_{out} \approx\ket{n_{p0}}_p\otimes\,\sum_{n=0}^{\infty} \, c^<_n(\tau) \,\ket{n_{s0}+n}_s\ket{n}_{\bar{i}} \,
\sum_{m=0}^{\infty} \, c^<_m(\tau) \,\ket{n_{\bar{s}0}+m}_{\bar{s}}\ket{m}_{i},$
$\equiv \ket{n_{p0}}_p\otimes\,\ket{\psi_{<}(\tau)}_{s,\bar{i}} \,\otimes\, \ket{\psi_{<}(\tau)}_{\bar{s},i}$ with separable density matrix $\rho^<_{(s,\bar{i}),(\bar{s},i)} = \rho_p\otimes\rho_{(s,\bar{i})}\otimes\rho_{(\bar{s},i)}$. Taking the partial transpose on the
$(\bar{s},i)$ subsystem yields the positive density matrix
$\rho^{PT_{(\bar{s},i)}}_{(s,\bar{i}),(\bar{s},i)} = \rho_p\otimes\rho_{(s,\bar{i})}\otimes\rho^T_{(\bar{s},i)}$ with no negative eigenvalues and hence a zero log-negativity, $E^{(s,\bar{i}),(\bar{s},i)}_{\mathcal{N}}(\rho_<)=0$.

However, a state of the form given in \Eq{psi:out:full2} formally has the density matrix
\bea{rho:correlated}
 \rho_{(s,\bar{i}),(\bar{s},i)} &=& Tr_p\left[\,\ket{\psi}_{out}\bra{\psi}\,\right],\no
 &=& \sum_{n,m}\,\sum_{n',m'}\, c_{n,m}(t)\,c^*_{n',m'}\,\delta_{n'+m',n+m}
\ket{n}_{s,\bar{i}}\ket{m}_{\bar{i},s} \,\, {}_{s,\bar{i}}\bra{n'}\,{}_{\bar{i},s}\bra{m'}, \\
 \ket{n}_{s,\bar{i}} &=& \ket{\ns0+n}_s \,\ket{n}_{\bar{i}}, \quad
\ket{m}_{\bar{s},i} = \ket{\nsbar0+m}_{\bar{s}} \,\ket{m}_{i},\nonumber
\eea
where the entangling coefficient arises in the trace over the pump from
$\delta_{n'+m',n+m} = {}_p\langle \np0-n'-m'|\np0-n-m\rangle_p$. If we were to make the short-time approximation $\np0-n-m \approx \np0$ this factor would be come unity ${}_p\langle \np0|\np0\rangle_p=1$, and the state in \Eq{rho:correlated} would become separable.
In the following, we keep the correlated form of $\rho_{(s,\bar{i}),(\bar{s},i)}$ in \Eq{rho:correlated} and examine the log-negativity for both short-times and long-times.

\begin{figure}[ht]
\begin{center}
\begin{tabular}{cc}
  \includegraphics[width=3in,height=1.75in]{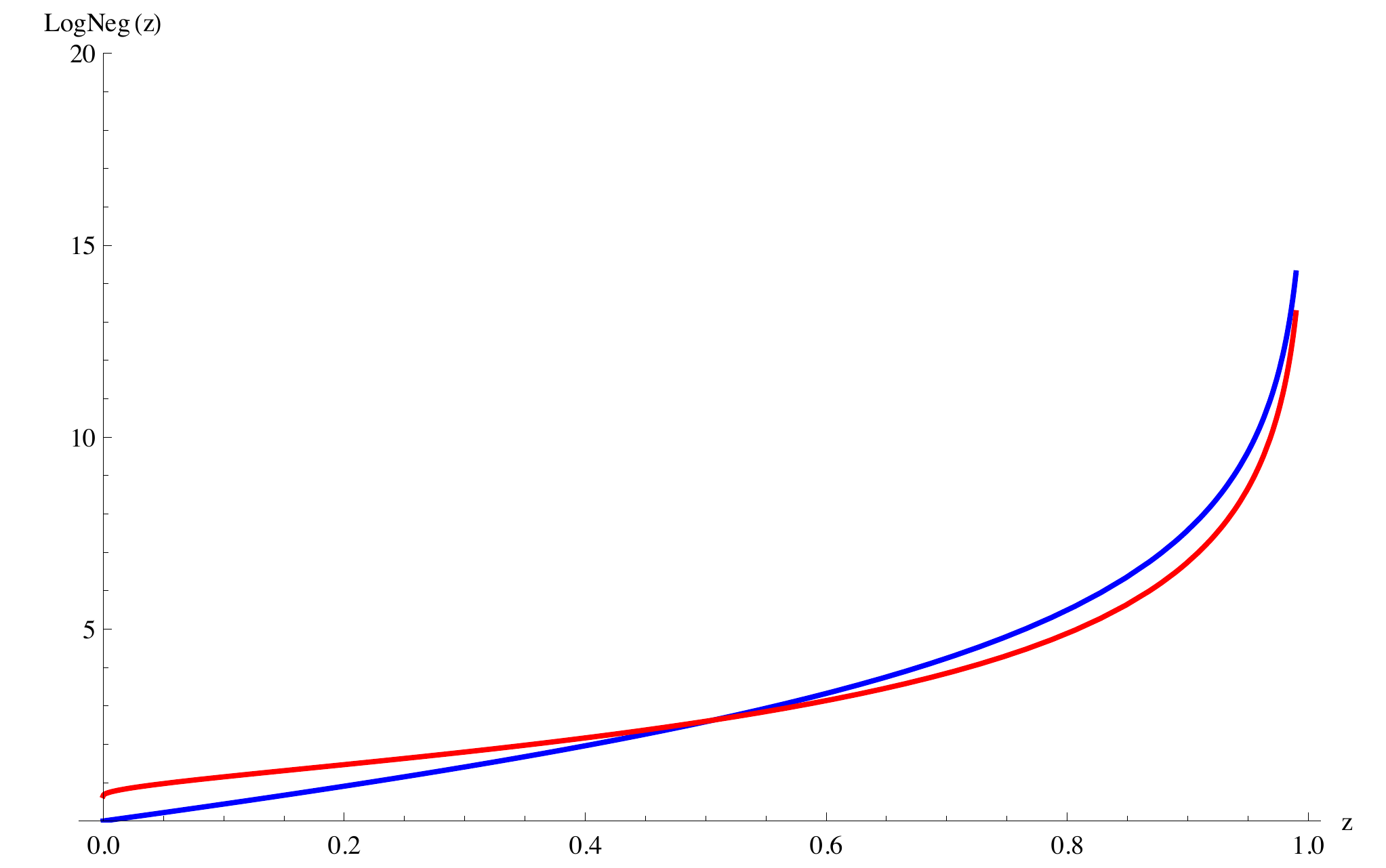} &
  \includegraphics[width=3in,height=1.75in]{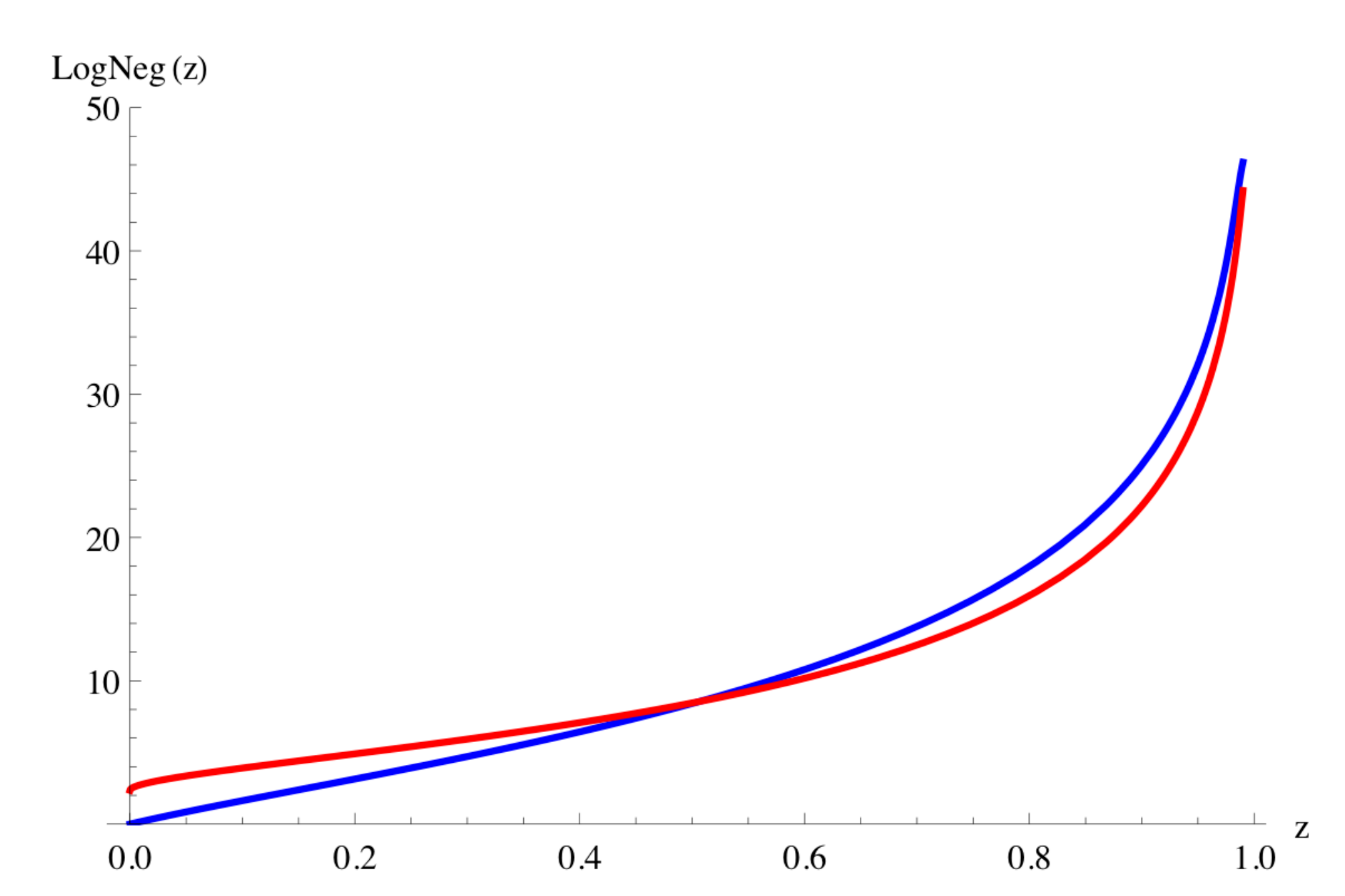}
\end{tabular}
\caption{\label{LogNeg:short:long:ns0:0-1} Log-Negativity: $E^{(s,\bar{i}),(\bar{s},i)}_{\mathcal{N}}(\rho_<)$ short-times (blue curve) and $E^{(s,\bar{i}),(\bar{s},i)}_{\mathcal{N}}(\rho_>)$ long-times (red curve) plotted for $0\le z=\tanh^2\tau \lesssim 1$ for $\ns0=\nsbar0=0$ (left) and $\ns0=10, \nsbar0=0$ (right). (color online)
}
\end{center}
\end{figure}

\begin{figure}[ht]
\begin{center}
\includegraphics[width=4.0in,height=2.0in]{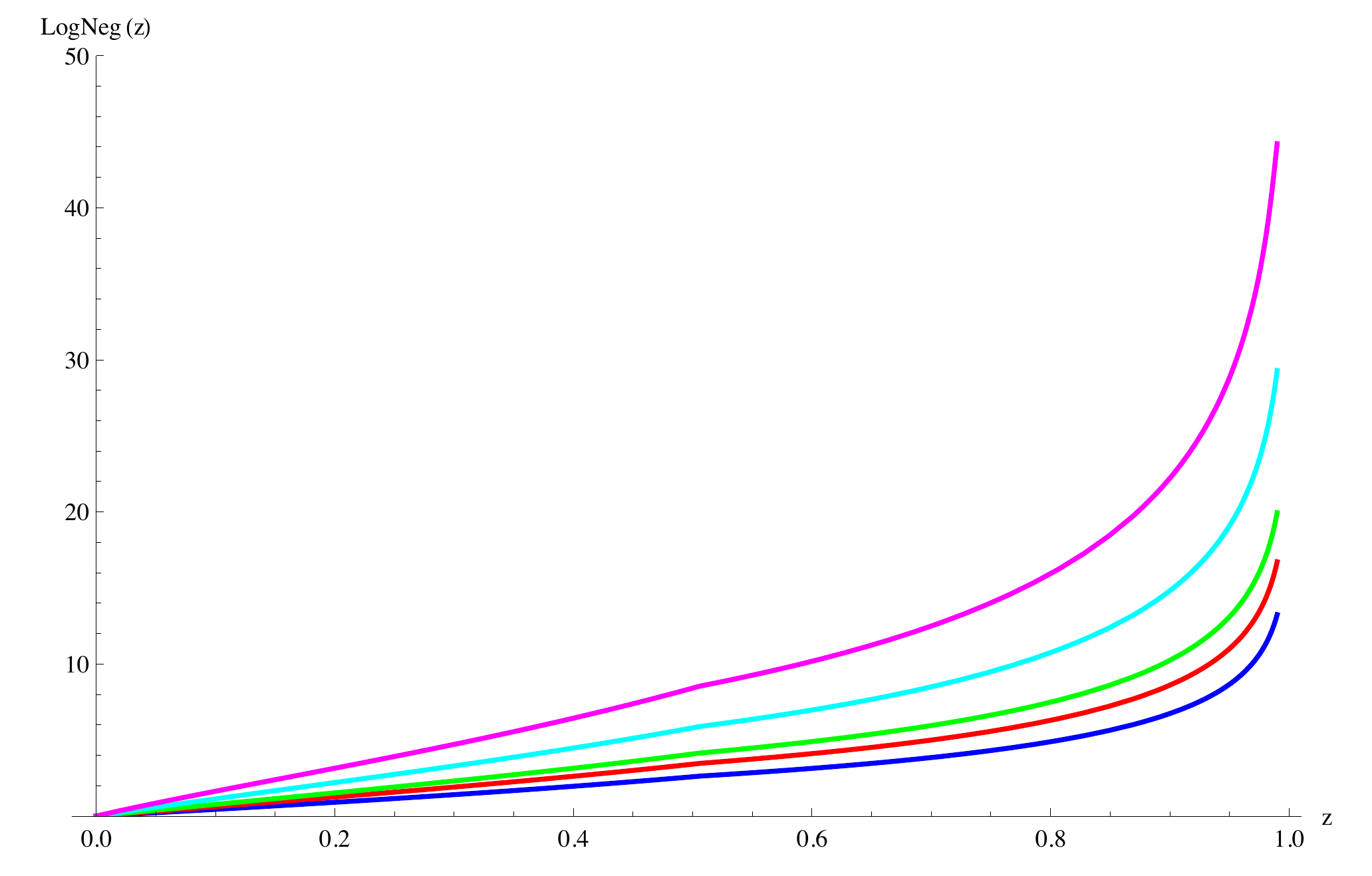}
\caption{\label{LogNeg:combined:ns0:0-10} Log-Negativity: combined  $E^{(s,\bar{i}),(\bar{s},i)}_{\mathcal{N}}(\rho_<)$ short-time and $E^{(s,\bar{i}),(\bar{s},i)}_{\mathcal{N}}(\rho_>)$ long-time formulas with crossover time $z^*=0.506407$ \Eq{zstar} in the limit $\np0\to\infty$ for $\ns0=(0,1,2,5,10)$ (lowest to highest curve) and $\nsbar0=0$. (color online)
}
\end{center}
\end{figure}

Repeating the procedure for developing a formula for the log-negativity in Section \ref{section:Ent:Hpsibar}
we consider the Hermitian combinations that arise in the partial transpose on subsystem $(\bar{s},i)$,
$c_{n,m}\,c^*_{n',m'}\,
\ket{n}_{s,\bar{i}}\ket{m'}_{\bar{i},s} \,\, {}_{s,\bar{i}}\bra{n'}\,{}_{\bar{i},s}\bra{m} + h.c.$ for $(n,m)\ne(n',m')$.
Assuming that the quantum amplitude factorization $c_{n,m} = c^{(s,\bar{i})}_n \,\, c^{(\bar{s},i)}_m$ holds for late-times as well as short-times as in \Eq{cnm:shorttime},
%
%Noting from \Eq{cnm:shorttime} that the amplitude $c_{n,m} = c^{(s,\bar{i})}_n \,\, c^{(\bar{s},i)}_m$
%has a factorized form for both short and long times,
%
we can again write the previous Hermitian combinations of states in diagonal form with a negative eigenvalue $|c^{(s,\bar{i})}_n \, c^{(\bar{s},i)}_m \, c^{(s,\bar{i})}_{n'} \, c^{(\bar{s},i)}_{m'}|$
appearing in front of the state
$\ket{\phi_-}\bra{\phi_-}$ where $\ket{\phi_{\pm}} =1/\sqrt{2} \left(\ket{n}_{s,\bar{i}}\ket{m'}_{\bar{i},s}
\pm e^{-i(\theta_n + \theta_m -\theta_{n'}-\theta_{m'})}\, \ket{n'}_{s,\bar{i}}\ket{m}_{\bar{i},s} \right)$, with
$c^{(s,\bar{i})}_n = |c^{(s,\bar{i})}_n|\,e^{i\theta_n}$, etc\ldots The state $\ket{\phi_-}$ vanishes (and hence produces no negative eigenvalue) only for the case $(n,m)\equiv(n',m')$, i.e. if the `correlating' coefficient in \Eq{rho:correlated} factorizes to $\delta_{n',n}\,\delta_{m',m}$. Thus, the log-negativity is given by the expression
$E^{(s,\bar{i}),(\bar{s},i)}_{\mathcal{N}}(\rho)
=\log_2\left[1+\sum_{n,m}\sum_{n',m'}\,|c_{n,m}\,c_{n',m'}|\delta_{n'+m',n+m}\,(1-\delta_{n',n}\,\delta_{m',m})\right]$.
Using the fact that $\sum_{n,m}\,|c_{n,m}|^2=1$ this can be reduced to
\bea{logneg:fullHamil}
E^{(s,\bar{i}),(\bar{s},i)}_{\mathcal{N}}(\rho)
&=&\log_2\left[\sum_{n,m}\sum_{n',m'}\,
|c^{(s,\bar{i})}_n \, c^{(\bar{s},i)}_m \, c^{(s,\bar{i})}_{n'} \, c^{(\bar{s},i)}_{m'}|\,\delta_{n'+m',n+m}\,\right], \no
&=&\log_2\left[\sum_{m=0}^{\infty}\left(\sum_{n=0}^{m}\,|c^{(s,\bar{i})}_n \, c^{(\bar{s},i)}_{m-n}|\right)^2\,\right],
\eea
Using the results of the previous section we obtain the expressions for the short-time and long-time log-negativity as

%==============================
\bea{logneg:shorttime:fullHamil}
\fl E^{(s,\bar{i}),(\bar{s},i)}_{\mathcal{N}}(\rho_<) &=&
\log_2\left( (1-z)^{\ns0+\nsbar0+2}\,\sum_{m=0}^{\infty} \,\lambda^2_m\,z^m , \right), \quad  0 \le z\equiv \tanh^2(\tau) \le z^*\\
\fl \lambda_m &=& \sum_{n=0}^{m} \sqrt{\binom{\ns0+n}{n} \,\binom{\nsbar0+m-n}{m-n}},
\eea

%==============================
%\begin{subequations}
\bea{logneg:longtime:approx:fullHamil}
\fl E^{(s,\bar{i}),(\bar{s},i)}_{\mathcal{N}}(\rho_>) &=&
\log_2\left( \frac{1}{(1+f(z))^{\ns0+\nsbar0+2}}\,\sum_{m=0}^{\infty} \,\lambda^2_m\,\left(\frac{f(z)}{1+f(z)}\right)^m\right),
\,\, k_e\approx 1, \,\, z^* \le z \le 1, \qquad \\
\fl f(z) &=& 4\,e^{-\pi} \,\left( \frac{1+\sqrt{z}}{1-\sqrt{z}} \right) = 4\,e^{-\pi}\,e^{2\tau}.
\eea
%\end{subequations}
%==============================
Again, due to the presence of the square root of the binomial coefficients, these formulas can only be computed analytically in the case $\ns0=\nsbar0=0$ yielding
%\begin{subequations}
\bea{logneg:shortlongtime:ns0nsbar0zero}
\fl \left. E^{(s,\bar{i}),(\bar{s},i)}_{\mathcal{N}}(\rho_<)\right|_{\ns0=\nsbar0=0} &=& \log_2\left(\frac{1+z}{1-z}\right)
=  \log_2\,\cosh 2\tau \myover{{\rightarrow} {\tau\gg 1}}  \frac{2\tau}{\ln 2} -1, \\
\fl \left. E^{(s,\bar{i}),(\bar{s},i)}_{\mathcal{N}}(\rho_>)\right|_{\ns0=\nsbar0=0} &=& \log_2\left(1+2\,f(z)\right)
\myover{{\rightarrow} {\tau\gg 1}}  \log_2\,\left( 8 e^{2\tau-\pi} \right) = \frac{2\tau}{\ln 2} + (3-\pi/{\ln 2}).\,\qquad
\eea
%\end{subequations}
In \Fig{LogNeg:short:long:ns0:0-1} and in \Fig{LogNeg:combined:ns0:0-10} below we plot the log-negativity $E^{(s,\bar{i}),(\bar{s},i)}_{\mathcal{N}}(\rho_)$ for both short-times, long-times and using the combined formula with the crossover time $z^*=0.506407$.

%============================================================
%
%\newpage
%
%
%
%\vspace{-.25in}

The quantum amplitude factorization $c_{n,m}(\tau) = c^{(s,\bar{i})}_n(\tau) \,\, c^{(\bar{s},i)}_m(\tau)$ indicates that the Hawking pairs $(s,\bar{i})$ and $(\bar{s},i)$ are essentially produced independently, though from a common source, the BH `pump' mode $p$. This leads (when tracing over mode $p$) to an entangling term
$_p\langle \np0-n'-m'\ket{\np0-n-m}_p = \delta_{n'+m',n+m}$ which entangles the two pairs.
The number of particles $n+m$ removed from the BH mode $p$ can appear in many combinations in the two pairs
$\{
\left[(n+m)_{s,\bar{i}},{0}_{\bar{s},i}\right],
\left[(n+m-1)_{s,\bar{i}},{1}_{\bar{s},i}\right],
\ldots,
[1_{s,\bar{i}},{(n+m-1)}_{\bar{s},i})],
[0_{s,\bar{i}},{(n+m)}_{\bar{s},i})]
\}$
such that the sum $n+m$ remains constant for fixed $n$ and $m$.
Thus, it is the common BH source mode $p$ that is the source of the entanglement of the
two pairs $(s,\bar{i})$ and $(\bar{s},i)$, a feature not found in the usual considerations of Hawking radiation.
Such entanglement would be difficult to observe since each pair incorporates the standard entanglement across the BH horizon for individual pair, as well as a `cross-entanglement' due to the common source mode $p$.

The fact that the full pure quantum state involving all modes $p,s,\bar{i},\bar{s},i$ is entangled across the horizon, vs. a product (or in general, a separable) state, argues against the necessity for the concept of a BH firewall at the horizon \cite{Almheiri:2013,Braunstein:2009,Braunstein:2013}. A further trace (as described in the next section) over the inaccessible degrees of freedom, the conventionally considered interior region $II$ modes $\bar{i},i$, and now, also including the BH `pump' source mode $p$, leads to a separable reduced  region $I$ density matrix
$\rho_{s,\bar{s}} = \rho_{s}\otimes \rho_{\bar{s}}$ for the outgoing Hawking radiation, which is in fact a two-mode thermal state.

%=========================================================

\section{Holevo capacity $\chi_{s,\bar{s}}(z)$ }\label{HolevoCapacity}
We now have all the components necessary to compute the channel (Holevo) capacity of Adami and Ver Steeg \cite{Adami:2014} as described in Section \ref{AvS:review}.
Here we are interested in the reduced density matrix $\rho_{s,\bar{s}}$ of the emitted particle/anti-particles in region $I$.
If we take the density matrix $\rho_{(s,\bar{i}),(\bar{s},i)}$ of \Eq{rho:correlated} and further trace out over the particle and anti-particle idler modes $(i,\bar{i})$ in region $II$
we obtain
\bea{rho:s:sbar}
\rho_{(s,\bar{s})} &=& Tr_{p,i,\bar{i}}\left[\,\ket{\psi}_{out}\bra{\psi}\,\right],\no
&=& \sum_{n,m}\,\sum_{n',m'}\, c_{n,m}(t)\,c^*_{n',m'}\,\delta_{n+m,n'+m'}\,\delta_{n,n'}\,\delta_{m,m'}, \no
&=& \sum_{n}\,|c^{(s)}_{n}|^2\,\ket{\ns0+n}_s\bra{\ns0+n} \,\otimes\,\sum_{n'}\,|c^{(\bar{s})}_{n'}|^2\,\ket{\nsbar0+n'}_{\bar{s}}\bra{\nsbar0+n'},\,\qquad\\
&\equiv& \rho_{s} \otimes \rho_{\bar{s}}, \nonumber
\eea
where the extra delta functions have come from $\delta_{n',n}\,\delta_{m',m} = {}_i\langle n'|n\rangle_i \,\, {}_{\bar{i}}\langle m'|m\rangle_{\bar{i}}$.
Here $\rho_s$ and $\rho_{\bar{s}}$ are given by \Eq{AvS:20} which Adami and Ver Steeg denoted as $\rho_{k|m}$ (where $m\to\ns0,\,\nsbar0$ in our notation).
The additional trace over the inaccessible (to the outside observer in region $I$) region $II$ modes $(\bar{i},i)$
yields the conventional result that the observable Hawking radiation (in region $I$) is uncorrelated (i.e. in a product of thermal states).

For our formulation, we use the short-time and long-time probabilities given in \Eq{logneg:shorttime} and \Eq{logneg:longtime:approx}
%
%\begin{subequations}
\bea{probs:shorttime:longtime}
\fl p_<(n,z) &=& |c^<_n(z)|^2 = (1-z)^{n_{s0}+1}\, z^n \,\binom{n_{s0} + n}{n}, \quad, 0 \le z \le z^*,\\
\fl p_>(n,z) &=& |c^>_n(z)|^2 =\frac{1}{\big(1+f(z)\big)^{n+\ns0+1}} \, \left(\frac{f(z)}{1+f(z)}\right)^n\, \binom{\ns0+n}{n}  \quad z^* \le z \le 1,\\
\label{fz}
\fl f(z) &=& 4\,e^{-\pi} \,\left( \frac{1+\sqrt{z}}{1-\sqrt{z}} \right) = 4\,e^{-\pi}\,e^{2\tau}.
\eea
%\end{subequations}
where we note that
\be{prob:longtime:from:shorttime}
p_>(n,z)=p_<\big(n, f(z)/\left[1+f(z)\right]\big).
\ee

%%==========================
%%==========================

\begin{figure}[ht]
\begin{center}
\begin{tabular}{cc}
  \includegraphics[width=3in,height=2.0in]{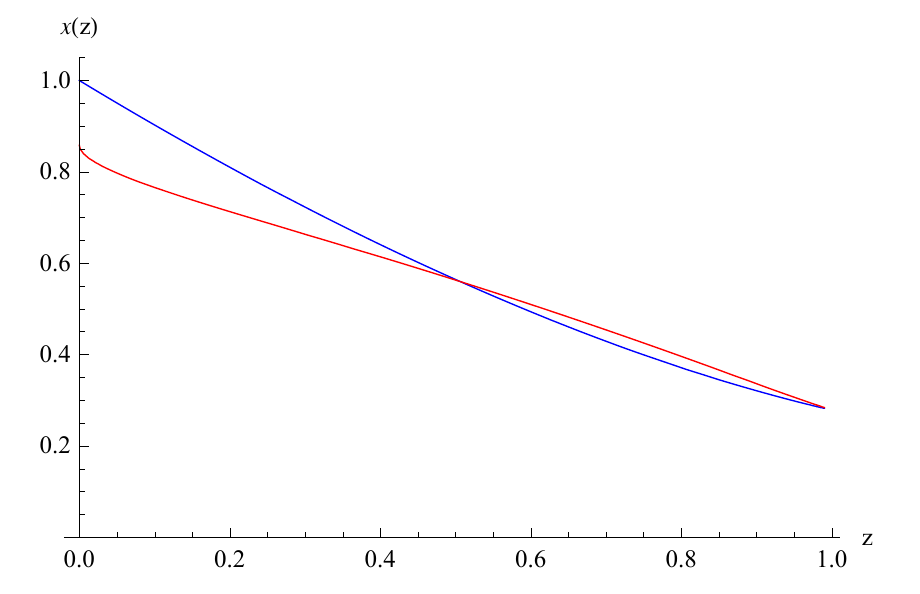} &
  \includegraphics[width=3in,height=2.0in]{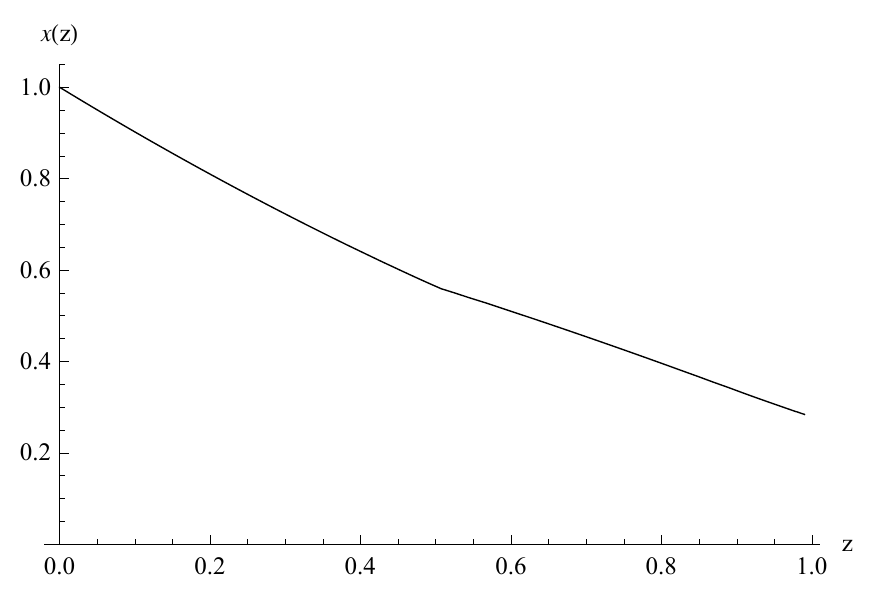}
\end{tabular}
\caption{\label{chi:figs} (Left): $\chi^<_{s,\bar{s}}(z)$  short-times (blue curve -also Adami and Ver Steeg Eq.(40) \cite{Adami:2014}) and $\chi^>_{s,\bar{s}}(z)$ long-times  (red curve),  plotted for $0\le z=\tanh^2\tau \lesssim 1$. (Right) $\chi_{s,\bar{s}}(z)$  combined formula with crossover time $z^*=0.506407$ \Eq{zstar} in the limit $\np0\to\infty$.
}
\end{center}
\end{figure}

In \Fig{chi:figs} below we plot the Holevo capacity $\chi_{s,\bar{s}}(z)$ for the reduced particle/anti-particle density matrix $\rho_{s,\bar{s}}$ in region $I$. For short times, the formula is the same as  Eq.(40) $\chi^{(AVS)}_{s,\bar{s}}(z)$
in Adami and Ver Steeg \cite{Adami:2014}
\be{chi:shorttime}
\fl \chi^<_{s,\bar{s}}(z) = 1 -\frac{1}{2} (1-z)^3 \sum_{m=0}^{\infty} z^m (m+1) (m+2) \log(m+1)
+ (1-z)^2 \sum_{m=0}^{\infty} (m+1) \log(m+1).
\ee
For long-times, $\chi^>_{s,\bar{s}}(z)$ is obtained from $\chi^<_{s,\bar{s}}(z)$ by the substitution $z\to f(z)/\big(1+f(z)\big)$ with $f(z)$ given by \Eq{fz}
\be{chi:shorttime}
\chi^>_{s,\bar{s}}(z) = \chi^<_{s,\bar{s}}\left(f(z)/\big[1+f(z)\big]\right), \\
%f(z) &=& 4\,e^{-\pi} \,\left( \frac{1+\sqrt{z}}{1-\sqrt{z}} \right) = 4\,e^{-\pi}\,e^{2\tau}.
\ee
Most curious is that while the short-time and long-time expression for the Holevo capacity $\chi$ cross at $z^*=0.506407$ \Eq{zstar} in the limit $\np0\to\infty$ as seen in the left figure in \Fig{chi:figs}, they both also reach the same terminal value at $z=1$ due to the factors of $1-z$ in $\chi^<_{s,\bar{s}}(z)$ in \Eq{chi:shorttime} and the fact that both $z$ and $f(z)/\big(1+f(z)\big)$ approach unity as $z\to 1$.

The main conclusion to be drawn from these results is that even treating the BH particle production/evaporation as analogous to parametric down conversion with a depleted pump mode (the later modeling the state of the evaporating BH), the results of the Holevo capacity calculation for the $\rho_{s,\bar{s}}$ is essentially the same as that found by the work Adami and Ver Steeg \cite{Adami:2014}, namely that at infinite rapidity $z=1$, the Holevo capacity is non-zero. Thus, while the particle/anti-particle $(s,\bar{i})$  and  anti-particle/particle $(\bar{s},i)$ production in region $I$/$II$ is entangled through the BH `pump' mode $p$ (and hence have a non-zero log-negativity), the outgoing particle/anti-particle $(s,\bar{s})$ pairs in region $I$ remain separable (i.e. zero log-negativity). The result $\chi^<_{s,\bar{s}}(z)$  of this section is identical to that of the calculation $\chi^{(AVS)}_{s,\bar{s}}(z)$  by Adami and Ver Steeg in the regime $0\le z \le z^*$, while the result $\chi^>_{s,\bar{s}}(z)$ is slightly larger than that of Adami and Ver Steeg's result in the regime $z^*\le z \le 1$, with both having the same value at $z=1$. The implication here is equivalent to that of
Adami and Ver Steeg \cite{Adami:2014} that the infalling matter is reradiated back out as a stimulated emission in combination with the spontaneously emitted Hawking radiation.

\section{Gray-body factors with beam splitter Hamiltonian}\label{GrayBody}
In this section we want to repeat the channel capacity for $\chi$, but now in the presence of a beam splitter Hamiltonian
\be{HBS}
H^{s,c}_{bs} =  \theta_k (a^{\dagger}_{s}\, a_c + a_{s} \, a^{\dagger}_c)
\ee
which affects the non-perfect absorption of the BH, by scattering in-coming late-time region $I$ modes $c$ from outgoing region $I$ modes $a$ just outside the horizon.
As discussed previously, the full Hamiltonian (changing from the $(k,-k)$ to $p,\, (s,\bar{i}),\, (\bar{s},i)$ mode notation)
\Bea
\mathcal{H} &=& H_{p,s,\bar{i}} + H_{p,\bar{s},i} + H^{s,c}_{bs} \equiv H^{s,\bar{i}}_{sq} + H^{\bar{s},i}_{sq} + H^{s,c}_{bs} \no
&=& r (a_p \, a_s^\dagger \, a_{\bar{i}}^\dagger + a_p^\dagger \, a_s \, a_{\bar{i}} ) + r' (a_p \, a_{\bar{s}}^\dagger \, a_i^\dagger + a_p^\dagger \, a_{\bar{s}} \, a_i )
+   \theta_k (a^{\dagger}_{s}\, a_c + a_{s} \, a^{\dagger}_c)
\Eea
treats the situation where both the squeezing ($H_{sq}$) between the pump mode $p$ and the two pair of modes $(s,\bar{i})$ and $(\bar{s},i)$ takes place simultaneously with the gray-body scattering ($H_{bs}$) between
the region $I$ modes $(a,c)$ and yields a late-time (on ${\mathcal{J}_+}$) observed mode
\bea{AvS:41:v2}
A = e^{-i {\mathcal{H}}} a_s e^{i {\mathcal{H}}} &=& \alpha \, a_s - \beta \,a^\dagger_{\bar{i}} + \gamma \, a_c \no
&=& \cosh r \, \cos\theta \, a_s - \sinh r \,a^\dagger_{\bar{i}} + \cosh r \, \sin\theta \, a_c.
\eea
However, we can take a much simpler approach and consider successive (in time) transformations
%==========================
%\cite{Adami:note}
\footnote{The successive squeezing followed by beam-splitter transformations reproduce the gray-body results of Adami and Ver Steeg \cite{Adami:2014}.}
%==========================
%$H'_k = i \theta_k (a^{\dagger}_{k}\, c_k - a_{k} \, c^{\dagger}_k)$
\be{AvS:eq42_v2}
A = e^{-i H^{s,c}_{bs}} \left( e^{-i H^{s,\bar{i}}_{sq}} a_s e^{-i H^{s,\bar{i}}_{sq}} \right) e^{i H^{s,c}_{bs}}
\ee
which also reproduce the late-time observed mode $A_k$ in \Eq{AvS:41:v2}.
Physically, one is considering the case in which the signal/idler pairs are first produced by the interaction with the BH, and subsequently the mode $a$ undergoes a region $I$ scattering by $H_{bs}$ with a in-coming late time mode $c$.
The beam splitter transformation affects the transformations
%\begin{subequations}
\bea{BStransf:v2}
e^{-i H^{s,c}_{bs}} a_s e^{ i H^{s,c}_{bs}} &=& \cos\theta \, a_s + i \sin\theta a_c, \\
e^{-i H^{s,c}_{bs}} a_c e^{ i H^{s,c}_{bs}} &=& \cos\theta \, a_c + i \sin\theta a_c.
\eea
%\end{subequations}
Under $H^{s,c}_{bs}$ Fock states $\ket{n}_a\ket{n'}_c$ are transformed into states of the form $\sum_{p=0}^{n+n'}\, f_p(n,n') \,\ket{p}_s\ket{n+n'-p}_c$ since the total particle number $a^\dagger_s a_s + a^\dagger_c a_c$ is preserved.
Here $f_p(n,n')$ is given by \cite{Agarwal:2013} $f_p(n,n') = \sum_{q=0}^{n}\,\sum_{q'=0}^{n'} \, \delta_{p,q+q'}\binom{n}{q}\binom{n'}{q'}\big((q+q')! (n+n'-q-q')!/(n! n'!)\big)^{1/2} (\cos\theta)^{n'+q-q'} \, (-i\sin\theta)^{n-q+q'}$.

%%======================================================
%% Begin: material from orig Sec7.1 v5 for appendix
%%======================================================
\subsection{Quantum states for sending a `0' and a `1'}\label{Sec:7.1}
For the channel capacity calculation we now send a `0' with in-state $\ket{\psi}^{(0)}_{in}$ and send a `1' with in-state $\ket{\psi}^{(1)}_{in}$ given by
%\begin{subequations}
\bea{in_states}
\label{in_states:0}
\ket{\psi}^{(0)}_{in} = \ket{\np0}_{p}\ket{0}_{s}\ket{0}_{\bar{i}}\ket{1}_{c}\ket{0}_{\bar{s}}\ket{0}_{\bar{i}}, \\
\label{in_states:1}
\ket{\psi}^{(1)}_{in} = \ket{\np0}_{p}\ket{0}_{s}\ket{0}_{\bar{i}}\ket{0}_{c}\ket{1}_{\bar{s}}\ket{0}_{\bar{i}},
\eea
%\end{subequations}
where in \Eq{in_states:0} there is now 1 input boson in mode $c$ vs 1 in the early-time mode $a$ (both region $I$), and in \Eq{in_states:1} there is still 1 input boson in the region $I$ mode $\bar{s}$, as considered
previously in section \ref{AvS:review}.
The construction of the output quantum states and their corresponding probability distributions is straightforward but somewhat involved due to the beam splitter transformation acting on Fock states discussed at the end of the previous section. The details are relegated to \ref{appendix:Sec:7.1_v5}.

\subsection{Holevo capacity with beam splitter}
\begin{figure}[ht]
\begin{center}
%\includegraphics[width=5in,height=3.5in]{AvS:fig1}
%\begin{tabular}{cc}
%  \includegraphics[width=3in,height=2.0in]{fig_8}
    \includegraphics[width=3in,height=2.0in]{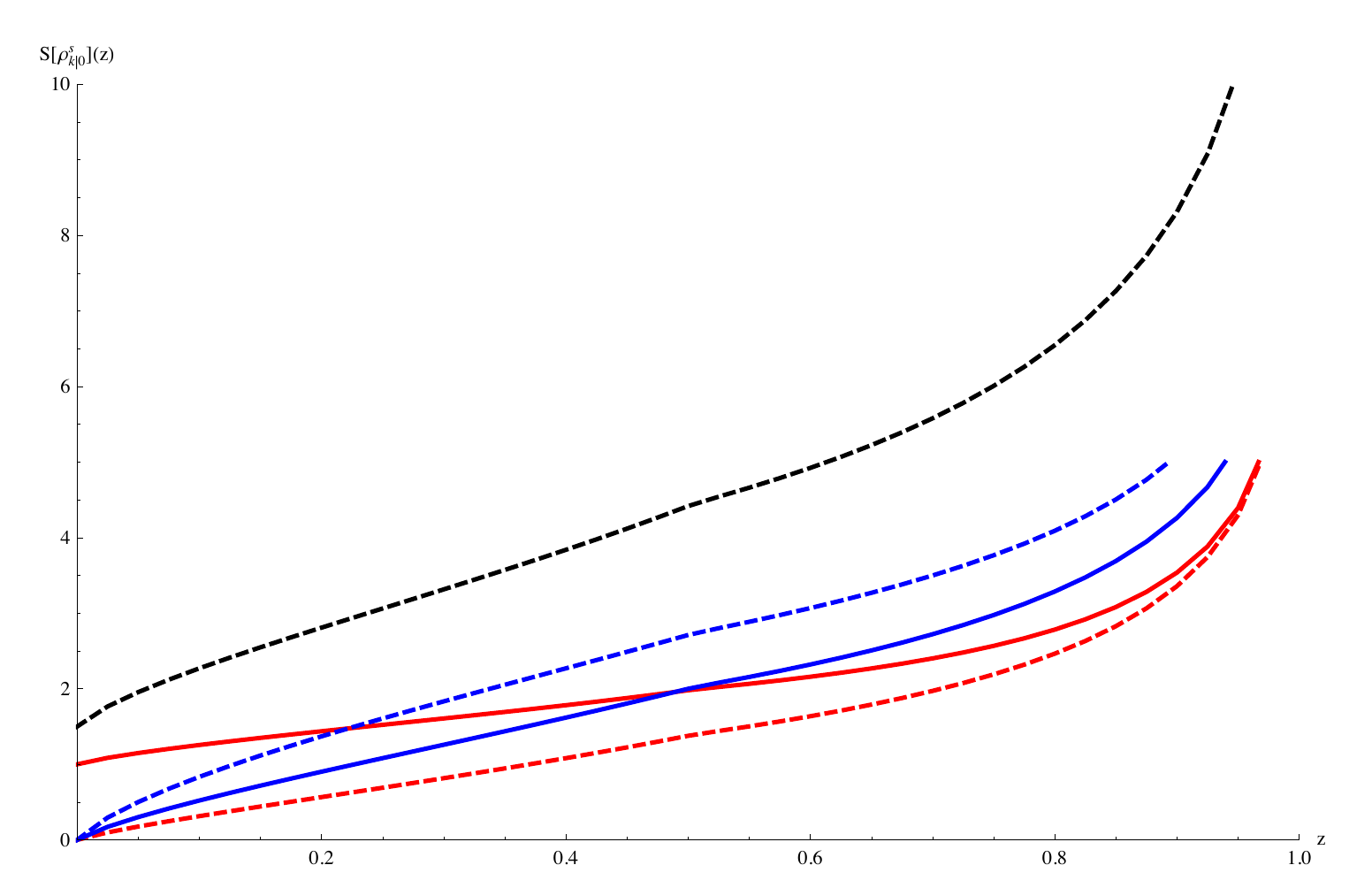}
  %\includegraphics[width=3in,height=2.5in]{Entropies_all_shorttime_only_theta_0_p_0p5} % &
%  \includegraphics[width=3in,height=2.4in]{chi_combined}
%\end{tabular}
\caption{\label{entropy:figs} Entropies (combined short and long time formula with crossover time $z^*=0.506407$ \Eq{zstar} in the limit $\np0\to\infty$) for various probability distribution involved in the computation of $\chi_{s,\bar{s}}(z,\theta=\pi/4)$ (where $\rho_I(0)$ is used to send `0,' and  $\rho_I(1)$ to send `1').
(blue): (solid) $S[\rho_I^{\bar{s}}(0)]$,
(dashed) $S[\rho_I^{\bar{s}}(1)]$;\,
(red):(solid) $S[\rho_I^{s}(0)]$ ,
(dashed) $S[\rho_I^{s}(1)]$; and
(black): $\left. S[\rho_I^{s,\bar{s}}]\right|_{p=1/2}$ (maximized at $p=1/2$). (color online)
}
\end{center}
\end{figure}

With the probabilities $\{\,p^{(s)}_k(0), \,p^{(s)}_k(1)\}$ and $\{p^{(\bar{s})}_m(0), p^{(\bar{s})}_m(1)\}$
utilized to send a `0' and a `1'
computed in \ref{Sec:7.1_v5}, we can now compute the Holevo channel capacity in the presence of the beam splitter as
\bea{chi:formula:BS}
\fl \chi^{(s,\bar{s})}(z,\theta) &=& \myover{{\textrm{max}} {p}}S \left[ p\,\rho^{(s)}(0)\otimes\rho^{(\bar{s})}(0) + (1-p)\,\rho^{(s)}(1)\otimes\rho^{(\bar{s})}(1) \right] \no
\fl &-& p\,\left( S[\rho^{(s)}(0)] + S[\rho^{(\bar{s})}(0)] \right)
- (1-p)\,\left( S[\rho^{(s)}(1)] + S[\rho^{(\bar{s})}(1)] \right), \no
&\equiv&
H \left[ p \,p^{(s)}_k(0) \,p^{(\bar{s})}_m(0)  + (1-p)\,\,p^{(s)}_k(1) \,p^{(\bar{s})}_m(1) \right] \no
&-& p\,\left( H[p^{(s)}_k(0)] + H[p^{(\bar{s})}_m(0)] \right)
-(1-p)\,\left( H[p^{(s)}_k(1)] + H[p^{(\bar{s})}_m(1)] \right).
\eea
The Holevo or channel capacity is the value of $\chi(z,\theta,p)$ maximized over $p$, which occurs for the value $p=1/2$.
The entropies for the various component density matrices are plotted in \Fig{entropy:figs} for $\theta=\pi/4$ (a 50:50 beam splitter) for illustration and for $p=1/2$.
Here, we again use \Eq{zstar} and \Eq{f:formula}
\Bea
\label{f:formula:v2}
f(z) &=& 4\,e^{-\pi} \,\left( \frac{1+\sqrt{z}}{1-\sqrt{z}} \right) = 4\,e^{-\pi}\,e^{2\tau}, \\
z^* &\equiv& \zeta^{*2}=\tanh^2\tau^* \myover{{\longrightarrow}{\np0\to\infty}} \frac{1}{(e^{\pi/2}/2-1)^2}\approx 0.506407,
\Eea
to plot the long-time behavior of $\chi$ as a function of $z$ for $z>z*$ in the limit $\np0\gg 1$.
%=====================================================
%=====================================================

\begin{figure}[ht]
\begin{center}
%\includegraphics[width=5in,height=3.5in]{AvS:fig1}
%\begin{tabular}{cc}
%   \includegraphics[width=3in,height=2.0in]{fig_9}
    \includegraphics[width=3in,height=2.0in]{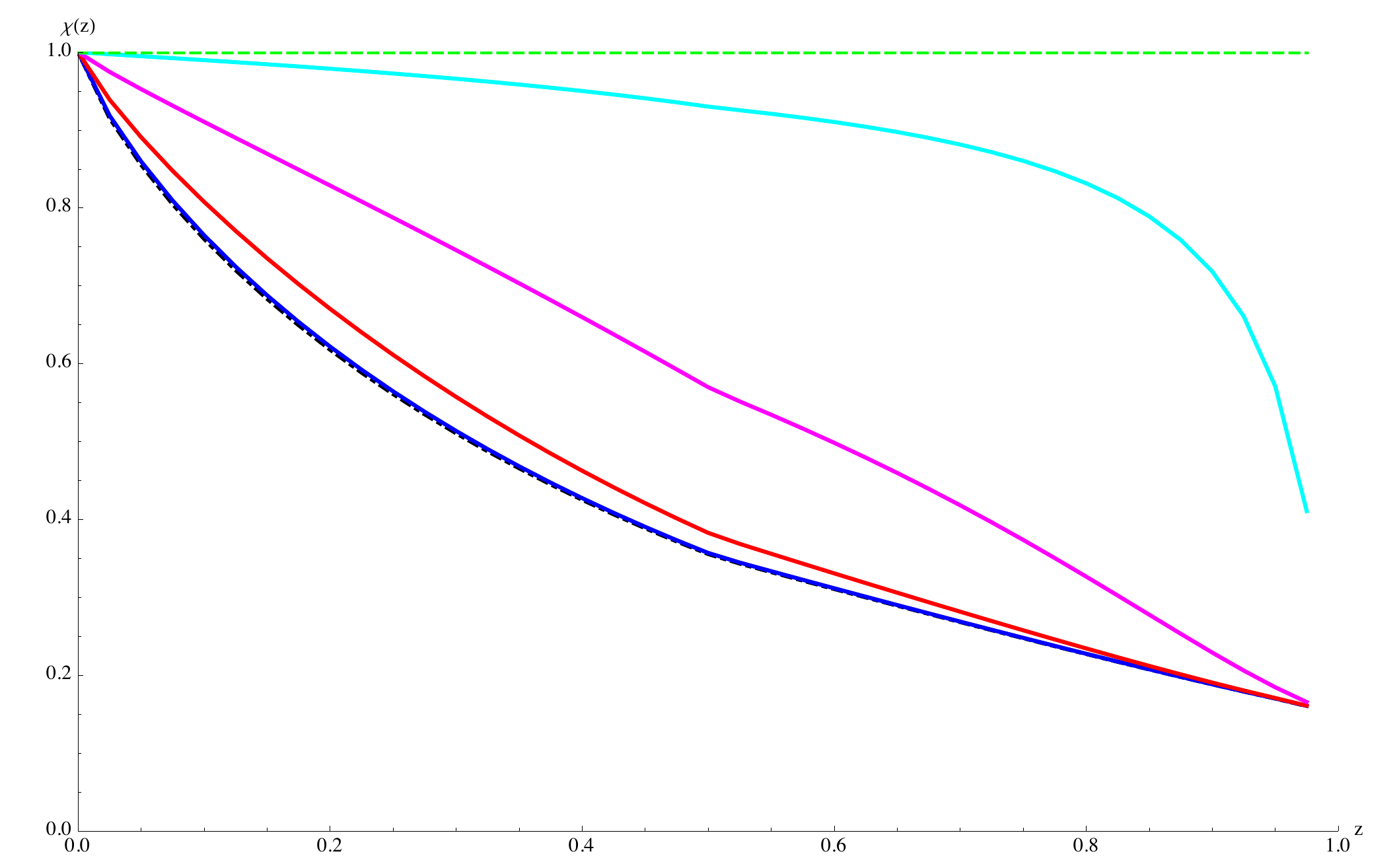}
  %\includegraphics[width=3in,height=2.5in]{X_longtime_various_theta} % &
%  \includegraphics[width=3in,height=2.4in]{chi_combined}
%\end{tabular}
\caption{\label{chi_theta:figs}  $\chi_{s,\bar{s}}(z,\theta)$  combined formula with crossover time $z^*=0.506407$ \Eq{zstar} in the limit $\np0\to\infty$ and maximized for $p=1/2$, for various values (bottom to top) of
$\theta = (0,\pi/8, \pi/4, 3\pi/8, 3.75\pi/8, \pi/2) \leftrightarrow$ (black, blue, red, magenta, cyan, green). The beams splitter transmissivity is $\cos^2\theta$; the  BH absorptivity is $|\alpha|^2 = \cos^2\theta\cosh^2 r$. (color online)
}
\end{center}
\end{figure}

In \Fig{chi_theta:figs} we plot $\chi(z,\theta)$ for $\theta = (0,\pi/8, \pi/4, 3\pi/8, 3.75\pi/8, \pi/2)$.
%=====================================================
%=====================================================
From equation \Eq{AvS:41:v2} for the observed late-time mode $A$ on $\mathcal{J}_+$  the single-quantum absorption probability of the BH is given by
$\alpha^2 = \cos^2\theta\,\cosh^2 r$ for a beam splitter transmissivity of $\cos^2\theta$.
In \Fig{chi_theta:figs}, the minimum $\chi$ occurs for $\theta=0$ (lower black dashed curve) corresponding to  a the maximum absorption probability (zero reflectivity) for a given value of the rapidity $r$ (which determines the BH temperature via $z=\tanh^2 \tau = e^{-\omega/T}$. It is interesting to note that his curve is essentially indistinguishable from the (blue) curve for the value $\theta=\pi/8$, with non-zero reflectivity.
Note that any non-zero value of the beam splitter reflectivity $\theta>0$ increases the channel capacity $\chi$. For the case $\theta=\pi/2$ we obtain the maximum and constant value of $\chi=1$ for all $z$.
This corresponds to the very special case of a perfectly reflecting beam splitter, i.e. effectively a mirror (though the states in \Eq{out_state:0} and \Eq{out_state:1} are entangled), which effects the transformation $a_s^\dagger \to -i\,a_c^\dagger$ and $a_c^\dagger \to -i\,a_s^\dagger$, which perfectly reflects the incoming $c$ mode into the outgoing $a$ mode in region $I$ and visa versa - a very unusual BH (essentially a `white hole' \cite{Bradler_Adami:2014}).

\subsection{Discussion}
%\bea{AvS:41:v2}
%A = e^{-i {\mathcal{H}}} a_s e^{i {\mathcal{H}}} &=& \alpha \, a_s - \beta \,a^\dagger_{\bar{i}} + \gamma \, a_c \no
%&=& \cosh r \, \cos\theta \, a_s - \sinh r \,a^\dagger_{\bar{i}} + \cosh r \, \sin\theta \, a_c.
%\eea
%which also reproduce the late-time observed mode $A_k$ in \Eq{AvS:41:v2}.
%

%
The above result that $\chi=1$ at $\theta=\pi/2$ for all $z$ simply indicates that the infalling late-time mode $c$ is perfectly reflected into the observed outgoing mode $s$ by the `beam-splitter' scattering process with transmittance $\cos^2\theta$. For the opposite end of unit transmittance $\theta=0$, we obtain a result similar to that of Adami and Ver Steeg \cite{Adami:2014}, namely that the channel capacity is non-zero at infinite time $(z=1)$, even when evaporation of the BH is taken into account by our dynamical BH-as-PDC model.
For the case $\theta=0$ (unit `beam-splitter' transmittance) our result differs from that of Br\'{a}dler and Adami \cite{Bradler_Adami:2014} who find that a perfectly absorbing BH ($\alpha=0$) has zero channel capacity at infinite time. The difference stems from the definition of the BH absorptivity $\alpha^2$ considered.
Both sets of author use Sorkin's \cite{Sorkin:1987} definition of the outgoing mode
in the presence of the unitary `beam-splitter' scattering process as
$A = e^{-i {\mathcal{H}}} a_s e^{i {\mathcal{H}}} = \alpha \, a_s - \beta \,a^\dagger_{\bar{i}} + \gamma \, a_c$,
such that $\alpha^2-\beta^2+\gamma^2=1$.
However, we further use the definition in \Eq{AvS:41:v2} that
$A=\cosh r \, \cos\theta \, a_s - \sinh r \,a^\dagger_{\bar{i}} + \cosh r \, \sin\theta \, a_c$ with
$\left.\alpha^2=\cosh^2 r \, \cos^2\theta\right|_{\theta=0}=\cosh^2 r$
%\myover{{\rightarrow} {\theta=0}}\cosh^2 r$
yielding a finite BH absorptivity for unit transmittance of the `beam-splitter.' Thus, in our definition \Eq{AvS:41:v2} we can never obtain the limit of a perfectly absorbing BH $\alpha=0$ even at unit `beam-splitter' transmittance scattering. Our results are consistent with those of Br\'{a}dler and Adami for an imperfectly absorbing BH, $\alpha > 0$.

\section{On the Page Information in the BH radiation}\label{Page}
The generally accepted conventional wisdom for when information leaks out of the BH stems from the seminal 1993 work of Page  in which he calculated (i) the average information in a subsystem \cite{Page:1993a} and (ii) then applied this to the information in the BH radiation \cite{Page:1993b}. The main result of this work is the \textit{Page time} which is roughly half the evaporation time of the BH  when the information in the outgoing Hawking radiation becomes appreciable. Here we briefly review the main results of Page's work and then compute the same relevant quantities for the case of the Hamiltonian
$H_{p,s,\bar{i}} = r (a_p \, a_s^\dagger \, a_{\bar{i}}^\dagger + a^\dagger_p \, a_s \, a_{\bar{i}})$.

\subsection{Review of Page's results}
In the first paper, \textit{The average entropy of a subsystem} \cite{Page:1993a}, Page considers a random pure state of fixed dimension $N= m n$ selected from the Haar measure, which is proportional to the standard geometric hypersurface volume on the unit sphere
$S^{2 m n}-1$ which those unit vectors give when the $m n$-complex-dimensional Hilbert
space is viewed as the $2 m n$-real-dimensional Euclidean space. The integer divisors $m,n$ of $N$ are considered as the dimensions subsystems $A$ and $B$ respectively, taking without loss of generality,
$m < n$. The goal is to compute the average entropy
$\langle S_A \rangle = S_{m,n}$ (with respect to the Haar measure) and the information $I_{m,n} = (S_A)_{max} - \langle S_A \rangle = \ln m - S_{m,n}$. The end result of this work is the following \cite{Page:1993a}:
$S_{m,n} = \sum_{k=n+1}^{m n} 1/k - (m-1)/(2n)\approx \ln m - m/(2n)$ for $1\ll m \le n$.
The conclusion drawn from this work is that for a typical pure state of the composite system, (i) very little of the information, roughly $m/(2n)$ units, are in the correlations within the smaller subsystem A itself, (ii) roughly $\ln n- \ln m + m/(2n)$ units are in the correlations within the larger subsystem B itself, and (iii) the remaining roughly $2 \ln m - m/n$ units of information are in the correlations between the larger
and smaller subsystems.

%====================================
In the second paper \textit{Information in black hole radiation} \cite{Page:1993b} subsystem $A$
of dimension $m$ is taken to be the Hawking radiation of the BH, system $B$ of dimension $n$. The  subsystems $A$ and $B$ are assumed to be correlated via the composite random pure state $\ket{\psi}_{A B}$, such that $\rho_A = Tr_B[\ket{\psi}_{AB}\bra{\psi}]$, and
$\rho_B = Tr_A[\ket{\psi}_{AB}\bra{\psi}]$.
 To model this, Page considers a large integer $N = m n = 291,600$, which has 105 integer factors $m$.
 For $m\le n$ he computes $S_{m,n} = \sum_{k=n+1}^{m n} 1/k - (m-1)/(2n)$, while for $m > n$ he uses
 $S_{m,n} = \sum_{k=m+1}^{m n} 1/k - (n-1)/(2 m)$, while in both regimes $I_{m,n} = \ln m - S_{m,n}$. This is plotted in \Fig{Page:S_I_vs_lnm} below.
%===================================
\begin{figure}[ht]
\begin{center}
%\includegraphics[width=5in,height=3.5in]{AvS:fig1}
%\begin{tabular}{cc}
%  \includegraphics[width=4in,height=2.25in]{fig_10}
    \includegraphics[width=4.0in,height=1.95in]{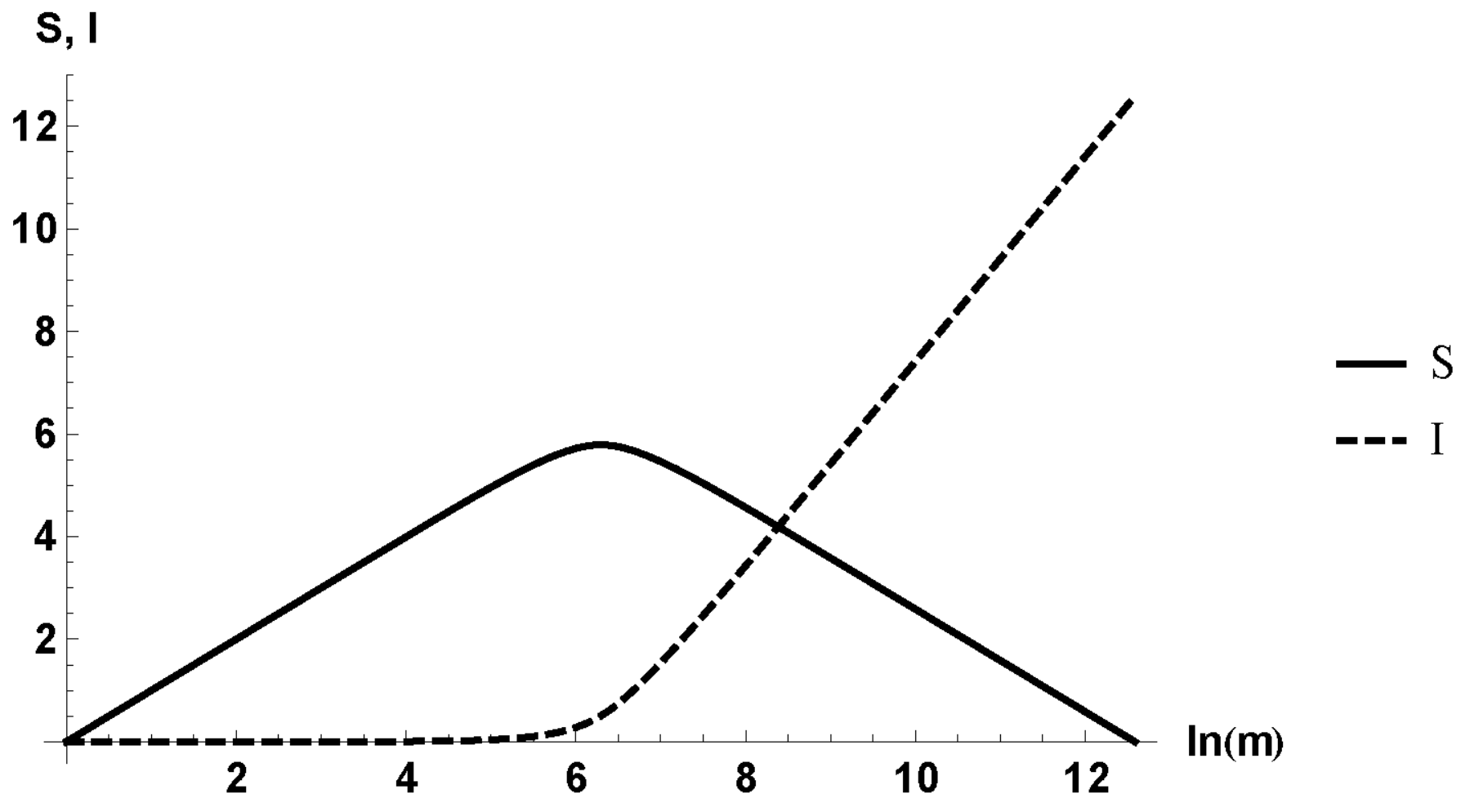}
  %\includegraphics[width=5in,height=3.0in]{fig_bw_Page_S_I_vs_lnm} % &
%  \includegraphics[width=3in,height=2.4in]{chi_combined}
%\end{tabular}
\caption{\label{Page:S_I_vs_lnm} Page entropy of BH radiation (solid) $S_{m,n}$ and information (dashed) $I=\ln m - S_{m,n}$ vs $\ln m$ for $m n = 291,600$ after \cite{Page:1993b}).
The interpretation is that the information in the BH leaks out into the outgoing radiation at roughly half the evaporation (Page) time of the BH.
}
\end{center}
\end{figure}
%==================================================
%%====================================
%In the second paper \textit{Information in black hole radiation} \cite{Page:1993b} subsystem $A$
%of dimension $m$ is taken to be the Hawking radiation of the BH, system $B$ of dimension $n$. The  subsystems $A$ and $B$ are assumed to be correlated via the composite random pure state $\ket{\psi}_{A B}$, such that $\rho_A = Tr_B[\ket{\psi}_{AB}\bra{\psi}]$, and
%$\rho_B = Tr_A[\ket{\psi}_{AB}\bra{\psi}]$.
%To model this, Page considers a large integer $N = m n = 291,600$, which has 105 integer factors $m$. For $m\le n$ he computes $S_{m,n} = \sum_{k=n+1}^{m n} 1/k - (m-1)/(2n)$, while for $m > n$ he uses $S_{m,n} = \sum_{k=m+1}^{m n} 1/k - (n-1)/(2 m)$, while in both regimes $I_{m,n} = \ln m - S_{m,n}$. This is plotted in \Fig{Page:S_I_vs_lnm} below.
%%===================================

\subsection{Page information with the trilinear Hamiltonian $H_{p,s,\bar{i}}$}
\begin{figure}[ht]
\begin{center}
\begin{tabular}{cc}
   \includegraphics[width=3.25in,height=2.0in]{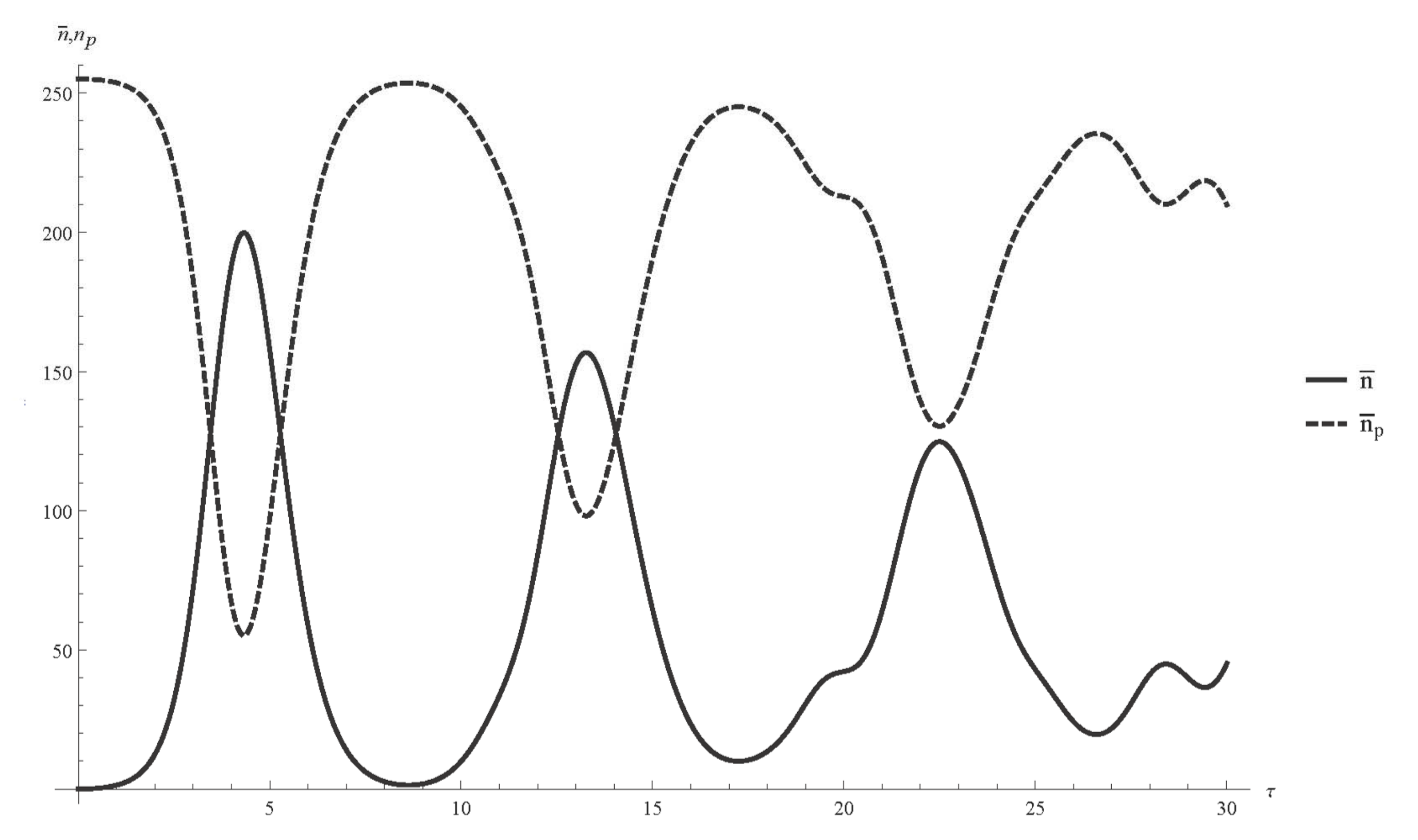}   &
   \includegraphics[width=3.25in,height=2.0in]{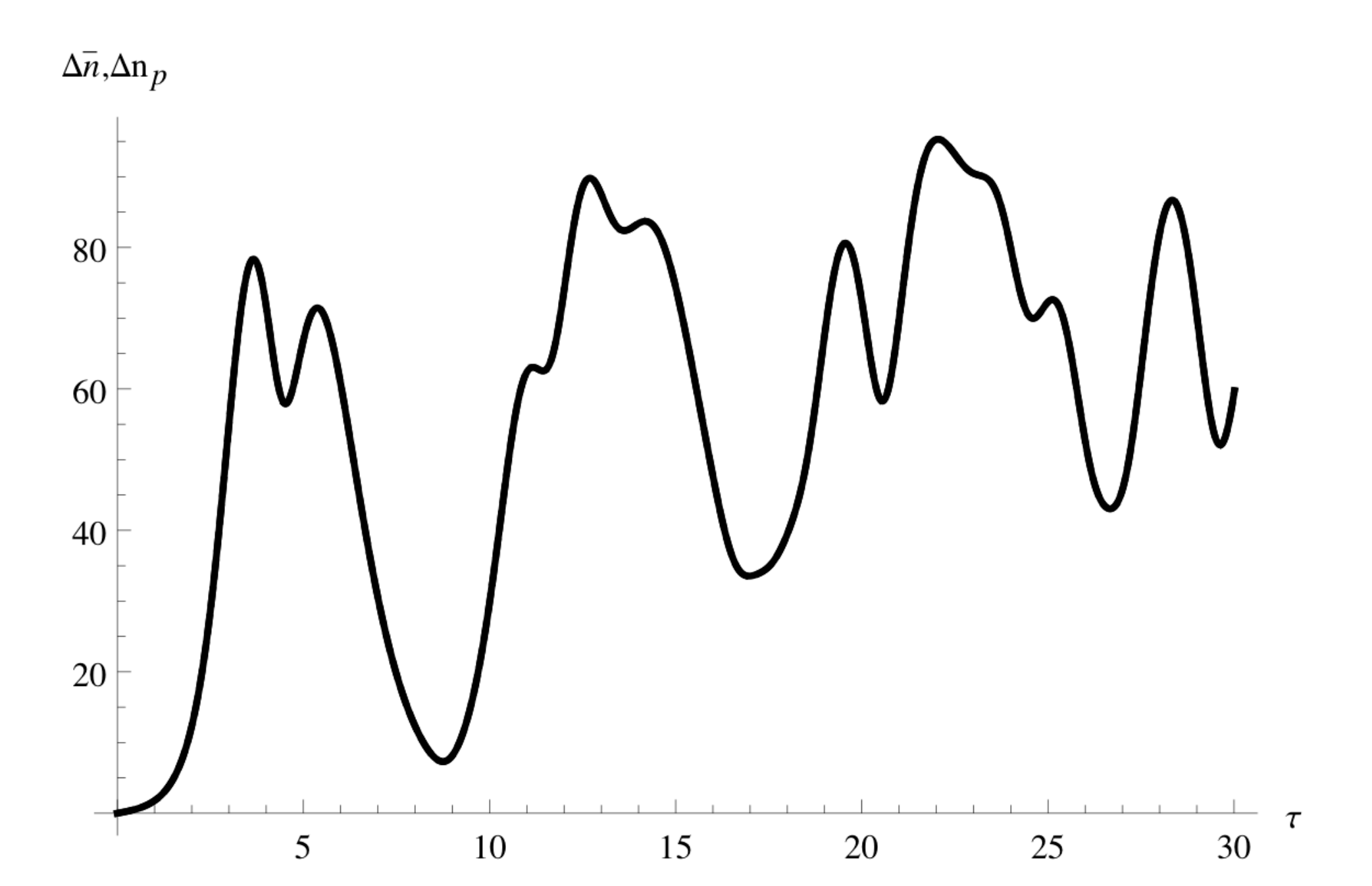}
\end{tabular}
\caption{\label{n_np_dn_dnp_vs_tau}  (left) Mean number $\bar{n}_s$, $\bar{n}_p$ of (solid) signal  and (dashed) pump particles, (right) variances $\Delta{n}_s=\Delta{n}_p$, vs $\tau$
from numerical evolution of quantum amplitude equations \Eq{cn:eqn}.
}
\end{center}
\end{figure}

For the dynamical model considered in this present work we plot
(\Fig{n_np_dn_dnp_vs_tau}, left)
the mean occupations number $\bar{n}_s$, $\bar{n}_p$ of signal and pump particles
%the corresponding formulas of $S$ and $I$ developed by Page above
using the state $\ket{\psi}_{out} = \sum_n c_n \ket{n}_L = \sum_n c_n \ket{\np0-n}_p\ket{\ns0+n}_s\ket{n}_{\bar{i}}$ for the Hamiltonian $H_{p,s,\bar{i}} = r (a_p \, a_s^\dagger \, a_{\bar{i}}^\dagger + a^\dagger_p \, a_s \, a_{\bar{i}})$. For the BH radiation subsystem we take $\rho_A = \rho_s = \sum_{n=0}^{\np0} |c_n|^2 \ket{\ns0+n}_s\bra{\ns0+n}$, and set $\ns0=0$ for simplicity. We numerically solve the quantum amplitude equations \Eq{cn:eqn} with $\np0=255=2^8-1$ and take $\ket{\psi(0)}_{in}=\ket{0}_L = \ket{\np0}_p\ket{0}_s\ket{0}_{\bar{i}}$, although the behaviors exhibited below are qualitatively similar once $\np0 \gg \ns0$, involving essentially a change in scale depending on the choice of $\np0$. We also plot results vs $\tau \equiv r t$ instead of vs. $z=\tanh^2 \tau$ since the latter tends to compress details for large times (roughly $\tau > 5$, corresponding to $z = 0.999818$, for $\np0=255$).
In the right plot of \Fig{n_np_dn_dnp_vs_tau} we plot the variances $\Delta{n}_s$, $\Delta{n}_p$,
where $\bar{n}_s = \sum_n n \, p_n$,  $\bar{n}_p = \sum_n (\np0-n) \, p_n$,
$\Delta{n}^2_s = \sum_n (n-\bar{n})^2 \, p_n$, and
$\Delta{n}^2_p = \sum_n \big((\np0-n)-\bar{n}_p\big)^2 \, p_n$.
%=============================================================
% NEW TEXT 1Jan2015
%=============================================================
%=========================================
%=========================================

In the left plot \Fig{n_np_dn_dnp_vs_tau} we see the oscillatory behavior of the mean number of signal and pump particles. For $\tau \le 1, (z\le0.58)$ and even up to times $\tau \le 2, (z\le0.93)$
we are in the short-time regime where $\bar{n}_p(\tau)\sim\np0\gg \bar{n}_s(\tau) $, and $d\bar{n}_s/d\tau$ and $d\bar{n}_p/d\tau$ are relatively flat. At $\tau=3.49, (z=0.996)$ the populations become equal
$\bar{n}_p(\tau)=\bar{n}_s(\tau)$.
During times $2 \le \tau \le 4.39, (0.93 \le z\le 0.999434)$ the BH pump rapidly depletes roughly $80\%$ (a phenomena known in the quantum optics community, see also \cite{Bandilla:2000}) of its particles into the signal, with an almost complete reversal of roles population-wise of the pump and the signal at $\tau = 4.39$ where $d\bar{n}_p/d\tau=0$. Beyond this time the dynamical model is no longer a valid physical representation of BH evaporation since $d\bar{n}_p/d\tau>0$ corresponds to the situation where the Hawking radiation flows back into the BH.
In the right plot of \Fig{n_np_dn_dnp_vs_tau} we observe the variances $\Delta{n}_s$, $\Delta{n}_p$
rise rapidly, reaching a peak at $\tau=3.49$, and their first minimum at $\tau = 4.39$.

To investigate the Page information we follow Page's second paper \cite{Page:1993b} and \cite{Nation:2010} and define the information as
\be{PageInfo}
I(\tau) = S_{thermal}(\tau) -  S\big(\rho_s(\tau)\big),
\ee
where $S_{thermal}$ is the effective thermal distribution $\rho_{thermal}(z_{thermal})$ with probability distribution
given by
$p_n^{thermal}=\left(1-z_{thermal}\right) z_{thermal}^n$ (\Eq{pn:shortime} for $\ns0=0$) with $z_{thermal}= \bar{n}_s/(\bar{n}_s+1)$, and $\bar{n}_s$
computed from $\rho_s(\tau) = \Tr_{p,\bar{i}}\left[\ket{\psi(\tau)}\bra{\psi((\tau))}\right]$.
We further model the evolution utilizing $\ket{\psi(0)}_{in} = \ket{\alpha}_p\ket{0}_s\ket{0}_{\bar{i}}$ with the pump in an initial coherent state $\ket{\alpha}_p = e^{-|\alpha|^2/2}\,\sum_{m=0}^\infty \alpha^m/\sqrt{m!}\,|m>_p$, with corresponding new quantum amplitude equations for the pure state vector
\bea{eqn:cn_with_CS_pump}
\hspace{-2cm}
\ket{\psi}_{out} &=& \sum_{\np0=0}^\infty \sum_{n=0}^{\np0} c_{\np0,n} \ket{\np0,n}_L = \sum_{\np0=0}^\infty \sum_{n=0}^{\np0} c_{\np0,n} \ket{\np0-n}_p\ket{\ns0+n}_s\ket{n}_{\bar{i}}, \,\qquad\\
\hspace{-2cm}
\ket{\psi}_{in}&=&\ket{\alpha}_p\ket{0}_s\ket{0}_{\bar{i}},
\quad \bar{n}_p(0) = \alpha^2 = 35, \quad \Delta n_p(0) = \alpha = 5.92\nonumber
\eea
The above equations reveal that each Fock state $\ket{\np0-n}$,  with non-negligible probability distribution under the coherent state $\ket{\alpha}_p$, evolves independently, as in the previous sections.
This leads to evolution curves for each $\np0$ which oscillate in time at slightly different frequencies. The eventually leads to a 'collapse' phenomena (well studied in the quantum optics community, see e.g. \cite{Agarwal:2013}) when all the initially in-phase oscillating curves de-phase after some time (with longer, periodic Poincare' recurrences), with $\bar{n}_p$ reaching a long-time quasi-steady state of roughly $80\%$ of it's initial mean of $|\alpha|^2$, and $\bar{n}_s$ correspondingly reaching roughly $20\%$ of $|\alpha|^2$.

To define the effective dimension of the subsystem $\rho_s$ Nation and Blencowe \cite{Nation:2010} utilized the definition
from Popescu \textit{et al.} \cite{Popescu:2006}
\be{dim:NB}
\tilde{d}^{eff}_i = \frac{1}{Tr\left[\left(\rho_{i, thermal}\right)^2\right]}
= \frac{1+z_{thermal}}{1-z_{thermal}} = 2\,\bar{n}_i + 1,
\quad ,i\in\{p, s, \bar{i}\} ,
\ee
with $\rho_{i, thermal}(z_{thermal})$ defined above. Here $Tr[\rho^2]$ is the purity of the state $\rho$, and $\tilde{d}^{eff}_i$
is constructed to capture the number of states with non-negligible probability under the its variance.

We will instead utilize the definition
\be{dim:PMA}
d^{eff}_i = 1 + \Delta n_i
= 1 + \left[\bra{\psi} \hat{n}^2_i(\tau) \ket{\psi} - \bra{\psi} \hat{n}_i(\tau) \ket{\psi}^2\right]^{1/2}.
\ee
with $\hat{n}_i = a^\dagger_i\,a_i$ and variances
$(\Delta n_i)^2 = {}_{out}\bra{\psi} (n_i- \bar{n}_i)^2 \ket{\psi}_{out}$.
This definition directly measures the
number of states with non-negligible probability under the probability distribution. For a Fock state $\ket{n}$ we have
$d^{eff}_{Fock}=1$ and for the thermal state we have $d^{eff}_{thermal}=\sqrt{\bar{n}\,(\bar{n}+1)}\approx \bar{n}$ for $\bar{n}\gg 1$.

\begin{figure}[ht]
\begin{center}
\begin{tabular}{cc}
  \includegraphics[width=3.0in,height=2.0in]{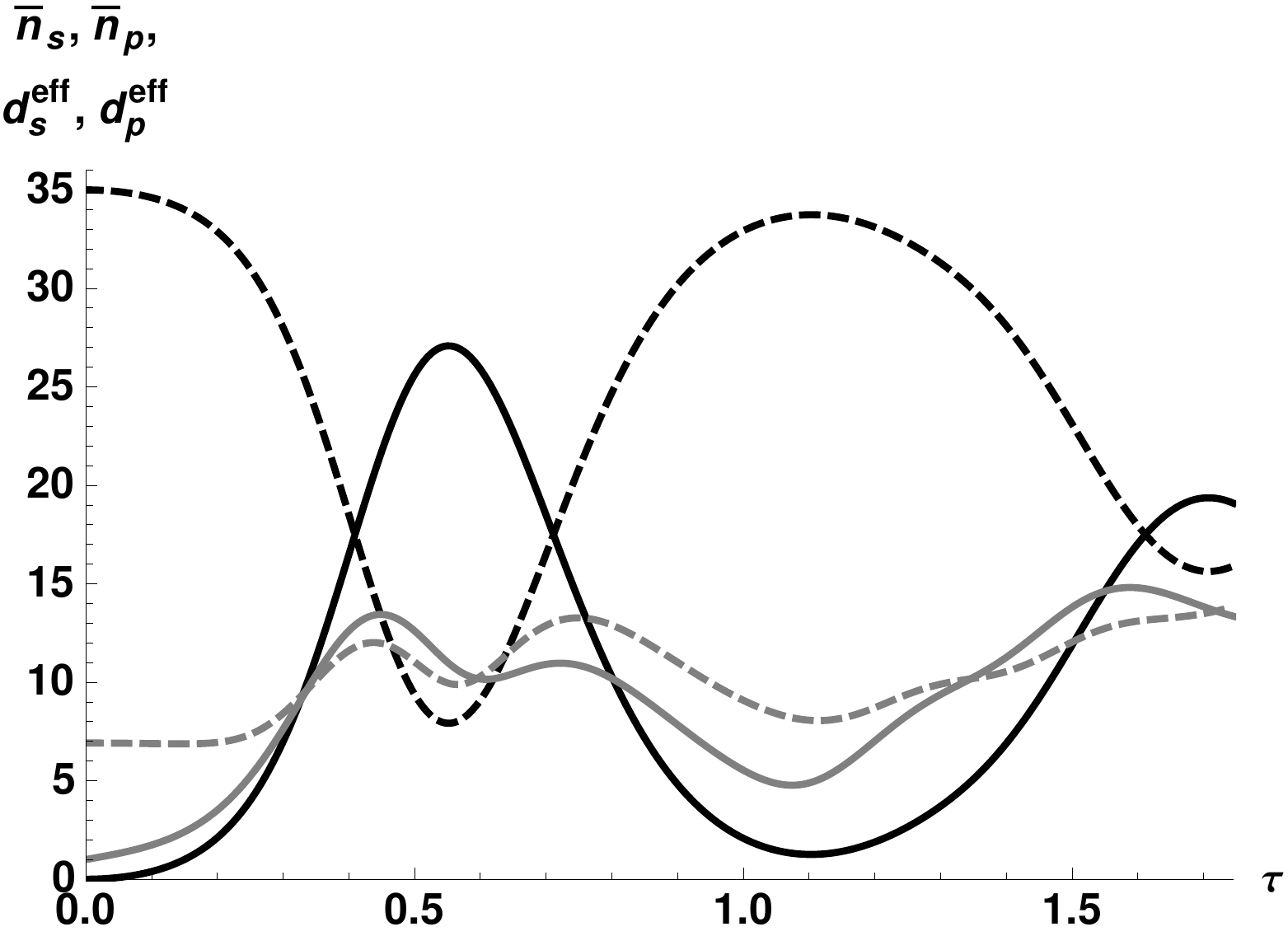}   &
  \includegraphics[width=3.0in,height=2.0in]{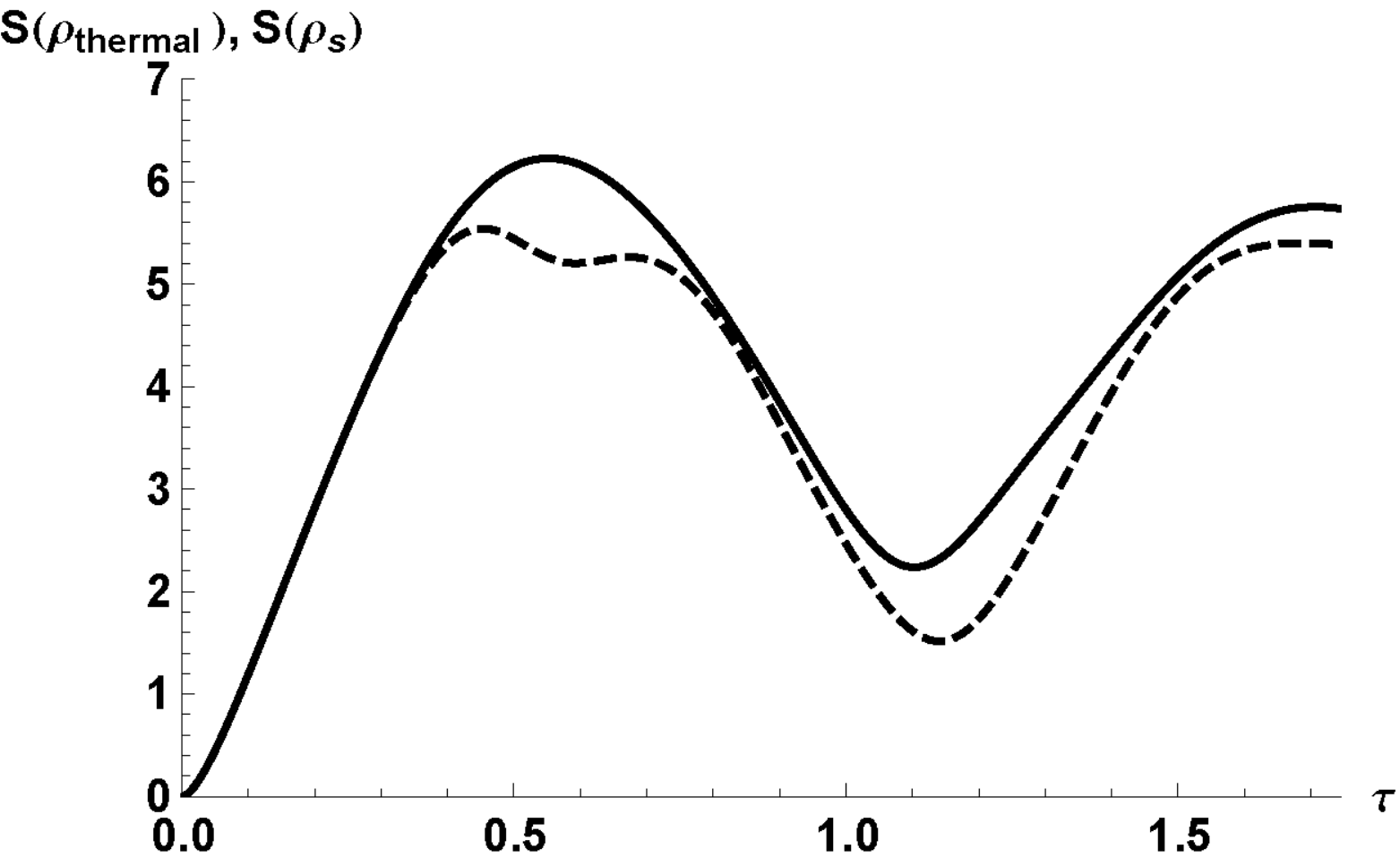}
\end{tabular}
\caption{\label{np_ns_variances_and_Stherm_S}  (Left) mean populations (black-solid) $\bar{n}_s$,
(black-dashed) $\bar{n}_p$, and effective dimensions (gray-solid) $d^{eff}_s = 1 + \Delta n_s$ and
(gray-dashed) $d^{eff}_p = 1 + \Delta n_p$, \Eq{dim:PMA}.
(Right) (black-solid) $S(\rho_{thermal})$ and (black-dashed) $S(\rho_s)$.
}
\end{center}
\end{figure}

In the left plot of \Fig{np_ns_variances_and_Stherm_S} we show the mean populations (black-solid) $\bar{n}_s$,
%=========================================
%=========================================
(black-dashed) $\bar{n}_p$, and the effective dimensions (gray-solid) $d^{eff}_s$ and
(gray-dashed) $d^{eff}_p$.  The effective dimensions, proportional to the variances of $\rho_s$ and $\rho_p$, are equal at $\tau\approx 0.34$.
The populations, essentially the variances of $\rho_{s,thermal}$ and $\rho_{p,thermal}$ for large $\bar{n}_s$, $\bar{n}_p$, are equal $\bar{n}_p = \bar{n}_p$ at $\tau\approx 0.41$.
The BH `pump' population $\bar{n}_p$ reaches its minimum value
$d\bar{n}_p/d\tau=0$ (again, with $\bar{n}_p$ about $80\%$ of its initial value) at $\tau\approx 0.55$, where the model ceases to be a physical representation of BH evaporation.
In the right plot of \Fig{np_ns_variances_and_Stherm_S} we show $S(\rho_{thermal})$ and $S(\rho_s)$ and observe that the latter begins to show deviations ($S(\rho_s) < S(\rho_{thermal})$) just around the time when
$d^{eff}_s\approx d^{eff}_p$, i.e. when the variances of $\rho_s$ and $\rho_p$ approach equality.

\begin{figure}[ht]
\begin{center}
\begin{tabular}{cc}
  \includegraphics[width=3.25in,height=2.5in]{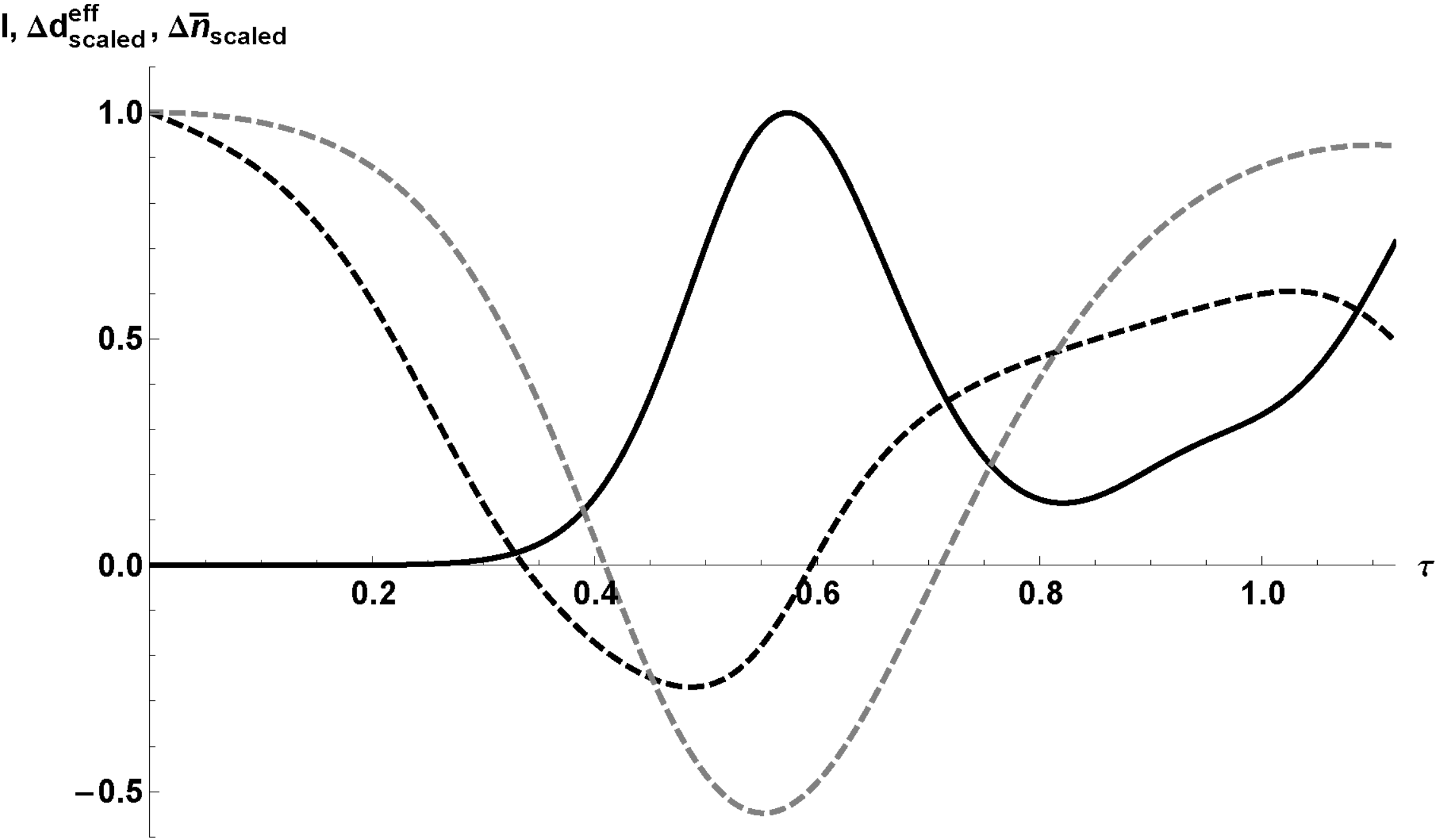} &
  \includegraphics[width=3.0in,height=2.5in]{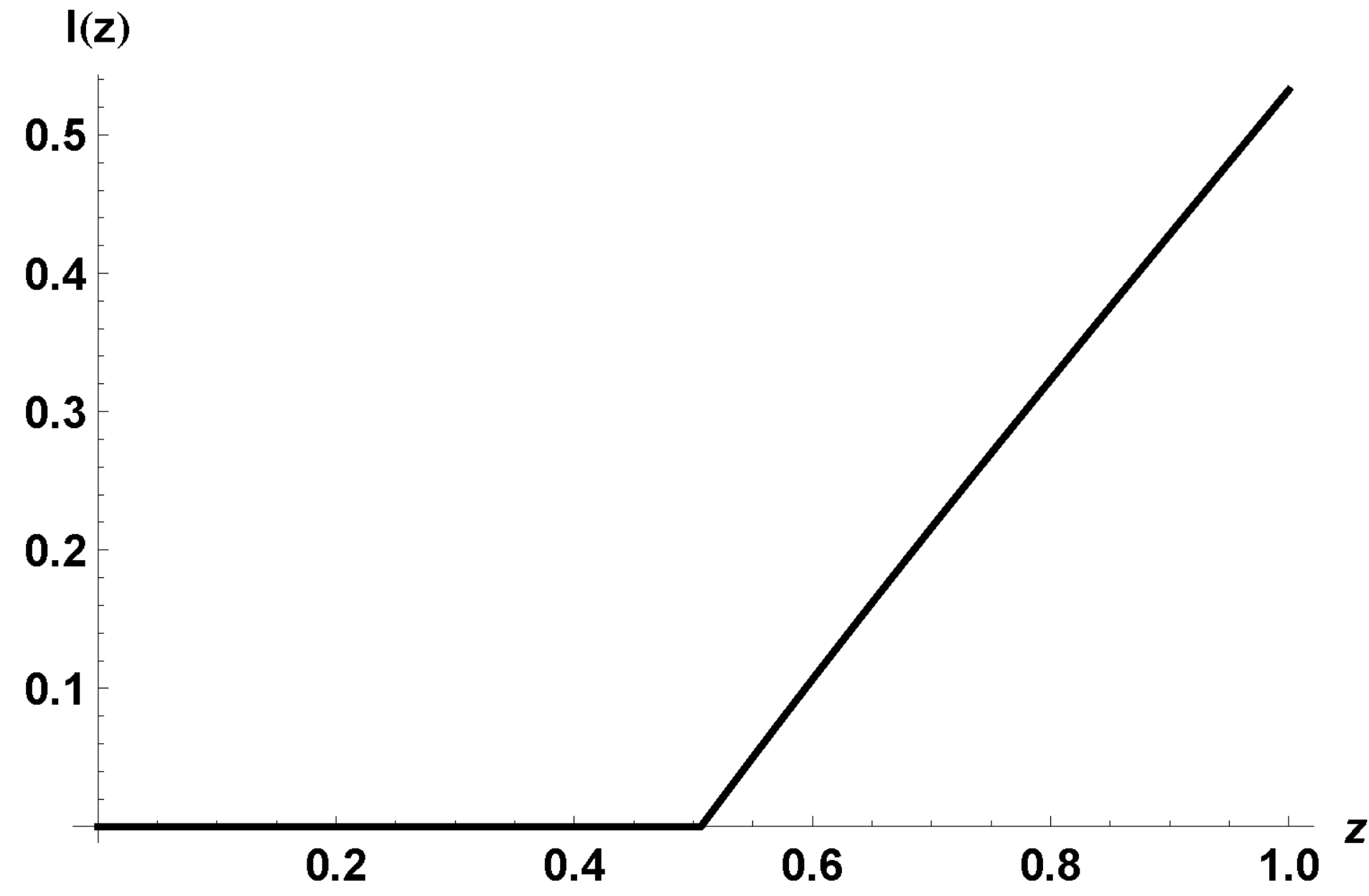}
\end{tabular}
\caption{\label{Info_scaled_dim_and_np_ns_diff} (left): (black-solid) Page information $I(\tau)$ \Eq{PageInfo}
from numerical simulation of \Eq{eqn:cn_with_CS_pump},
(black-dashed) $\Delta d^{eff}_{scaled} \equiv (d^{eff}_p-d^{eff}_s)/\textrm{max}(d^{eff}_p-d^{eff}_s)$ and
(gray-dashed) $\Delta \bar{n}_{scaled} \equiv (\bar{n}_p - \bar{n}_s)/\textrm{max}(\bar{n}_p - \bar{n}_s)$.
(right): Page information from single period analytic solutions (\Eq{pn:shortime} and \Eq{pn:longtime}) of $\rho_<(z)$ for $z\le z^*=0.506407$ and $\rho_>(z)$ for $z>z^*$, with $S_{thermal}\equiv S(\rho_<)$ for all $z$.
}
\end{center}
\end{figure}

In the left plot of \Fig{Info_scaled_dim_and_np_ns_diff} we show the Page Information (black-solid) \Eq{PageInfo}, which begins to become
non-negligible when $\Delta d^{eff}=d^{eff}_p - d^{eff}_s\approx 0$ (variance difference of $\rho_{thermal}$ and $\rho_s$), which we have scaled to its maximum value (black-dashed) to more
%=========================================
%=========================================
clearly note its zero crossing.
We also plot the scaled difference of the populations
$\Delta \bar{n}_{scaled} \equiv (\bar{n}_p - \bar{n}_s)/\textrm{max}(\bar{n}_p - \bar{n}_s)$ (gray-dashed) to more clearly note the zero-crossing of the approximate variance difference (for large $\bar{n}_p$, $\bar{n}_s$) of
$S(\rho_{p,thermal})$ and $S(\rho_{s,thermal})$. We also observe a peak in the information $I(\tau)$ at the time when
$d\bar{n}_p/d\tau$ ceases to be negative.
As a means of comparison, in the right plot of \Fig{Info_scaled_dim_and_np_ns_diff}
we show the Page information from single period analytic solutions (\Eq{pn:shortime} and \Eq{pn:longtime}) of $\rho_<(z)$ for $z\le z^*=0.506407$ and $\rho_>(z)$ for $z>z^*$. The abrupt turn on of the Page information at $z^*$ is a consequence of our utilization of the definition $S_{thermal}\equiv S(\rho_<)$ for all $z$.

\subsection{Discussion}
The conclusion from these results is that the trilinear Hamiltonian as a simplified (zero-dimensional) model of BH particle production/evaporation does reproduce the Page Information results \cite{Page:1993b} precisely because it
incorporates the non-thermal behavior of the long time Hawking radiation as the BH 'pump' source dynamically transfers populations into the signal and idler modes during evaporation (similar results were also obtained by Nation and Blencowe \cite{Nation:2010}). In addition, we conclude that Page's assertion that the  information becomes non-negligible essentially around the time the BH has deposited half its population into the Hawking radiation $\bar{n}_p=\bar{n}_s$ can also be interpreted as the time when the variance of the BH particle number becomes equal to the variance of the number of particles in the Hawking radiation $\Delta n_p = \Delta n_s$, under an evaporating BH (with a dynamical BH 'pump'). The widely known results (in the quantum optics community) that under the trilinear Hamiltonian
the pump rapidly depletes roughly $80\%$ of its population into the signal and idler modes at $d\bar{n}_p/d\tau=0$
\footnote{And also in steady state, at very long times where the validity of this model for BH particle production/evaporation ceases to be physical since $\bar{n}_p$ repeatedly passes through regimes of $d\bar{n}_p/d\tau>0$.}
would most likely have to be modified
in a more complete open-system  model (master equation see \cite{Walls:1994})
which more carefully models the transport of the Hawking radiation away
from the trilinear Hamiltonian interaction region. This would lead to a completely evaporated BH with $\bar{n}_p=0$ when $d\bar{n}_p/d\tau=0$ for the very first time (analogous
to the single laser `burst' discussed after \Eq{nbar:longtime:sech}).

The salient point of the above simulation is that the \textit{short-time} behavior utilizing $\ket{\psi(0)}_{in} = \ket{\alpha}_p\ket{0}_s\ket{0}_{\bar{i}}$ is effectively that of
$\ket{\psi(0)}_{in} =\ket{\np0}_p\ket{0}_s\ket{0}_{\bar{i}}$ with $\np0=|\alpha|^2$, the Fock number state corresponding to the mean of the coherent state $\ket{\alpha}_p$, since for early times all the Fock pump number states under $\ket{\alpha}_p$ remain approximately in-phase. Hence, the left plot of \Fig{np_ns_variances_and_Stherm_S} for short-times (until the first crossing $\bar{n}_s=\bar{n}_p$) and for long-times until $d\bar{n}_p/d\tau=0$ is nearly identical for both initial conditions, Fock and coherent state of the BH 'pump' source. Again, the dimension of both $\rho_s$ and $\rho_p$ rises rapidly to a common saturated value $\approx\mathcal{O}(\np0/2)$, as the variances steadily equilibrate in time. This means that the analytic results of the previous sections using a Fock number state
for the initial state of the BH  $\ket{\psi(0)}_{in} = \ket{\np0}_p\ket{0}_s\ket{0}_{\bar{i}}$ give qualitatively equivalent results when using a more reasonable  initial condition utilizing a coherent state pump.

%===================================================================
%===================================================================
\section{Monogamy of Entanglement Considerations}\label{EntMonog}
As a step toward considering issues concerning the monogamy of entanglement \cite{Lloyd:2014,Mathur:2009,Hayden:2013} during the process of BH evaporation, we consider two scenarios below of entanglement of the outgoing Hawking radiation mode $s$ with an external region $I$ mode $c$. Though we do not explicitly model scenarios in which two participants fall behind the BH horizon in an effort to observe entanglement (as in \cite{Lloyd:2014} and references therein), we do
purport that our model does have something germane to say on the issue of entanglement distributed amongst the various modes involved.

The principle of monogamy of entanglement \cite{Lloyd:2014,Terhal:2003,Koashi_Winter:2004} asserts that if
quantum systems A and B are maximally entangled, then neither can be correlated with any other system.
In the following we will take subsystem $A$ to be $s$, and subsystem $B$ to be the external mode $c$
under the scenarios
(i) $c$ is initially maximally entangled with $s$, the later of which falls inward and
contributes to the formation of the BH and
(ii) $c$ is initially separable with $s$, but at late-times $c$ falls inward and scatters off the formed BH.
In the first case (i) we will see that the initially maximal entanglement between mode $s$ and $c$ is degraded,
and in the second case (ii) the initial zero entanglement grows. In both cases the change from the
initial entanglement is due to the interaction with the common mode $s$, which in turn interacts with
modes $p$ and $\bar{i}$
(through $H_{p,s,\bar{i}}$) and with $c$ (either from its initial condition with $s$ in case (i), and
through the unitary `beam splitter late-time scatter process $H^{(s,c)}_{bs}$ in case (ii)).

\subsection{Entangled Initial State}\label{ent_IS}
%As a step toward considering issues concerning the monogamy of entanglement \cite{Lloyd:2014,Mathur:2009,Hayden:2013} %during the process of BH evaporation,
For our first case, we consider a maximally entangled initial state (\textit{ent\,I.S.}) between modes $s$ and $c$ in region $I$, i.e.
\be{psi:init_ent}
\hspace{-1cm}
\ket{\psi}^{(ent\,I.S.)}_{in} = \ket{\np0}_p\ket{0}_{\bar{i}}\,\left(\ket{\ns0}_s\ket{0}_c + \ket{0}_s\ket{\nc0}_c\right)/\sqrt{2} \equiv
\ket{\np0}_p\ket{0}_{\bar{i}}\ket{\phi}^{(ent\,I.S.)}_{s,c}\,\,
\ee
where mode $c$ stays forever outside and non-interacting with the BH (the evolution proceeds by $H_{p,s,\bar{i}}$).
%We consider the case when $\nc0 = \ns0$.
We are interested in the evolution of the entanglement of $\rho_{s,c}$ between mode $s$ which acts in the formation of the BH and mode $c$ with does not.

The partial transpose on the initial density matrix formed from $\ket{\phi}^{(ent\,I.S.)}_{s,c}$ in \Eq{psi:init_ent} has negative eigenvalue $-1/2$ and thus $E_{\mathcal{N}}\big(\rho_{s,c}(0)\big)=1$. As previously considered, we also have $E_{\mathcal{N}}\big(\rho_{(p,\bar{i}),s}(0)\big)=0$.
We can think of the state $\ket{\psi}^{(ent\,I.S.)}_{in}$ as arising as follows. Consider two early-time modes $(s,s')$ on ${\mathcal{J}_-}$ in \Fig{AvS:fig1} that are maximally entangled according to $\ket{\phi}^{(ent\,I.S.)}_{s,s'}$. We allow mode $s$ to fall into the BH as
usual, but require $s'$ to stay far away an non-interacting with the BH event horizon for all times. We can take the unitary evolution to be $e^{-i H_{p,s,\bar{i}}t}\otimes e^{-i a^\dagger_c a_c t}$.
While the state $\ket{0}_{s'}$ picks up no phase, the state $\ket{\nc0}_{s'}$ evolves to
$e^{-i \nc0 t}\ket{\nc0}_{s'} = e^{-i (\nc0/\sqrt{\np0})\, \tau} \ket{\nc0}_{s'} \approx \ket{\nc0}_{s'}$ where we have used
$\tau\sim\sqrt{\np0}t$ and $\np0 \gg \nc0$. We then take the mode $s'$ to be the late time mode $c$, which again remains non-infalling and non-interacting.

For transparency of discussion, we utilize only the short-time expressions for the quantum amplitudes in \Eq{cn:shorttime}
which we write as
\be{cn:shorttime:ns0}
\hspace{-1cm}
c^<_n(\tau) \equiv c^{(\ns0)}_n = \left[(1-z)^{\ns0+1}\,z^n\,\binom{\ns0+n}{n}\right]^{1/2}, \quad c^{(0)}_n = \sqrt{(1-z)\,z^n},
\ee
for all $0\le z \le 1$ (since the long-time solution for $z>z^*$ introduces only minor modifications). The initial state $\ket{\psi}^{(ent\,I.S.)}_{in}$ then evolves to
\be{psi:ent}
\hspace{-2cm}
\ket{\psi(\tau)}^{(ent\,I.S.)}_{out} =
\sum_n \,\ket{\np0-n}_p\ket{n}_{\bar{i}}\,\frac{1}{\sqrt{2}}\,\left(c^{(\ns0)}_n\,\ket{\ns0+n}_s\ket{0}_c + c^{(0)}_n\,\ket{n}_s\ket{\nc0}_c\right).
\ee

In forming the density matrix $\rho_{s,c}(\tau)$, we find the partial transpose on mode $c$ yields a negative eigenvalue
${\mathcal{N}_n^{(s,c)}} = -1/2\,|c_n^{(0)}\,c_n^{(\ns0)}|$ for $\ns0>0$ (for $\ns0=0$, $\ket{\phi}^{(ent\,I.S.)}_{s,c}$ evolves separably for all times) for each value of $n$ (arising from a re-writing of the off-diagonal terms of
$\rho^{PT_c}_{s,c}(\tau)$ as discussed in Section~\ref{section:Ent:Hpsibar}) leading to a log-negativity given by
\be{LogNeg:s-c:ent}
E^{(ent\,I.S.)}_{\mathcal{N}}(\rho_{s,c}) = \log_2\left[ 1 + (1-\delta_{\ns0,0})\sum_n\,|c_n^{(0)}\,c_n^{(\ns0)}| \right],
\ee
%=========================================
\begin{figure}[ht]
\begin{center}
\begin{tabular}{cc}
  \includegraphics[width=3.25in,height=2.5in]{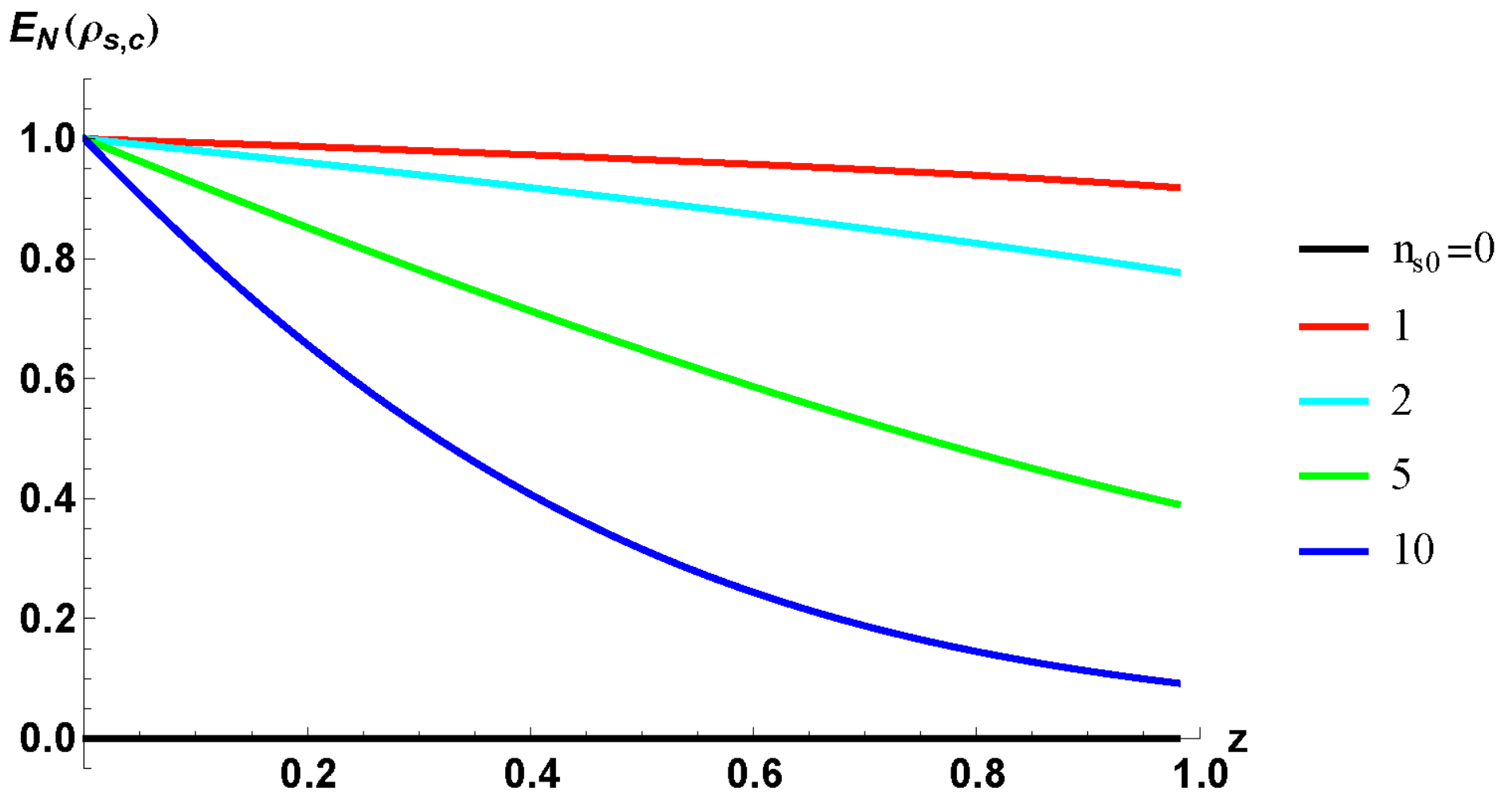}   &
  \includegraphics[width=3.25in,height=2.5in]{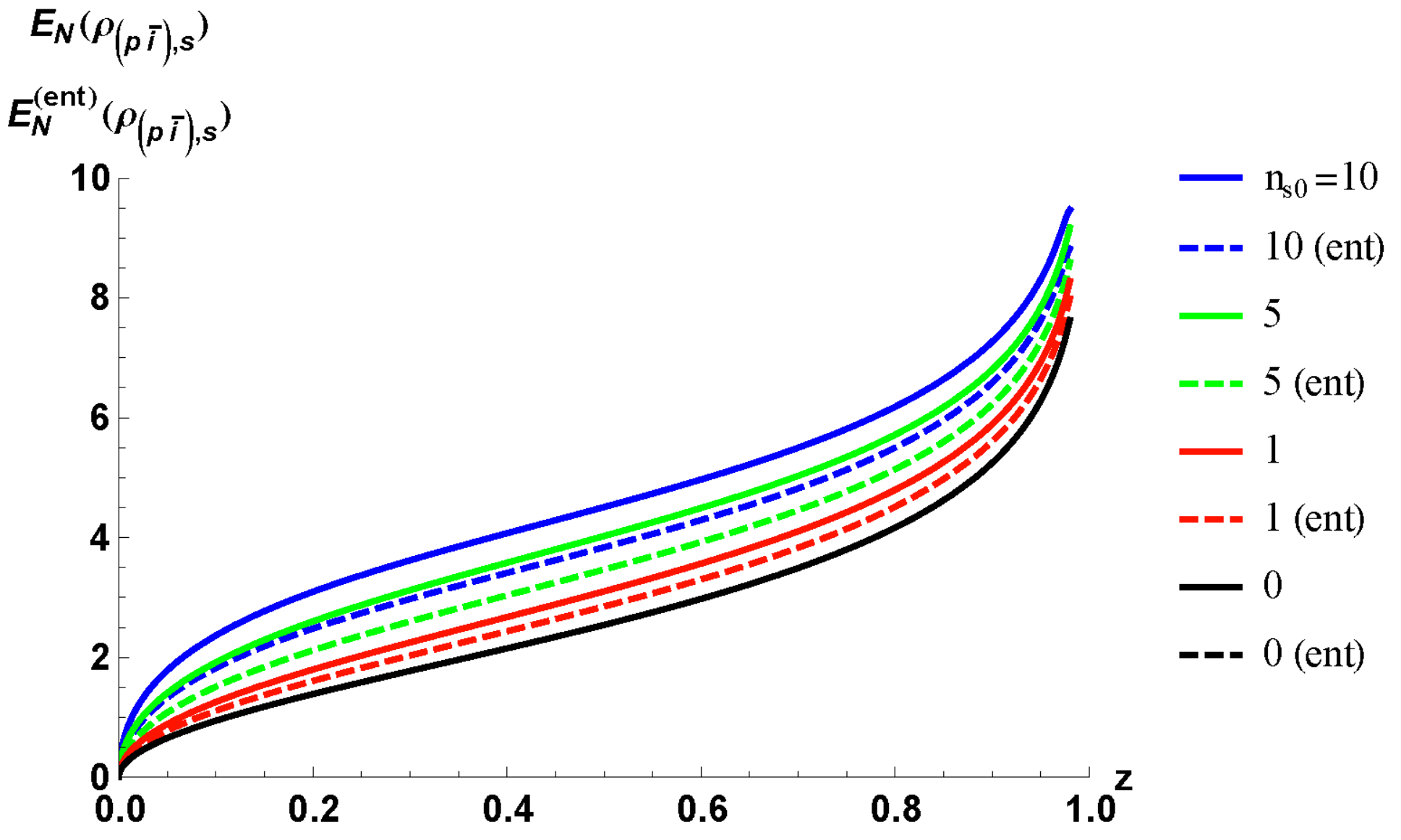}
\end{tabular}
\caption{\label{LogNeg:ent} (left) $E^{(ent\,I.S.)}_{\mathcal{N}}(\rho_{s,c})$
and (right)
(solid) $E^{(sep\,I.S.)}_{\mathcal{N}}(\rho_{(p,\bar{i}),s})$  and
(dashed) $E^{(ent\,I.S.)}_{\mathcal{N}}(\rho_{(p,\bar{i}),s})$
for initially maximally entangled $(s,c)$-mode  state
$\ket{\psi}^{(ent\,I.S.)}_{in}=(\ket{\ns0}_s\ket{0}_c + \ket{0}_s\ket{\nc0}_c)/\sqrt{2}$
with $\nc0=\ns0 = (0,1,2,5,10)$, using short-time amplitudes
$c_n^{(\ns0)} = \left[(1-z)^{\ns0+1}\,z^n\,\binom{\ns0+n}{n}\right]^{1/2}$ for $0 \le z \le 1$.
For $\ns0=0$ (blue curves), the initial state is separable for in modes $s$ and $c$, and remains so for all times. (color online)
}
\end{center}
\end{figure}
%=========================================
which is plotted in left figure of \Fig{LogNeg:ent} for various values of $\nc0 = \ns0$. The single summation over $n$ in \Eq{LogNeg:s-c:ent} arises from the
matrix elements ${}_p\bra{\np0-m} \np0 -n \rangle_p\,\,{}_{\bar{i}}\bra{m} n \rangle_{\bar{i}} = \delta_{m,n}$ that arise when tracing $\ket{\psi(\tau)}^{(ent\,I.S.)}_{out}\bra{\psi(\tau)}$ written as a double sum over $m,n$, over modes $(p,\bar{i},s)$.

As means of comparison, in the right plot of \Fig{LogNeg:ent} we plot
$E^{(ent\,I.S.)}_{\mathcal{N}}(\rho_{(p,\bar{i}),s})$ after forming the density matrix for the bipartite subdivision $(p,\bar{i}), s$ as in Section~\ref{section:Ent:Hpsibar}. Recall that in Section~\ref{section:Ent:Hpsibar} with
separable initial state ($sep\,I.S.$)
$\ket{\psi}^{(sep\,I.S.)}_{in} = \ket{\np0}_p\ket{0}_{\bar{i}}\,\ket{\ns0}_s\ket{\phi}_c$ (for some arbitrary state $\ket{\phi}_c$) we found the negative eigenvalue
${\mathcal{N}_n^{((p,\bar{i}),s)}} = -|c_n^{(\ns0)}\,c_{m}^{(\ns0)}|$ for the partial transpose on $s$ of $\rho_{(p,\bar{i}),s}$
leading to the log-negativity
\be{LogNeg:pibars:no-ent}
E^{(sep\,I.S.)}_{\mathcal{N}}(\rho_{(p,\bar{i}),s}) = \log_2\left[1+\sum_{m,n} \,|c_n^{(\ns0)}\,c_{m}^{(\ns0)}|\right] = \log_2\left[\left(\sum_n \, |c_n^{(\ns0)}|\right)^2\right],\,\,
\ee
upon utilizing $\sum_n \,|c_n^{(\ns0)}|^2 = 1$.
For the entangled initial state $\ket{\phi}^{(ent\,I.S.)}_{s,c}$ in \Eq{psi:init_ent}
we now find the incoherent sum of negative eigenvalues
 ${\mathcal{N}_n^{((p,\bar{i}),s)}} = -1/2\left( |c_n^{(0)}\,c_{m}^{(0)}| + |c_n^{(\nc0)}\,c_{m}^{(\nc0)}|\right)$ in the partial transpose, arising from the $\ket{0}_c$ and  $\ket{\nc0}_c$ portions of $\ket{\psi(\tau)}^{(ent\,I.S.)}_{out}$ when forming
 the partial transpose of $s$ for $\rho_{(p,\bar{i}),s}$. This leads to the log-negativity
\be{LogNeg:pibars:ent}
E^{(ent\,I.S.)}_{\mathcal{N}}(\rho_{(p,\bar{i}),s}) = \log_2\left[\,\frac{1}{2}\, \left(\sum_n\,|c_n^{(0)}|\right)^2
+ \frac{1}{2}\, \left(\sum_n\,|c_n^{(\ns0)}|\right)^2 \,\right].
\ee
In the right plot of \Fig{LogNeg:ent} the solid curves are $E^{(sep\,I.S.)}_{\mathcal{N}}(\rho_{(p,\bar{i}),s})$, while the dashed curves are $E^{(ent\,I.S.)}_{\mathcal{N}}(\rho_{(p,\bar{i}),s})$ which always lie below the former (except for $\ns0=0$, which are identical).

\subsection{Discussion}
The implication of this calculation is that even though particle modes $s$ and $c$ in region $I$ are initially maximally entangled, the coupling of $s$ to the modes $(p,\bar{i})$ (BH mode and emitted anti-particle in region $II$) by means of the evolution under $H_{p,s,\bar{i}}$, weakens the entanglement between $s$ and $c$ as the entanglement between $s$ and $(p,\bar{i})$ grows from an initial zero value. Likewise, the entanglement between the bipartite partition
$(p,\bar{i}), s$ is reduced from what it would be if mode $s$ were not initially entangled with mode $c$.
This is an alternative way to observe that the BH particle production degrades entanglement (see Br\'{a}dler and Adami \cite{Bradler_Adami:2014} for further in depth analysis)
as the entanglement is distributed between bipartite mode partitions $(s,c)$ and modes $(p,\bar{i}), s$ through the common Hawking radiation mode $s$. The fact that the full pure quantum state is entangled across the horizon, vs a separable state, argues against the necessity for the concept of a BH firewall at the horizon
\cite{Almheiri:2013,Braunstein:2009,Braunstein:2013}.

%========================================================================
\subsection{Entanglement by 'Beam-Splitter' Scattering}\label{Ent_by_BS}
We also consider, in a sense, a dual to the scenario of the previous section, namely taking the
initial state to be the separable `0' in-state $\ket{\psi}^{(0)}_{in}$ of \Eq{in_states:0} (dropping
the modes $(\bar{s},i)$)
\be{psi:init_no-ent}
\ket{\psi}^{(0)}_{in} = \ket{\np0}_p\ket{0}_{\bar{i}}\ket{0}_s\ket{1}_c,
\ee
and allowing it to evolve via the combined BH evaporation squeezing Hamiltonian $H_{p,\bar{i},s}$ and the $(s,c)$-mode 'beam-splitter' scattering Hamiltonian $H^{s,c}_{bs}$ as in Section~\ref{GrayBody}. Mode $c$ can be considered as
initially the same mode in the previous section \ref{ent_IS}, (though this time in a product state with the Hawking vacuum) which remains in region $I$ until late times, upon which it is allowed to infall and scatter with the formed BH.
We are again interested in the evolution of the entanglement of $\rho_{s,c}$ between mode outgoing Hawking radiation mode $s$,  and mode $c$   considered  to ultimately fall into region $II$.

\begin{figure}[ht]
\begin{center}
%\includegraphics[width=5in,height=3.5in]{AvS:fig1}
%\begin{tabular}{cc}
  \includegraphics[width=4.0in,height=2.5in]{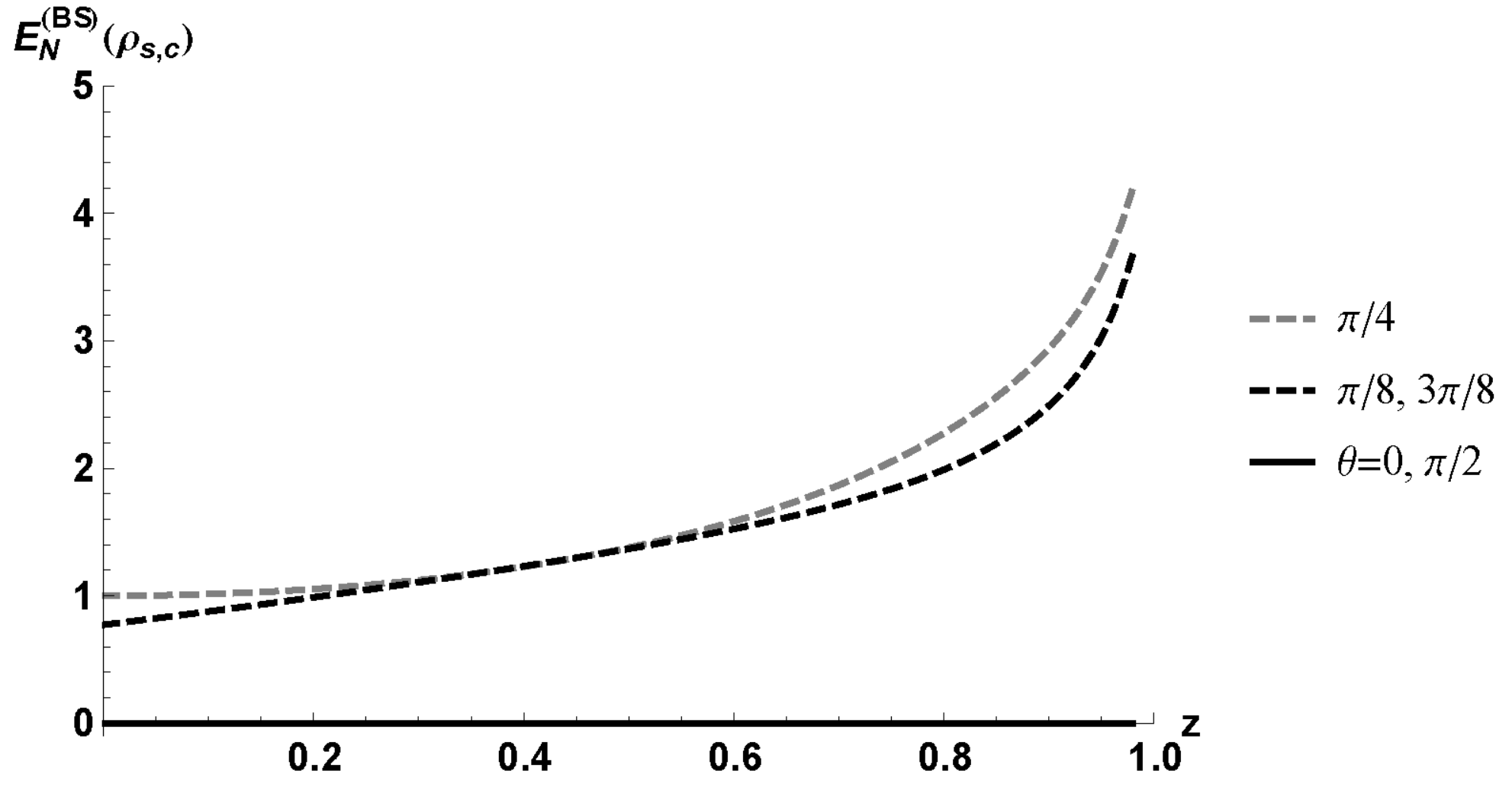}   %&
  %\includegraphics[width=3.5in,height=2.5in]{fig_15_right}
%\end{tabular}
\caption{\label{LogNeg:ent:BS}  $E^{(BS)}_{\mathcal{N}}(\rho_{s,c})$  for initially separable state
$\ket{\psi}^{(0)}_{in} = \ket{\np0}_p\ket{0}_{\bar{i}}\ket{0}_s\ket{1}_c$
evolving under both the BH evolution Hamiltonian $H_{p,\bar{i},s}$ and the mode $s,c$ 'beam-splitter' scattering Hamiltonian $H_{s,c}^{(BS)}$ with transmittance $\cos^2\theta$, using short-time amplitudes
$c_n^{(0)} = \left[(1-z)\,z^n\right]^{1/2}$ for $0 \le z \le 1$.
The curves are symmetric about $\theta=\pi/4$ ($50\%$ transmittance).
}
\end{center}
\end{figure}

In Section~\ref{GrayBody} the scattering between the region $I$ early-time mode $s$ and late-time infalling mode  $c$  was modeled as the unitary `beam-splitter' process (as we shall refer to it) generated by $H^{s,c}_{bs}$ \Eq{AvS:42}.
%However, one could also simply place a physical beam splitter near the event horizon in order to couple the late-time %infalling mode $c$ to the mode $s$ generated by BH particle production.
The beam-splitter Hamiltonian generates entanglement (\Eq{BStransf}) due to the indistinguishability of the reflecting and transmitted paths emerging from the unitary process. Utilizing \Eq{out_state:0}, $\ket{\psi}^{(0)}_{out}$ is now given by
%\begin{subequations}
\bea{out_state:0:monog}
\ket{\psi}^{(0)}_{out} &=&  \sum_{n=0}^\infty \, \sum_{k=0}^\infty \,c_{n,k}\,\ket{\np0-n}_p\ket{k}_s\ket{n}_{\bar{i}}\ket{n+1-k}_c, \\
c_{n,k} &\equiv& c^{(0)}_n \,f^{(0)}_k(n), \quad c^{(0)}_n = \sqrt{(1-z)\,z^n},\\
f^{(0)}_k(n) &=& \sqrt{n+1-k}\,\cos\theta f^{(1)}_k(n) - i \sqrt{k}\,\sin\theta f^{(1)}_{k-1}(n), \\
f^{(1)}_k(n) &=& \sqrt{\binom{n}{k}}\cos^k\theta\,(-i\sin\theta)^{n-k},
\eea
%\end{subequations}
where $\sum_{k=0}^{n} |f^{(1)}_k(n)|^2 = \sum_{k=0}^{n+1} |f^{(0)}_k(n)|^2 = 1$. Some straight forward algebra yields that the partial transpose of the reduced density matrix $\rho_{s,c}$
(written as the double sum $\sum_{n,k}\sum_{m,\ell}$) has negative eigenvalues
${\mathcal{N}}_{k,\ell} = -|c_{n,k}|\,|c_{n,\ell}|$ (where, again a $\delta_{m,n} ={}_p\bra{\np0-m} \np0 -n \rangle_p\,\,{}_{\bar{i}}\bra{m} n \rangle_{\bar{i}}$ has arisen when tracing over the modes $p,\bar{i}$).
This leads to the log-negativity
%\begin{subequations}
\bea{LogNeg:pibars:no-ent:BS}
\fl E^{(BS)}_{\mathcal{N}}(\rho_{s,c}) &=&
\log_2\left[
1+2\sum_{n=0}^{\infty} \sum_{k,\ell\ne k}^{n+1} \,|c_{n,k}|\,|c_{n,\ell}|
\right]
= \log_2\left[ \sum_{n=0} (1-z)\,z^n\,
\left(
\sum_{k=0}^{n+1} \, |f_k^{(0)}(n)|
\right)^2
\right], \qquad \\
\fl |f_k^{(0)}(n)| &=& \frac{1}{\sqrt{n+1}}\,\sqrt{\binom{n+1}{k}} \, (\cos\theta)^{k-1} \, (\sin\theta)^{n-k}
\left| (n+1-k)\,\cos^2\theta - k\,\sin^2\theta \right|,
\eea
%\end{subequations}
which is plotted in \Fig{LogNeg:ent:BS} for various values of $\theta$, where $\cos^2\theta$ is the transmittance of the beam-splitter, again using the short-time amplitudes $c_n^{(0)}(z) = \left[(1-z)\,z^n\right]^{1/2}$ for $0 \le z \le 1$.
%=========================================
%=========================================

For $z=0$ (with $c_n^{(0)}(z)=\delta_{n,0}$) the log-negativity is given by
$E^{(BS)}_{\mathcal{N}}\big(\rho_{s,c}(0)\big) = 2 \log_2( |\cos\theta + \sin\theta|)$ which is zero at $\theta=0,\pi/2$ and reaches its maximum value of unity at $\theta=\pi/4$. The non-zero log-negativity at $z=0$ arises from our model where under the 'squeezing' Hamiltonian the state $\ket{0}_s\ket{1}_c$ is unchanged at $z=0$, while the 'beam-splitter' Hamiltonian transforms the state to $\cos\theta\ket{0}_s\ket{1}_c + \sin\theta\ket{1}_s\ket{0}_c$, for which the partial transpose of the density matrix has negative eigenvalue $-\sin\theta\,\cos\theta$. For $\theta=0,\pi/2$ the beam splitter is either perfectly transmitting or perfectly reflecting and modes $s$ and $c$ remain separable for all times.

\subsection{Discussion}
The implication of the above calculation is that while the $s$ and $c$ modes are initially separable, they become entangled by $H^{s,c}_{bs}$ for each state $\ket{n}_{\bar{i}}\,\ket{n}_s\,\ket{1}_c$ of the two-mode ($s,\bar{i}$)  correlated (squeezed) state generated by the BH particle production in region $I$ as
$\ket{n}_{\bar{i}}\,\ket{n}_s\,\ket{1}_c\rightarrow \ket{n}_{\bar{i}}\,U^{(BS)}_{s,c}\ket{n}_s\,\ket{1}_c$.
Here the unitary $s,c$ mode scattering process is the source of the entanglement generation.
At the transmittance endpoints (unit transmittance: $\theta=0$, unit reflectivity: $\theta=\pi/2$)
the unitary scattering process transforms input product Fock states of modes $s$ and $c$ to output product Fock states, maintaining separability of the two modes for all time.
For  transmittance between $0<\cos^2\theta < 1$, the `beam splitter' scattering process creates the entangled output state of the form  $\sum_{k=0}^{n+1} \,c_{n,k}\,\ket{k}_s\ket{n+1-k}_c$ for each generated squeezed state Hawking radiation pair $\ket{n}_s\ket{n}_{\bar{i}}$ of fixed value $n$, and hence $\rho_{s,c}$ is entangled across the horizon.
Again, the entangled (vs separable) nature of the state $\rho_{s,c}$ across the BH horizon argues against the need for a BH firewall.

\section{Summary and Discussion}\label{Summary}
In this work we have explored the trilinear Hamiltonian for parametric down conversion (PDC) as a
simple, phenomenological unitary zero-dimensional model for pair production in the neighborhood of an evaporating black hole (BH). Here the signal and idler modes of PDC are analogous to the modes that propagate just outside (region $I$) and inside (region $II$) the BH, while the pump mode of PDC models the gravitational field energy degree of freedom as the source of pair production. The primary motivation for utilizing the trilinear Hamiltonian is that it is the simplest, most general unitary form enforcing quantized energy/particle number conservation (and hence back-action effects) between the BH source and emitted Hawking radiation particles.

%======================================================
% discussion of end state of the BH
%======================================================
As in previous investigations \cite{Adami:2014,Bradler_Adami:2014,Nation:2010} this present work does not directly address the ultimate fate of the information in the interior (region $II$), behind the BH horizon.
%It's main thrust goal was to address the post-thermal state that can arise under a dynamical model of BH particle production/evaporation incorporating the simplest energy/particle-number generalization of Hamiltonian employed in the standard treatments of Hawking radiation production.
%=========================================
\begin{figure}[ht]
\begin{center}
%\includegraphics[width=5in,height=3.5in]{AvS:fig1}
%\begin{tabular}{cc}
  \includegraphics[width=3.75in,height=2.25in]{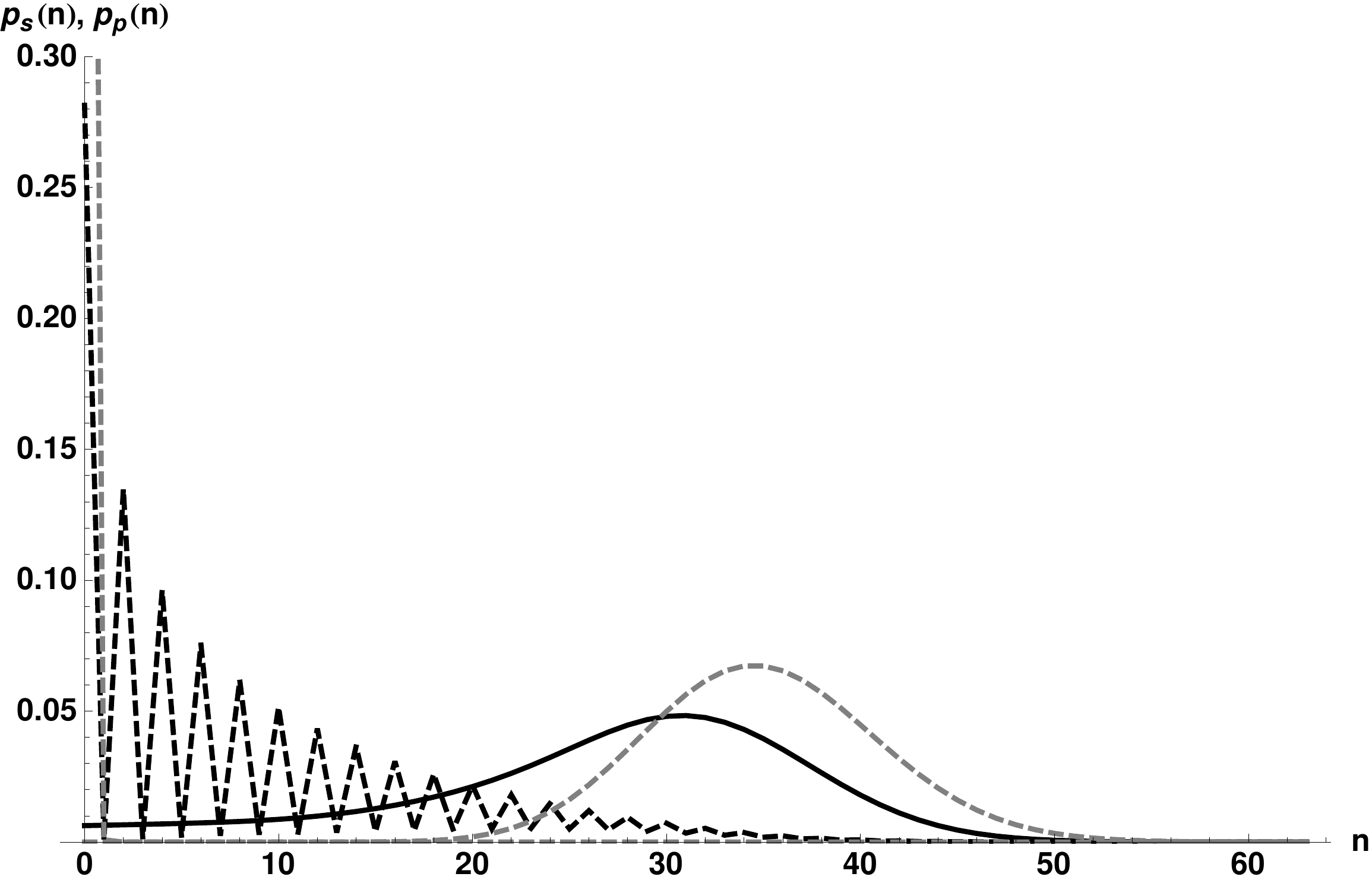}   %&
  %\includegraphics[width=3.5in,height=2.5in]{fig_15_right}
%\end{tabular}
\caption{\label{pns_pnp_in_comp_basis}  Probability distributions $p_s(n,\tau)$ and $p_p(n,\tau)$ in the computational basis (Fock states)from numerical integration of \Eq{eqn:cn_with_CS_pump}.
Initial distributions:
(gray-solid) $p_s(n,0)$,
(gray-dashed) $p_p(n,0)$
with BH $\bar{n}_p(0)=|\alpha|^2=35$.
Late time distributions $\tau\approx 0.55$ when $dn_p/d\tau=0$:
(black-solid) $p_s(n,\tau)$ and
(black-dashed) $p_p(n,\tau)$.
}
\end{center}
\end{figure}
%=========================================
However, the work presented here does suggest that the late-time deviations from the conventionally considered
Hawking radiation thermal state do contain information that can be used to partially reconstruct the initial state of the BH (see additionally \cite{Nation:2010}).
Qualitative evidence for this suggested in \Fig{pns_pnp_in_comp_basis} of the numerical simulations of \Eq{eqn:cn_with_CS_pump} for the
probability distributions $p_p(n,\tau)$ of the BH `pump' state, and $p_s(n,\tau)$ the outgoing Hawking radiation,
in the computational basis (Fock states) presented in Section \ref{Page}.
The gray-dashed curve in \Fig{pns_pnp_in_comp_basis} is the initial
coherent state probability distribution of the BH `pump' mode $n_p$ with $\bar{n}_p(0)=|\alpha|^2=35$
(the initial distribution $p_s(n,0)$ of the Hawking radiation is a delta function at $n=0$).
The black-solid curve is the probability distribution $p_s(n,\tau)$ of the outgoing Hawking radiation mode $s$ at time $\tau\approx 0.55$ when $dn_p/d\tau=0$. Is is qualitatively similar to $p_p(n,0)$.
The black-dashed curve is probability distribution $p_p(n,\tau)$ of the BH `pump' mode at $\tau\approx 0.55$.
The zeros (and near zeros) in $p_p(n,\tau)$ indicate that the BH `pump' is approximately
in a \textit{single-mode squeezed state} \cite{Walls:1994} characterized by the
pure state vector $\ket{\psi}^{(sq)}_p = \sum_n  \tilde{c}_{2n}\ket{2n}_p$
(with odd $\tilde{c}_{2n+1}\equiv 0$).
The implication of these results is that, at the limits of the validity of our model, when $dn_p/d\tau=0$,
the two-mode squeezed nature of the short-time Hawking radiation is impressed upon the BH `pump', while the initial coherent state probability distribution of the BH `pump` is effectively impressed upon the late-time Hawking radiation.
If this were the approximate endpoint of the BH evaporation, the initial information of the BH (mode $p$) has (at least partially) been transferred to the outgoing Hawking radiation (mode $s$).
 The results lends credence  to negation of the assertion quoted in the Introduction to this work (from Lloyd and Preskill \cite{Lloyd:2014}), that the infalling quantum state is cloned (by some supposedly unitary mechanism) in the outgoing radiation, which would violate the linearity of quantum mechanics.

%Initial distributions (gray-solid) $p_s(n,0)$
%(gray-dashed) $p_p(n,0)$ with BH $\bar{n}_p(0)=|\alpha|^2=35$. Late time distributions $\tau\approx 0.55$ when $dn_p/d\tau=0$ (black-solid) $p_s(n,\tau)$ and (gray-dashed) $p_p(n,\tau)$

%In future work we will also examine the numerical simulation of the various entanglement and channel capacities explored in this current work in addition to further quantum correlation effects, following the numerical eigenvalue approach of Walls and Barakat \cite{Walls:1970}. This approach involves the numerical eigenvalue solution of the tridiagonal matrix generated from the right hand side of the exact quantum amplitude equations \Eq{cn:eqn} and \Eq{Cn:eqn}. From our present work and that of Bonifacio and Preparata \cite{Bonifacio:1970} we estimate that one needs $\np0\gtrsim 10^2 - 10^3$ in order for the pump mode to appear nearly constant for early times, yet be small enough to observer pump depletion effects for long times. Preliminary results from direct numerical integration of the quantum amplitude equations has been presented in Section \ref{Page} for $\np0=255$ for cases with the BH (pump) in a (i) Fock, and (ii) a coherent state, which show qualitatively similar behavior, especially for short-times, even for much smaller values of $\np0$ utilized in this work. Future work will concentrate on more general initial states and analytic approximations, and BH information problem entanglement investigations.
%=========================================================================

We have used our unitary trilinear Hamiltonian model to investigate issues of entanglement between various bipartite divisions of the modes involved for both short-times when energy in the BH is roughly large and constant and at late-times when the BH is evaporating.
Specifically, we have examined the entanglement (via the log-negativity) across the BH horizon between the bipartite division of the (pump, idler) and signal modes, as well as entanglement between two separate set of pairs generated by the pump. We find entanglement between the emitted region $I/II$ pairs $(s,\bar{i})$ and $(\bar{s},i)$ through the common BH source 'pump' mode $p$. If one traces out all modes except the particle and antiparticle ($s$ in $\bar{s}$ in this work) in region $I$, we find a separable density matrix. We have also shown both analytically and numerically that this dynamical BH-as-PDC model reproduces the long held conjecture that the (Page) information \cite{Page:1993a,Page:1993b} essentially emerges from the evaporating BH when it has transferred roughly half its population into the Hawking radiation, precisely because of the thermal state deviations that arise in the outgoing radiation at late times.
In this work we also interpret this \textit{Page time} as the time when the variance of the BH `pump' population becomes
equal to the variance of the population of the outgoing radiation.
Further, the entangled, vs separable, nature of the full quantum state across the BH horizon, resulting from the coupling of the evaporating BH `pump' source mode with the emitted internal and external Hawking radiation modes argues against the need for concepts such as BH firewalls.

The work presented here extended the previous work of  Bonifacio and Preparata \cite{Bonifacio:1970} to develop both a short-time and long-time solution to the differential-difference equations \Eq{cn:eqn} for the quantum amplitudes of various relevant `in-states.' In the short-time limit defined by $\np0\gg, n, \ns0, 1$ the pump Fock state $\ket{\np0-n}_p$ essentially factors out from the rest of the signal/idler portion of the state, and the usual two-mode squeezed state with an un-quantized source of PDC is recovered. Since a Fock number state is a highly non-classical state, it might be more appropriate to have modeled the initial BH `pump' state by a coherent state $\ket{\alpha}_p = e^{-|\alpha|^2/2}\,\sum_{m=0}^\infty \alpha^m/\sqrt{m!}\,|m>_p$ with $|\alpha|^2=\np0$, which for a laser pump source is the quantum state most like the classical state of complex amplitude $\alpha$ (i.e. ${}_p\bra{\np0} (a_p + a^\dagger_p)/2\ket{\np0}_p = 0$, while ${}_p\bra{\alpha} (a_p + a^\dagger_p)/2\ket{\alpha}_p = \textrm{Re}(\alpha)$). However, the point of view taken in this paper is that since $\np0\gg 1$, the coherent state $\ket{\alpha}_p$ is very sharply peaked about the mean $\bar{n} = |\alpha|^2$ with \cite{Loudon:1983} standard deviation $\Delta n = \alpha = \sqrt{\bar{n}}$ implying a fractional uncertainty $\Delta n/\bar{n} = 1/\sqrt{\bar{n}}$, and hence $|\alpha|^2\sim\np0$. The extra summation in the definition of the coherent state could be accommodated in an analytical treatment   (as was performed numerically in Section \ref{Page} and will be explored further in future work \cite{Alsing:2015}), but in this present work would only serve to complicate and obscure the central analytical features discussed.

For long times, we have extended the methodology of  Bonifacio and Preparata \cite{Bonifacio:1970} to develop an approximate pde formulation of the exact differential-difference quantum amplitude equations \Eq{Cn:eqn} and showed that each scaled amplitude
$\tilde{\mathcal{C}}_n(t)$ is stationary in the long-time regime \Eq{Cn:gensoln}. To match the long-time solution to the short-time solution the later was used as the initial condition to the pde, which allowed us to compute the crossover time $z^*\approx 1/2$ \Eq{zstar} by (following the reasoning of \cite{Bonifacio:1970}) equating the short-time and long-time expression for the mean occupation number of particles  $\bar{n}(z) = \sum_{n=0}^\infty \, n\, p(n,z)$, where $p(n,z) = |c_n(z)|^2$ is the square of the quantum amplitudes.
We then argued that the stationarity of the quantum amplitudes $\tilde{\mathcal{C}}_n(t)$ in the long-time regime $z>z^*$ implied that the generalized thermal form of the probability $p(n,z)$ for short-times $z\le z^*$, written in terms of the mean number $\bar{n}(z)$ for short-times \Eq{pn:shortime}, holds for long times as well, if one replaces $\bar{n}(z)$ with it's appropriate longtime expression \Eq{nbar:longtime} in the expressions for the long-time probabilities. Bonifacio and Preparata \cite{Bonifacio:1970} found that the dispersion $\sigma(\bar{n}) \equiv (\Delta n)^2$ at any time is practically of a Bose type: $\sigma(\bar{n})\approx \bar{n}$ and further state \textit{We conclude by emphasizing that the chaotic (thermal) character of the short-time (photon) distribution is preserved by the dynamics of our system (the trilinear Hamiltonian) \ldots } Our results are consistent with their findings. However, it is precisely these deviations from
the thermal state arising from  BH evaporation explored here that lead to information emerging from the BH at late times and further, to more diverse entanglement distributions amongst the modes involved.
It is the stationarity of the long-time quantum amplitudes that ultimately leads to only a slight change in the channel (Holevo) capacity calculation (for $z>z^*$) as exhibited in \Fig{chi:figs}) over that of the calculation by Adami and Ver Steeg \cite{Adami:2014}.

A subtle example of the effect of BH evaporation due to
the inclusion of the quantized BH `pump' mode $p$ was given in Section \ref{EntHpsibarHpsbari}
where we explored entanglement between different modes in the long-time outgoing Hawking radiation field: particles $s$ and anti-particles $\bar{s}$ in region $I$. These outgoing modes are generated coherently in two region $I/II$ pairs, $(s,\bar{i})$ and $(\bar{s},i)$, from the same BH source mode $p$.  As discussed in Section~\ref{EntHpsibarHpsbari}, the out state is given by $\ket{\psi}_{out} = \sum_{n}\,\sum_{m} c_{n,m}(t)\,\ket{n,m}_L$ where
$0\le n+m \le \np0$ with $\ket{n,m}_L =  \ket{\np0-n-m}_p\ket{\ns0+n}_{s}\ket{n}_{\bar{i}}\ket{\nsbar0+m}_{\bar{s}}\ket{m}_{i}$. It is readily seen that a constant of the motion of this process is $a^\dagger_{p} a_{p} + 1/2(a^\dagger_{s} a_{s}+a^\dagger_{\bar{i}} a_{\bar{i}} ) + 1/2(a^\dagger_{\bar{s}} a_{\bar{s}}+a^\dagger_{i} a_{i}) = \np0 + \ns0 + \nsbar0$. Upon tracing over the pump we obtain entanglement between the bipartite subsystems of $(s,\bar{i})$ and $(\bar{s},i)$ through the inner product ${}_p\langle \np0-n'-m'|\np0-n-m\rangle_p = \delta_{n'+m',n+m}$ which states that any combination of emitted region $I/II$ pairs that gives rise to the same sum $n+m$ contributes. However, if we compute the density matrix for the outgoing radiation in region $I$ alone, i.e. modes $(s,\bar{s})$, the trace over the region $II$ idler modes
yields the additional delta functions
${}_{\bar{i}}\bra{n'} n\rangle_{\bar{i}} = \delta_{n',n}$ and
${}_{i}\bra{m'} m\rangle_{i} = \delta_{m',m}$ which renders
$\delta_{n'+m',n+m}\,\delta_{n',n}\,\delta_{m',m} =\delta_{n',n}\,\delta_{m',m}$ and
wipes out the entanglement between $s$ and $\bar{s}$, producing a separable region $I$ out-state $\rho_s\otimes\rho_{\bar{s}}$, appropriate for an uncorrelated thermal state.
Interestingly, in our model we find entanglement between the bipartite subdivision of modes $(p,\bar{i})$ and $s$, but no entanglement directly between the modes $\bar{i}$ and $s$. In the short-time regime where one can factor out the BH source 'pump' mode (or in the case usually considered in the literature where the pump mode occupation number is implicitly taken to be large and constant) the former case reduces to the later case.

If we push the analogy of BH pair production with  quantum optical process of PDC it is natural to ask `where is the `cavity' (crystal) in which the unitary process takes place and what is its length?'  Mottola and Vaulin \cite{Mottola:2013} take a complete different approach and model the interior of the BH as a Bose-Einstein condensate. This gives rise to an internal negative pressure much like dark energy which prevents the BH singularity from forming. From this work the authors estimate the 'thickness' of the BH event horizon as $\Delta\ell\approx\mathcal{O}(\sqrt{\ell_p R_s}) \approx 2\times10^{-16} m \sqrt{M/M_{\odot}} $ where $\ell_p = \sqrt{\hbar G/c^3}$ is the Planck length and $R_s=2G M_{bh}/c^2$ is the  Schwarzschild radius of the of the BH of mass $M_{bh}$.
\footnote{This result could be argued from dimensional analysis by assuming that
$\Delta\ell\sim\ell_p^{x}\,R_s^{(1-x)}$ so that both sides have units of length. The solution
$\Delta\ell\approx\mathcal{O}(\sqrt{\ell_p R_s})$ then represents the geometric mean $x=1/2$, the most equitable weighting of both lengths $\ell_p$ and $R_s$.}
As such, the `length' of the `non-linear crystal' in which the BH particle pair production occurs with depleted BH `pump' source could be taken to be $\mathcal{O}(\sqrt{\ell_p R_s})$. How long the modes propagate just outside (region $I$) and inside (region $II$) the BH governs how long (i.e. $z_{max}$) the PDC process continues to generate pair production via $a_p\,a^\dagger_s\,a^\dagger_{\bar{i}}$, which simultaneously feeds back into replenishing the BH `pump' via $a^\dagger_p\,a_s\,a_{\bar{i}}$. However, we have shown that up to the time for
which the BH population $n_p$ reaches its first minimum value $dn_p/d\tau=0$, the energy flow is from the
BH source to the Hawking radiation $n_s$ and $n_{bar{i}}$.
If this time were to be infinite (i.e. $z\rightarrow 1$) then it is apparent how a steady state arises to produce a stationarity in the quantum amplitudes $c_n(z)$ for large $z$. Yurke \cite{Yurke:1987,Alsing:2004} has pointed out that in the case of spontaneous parametric down conversion (SPDC) (i.e. $\ns0=\nsbar0=0$) an effective thermal temperature $T$ arises if one chooses to only observe the signal mode (and trace out over the idler mode, to mimic the inaccessible region $II$ modes just inside the BH), and is given effectively by the nonlinear interaction strength times the length of the non-linear crystal.

Of course, once the signal mode exits the `crystal' (at some $0\le z_{max} \le 1$) the PDC process stops, and this mode is available for external detection. In the case of the BH, how long  the idler modes hover just inside the horizon (i.e. the determination of $z_{max}$) is technically problematic since by classical general relativity an object, massive or massless, reaches the BH singularity in a finite proper time once it crosses the horizon. Thus, lacking a full quantum theory of gravity the modeling of the interior region $II$ quantum state is phenomenological at best.
Thus, something `more' is called for to address this question, possibly along the lines of investigation of Mottola and Vaulin \cite{Mottola:2013} (who, in fact advocate that their method eliminates the whole issue of information loss since their BEC model eliminates the BH singularity).
Nonetheless, the  unitary model put forth here explicitly describes an energy/particle-number preserving process to be
used as a simplified arena to explore BH particle production/evaporation.

In future work, we will include loss in the outgoing signal modes via an open-system quantum optical master equation \cite{Walls:1994} to better mimic their escape from the BH 'non-linear cavity' to infinity.
%(and hence, their non-feedback into the pump and idler modes).
Some researchers have modeled the production of Hawking radiation as arising from the curvature distortion resulting from
the connection of interior and exterior (relative to the BH horizon) spacelike surfaces as one constructs
a time-slicing of the BH spacetime.
As a consequence, the Hawking radiation emerges as a time ordered sequence of emitted wave packets (see e.g. \cite{Gerlach:1976,Mathur:2009} and references therein) each of which moves away from the subsequent region of curvature distortion (at reduced radius), and interacting only slightly with this new region. In the standard Hawking radiation formulation (the `non-depleted pump' in our terminology), this interaction time of the emitted wave packet with the new curvature region is (to lowest order) taken to be zero. However, precisely how long this finite interaction time is could
embody many of the BH source/Hawking radiation effects (information emergence, entanglement distributions, etc\ldots) explored in this work. As such, it would seem appropriate to switch to a temporal wavepacket description \cite{Hawking:1975} defined by \cite{Skaar:2004}
$a^\dagger_{i,t_0} = \int d\omega\, \xi(\omega,t_0) \, a^\dagger_i(\omega)$
for a energy mode $i$ centered about time $t_0$,
with the mode function $\xi(\omega,t_0)$ typically taken to be Gaussian
$\xi(\omega,t_0) = (2\pi\omega)^{-1/4} e^{i(\omega-\omega_0)- (\omega-\omega_0)^2/4\Delta^2}$ of frequency
bandwidth $\Delta\ll \omega_0$ about a central frequency $\omega_0$.
Here, the usual frequency modes satisfying $[a_i(\omega),a^\dagger_j(\omega')] = \delta_{i,j}\delta(\omega-\omega')$
gives rise to the commutation relation $[a_{i,t_0},a^\dagger_{j,t_1}] = \delta_{i,j}\,e^{-\Delta^2(t_1-t_0)^2/2}$
which treats two temporal modes as independent if sufficiently separated in time, $t_1-t_0\gg 1/\Delta$.
Thus, one could model the BH particle production/evaporation process similar to the work presented here, but now with a set of time-bin modes to carry away the successive particle production pairs. However,  it remains reasonable that a multi-temporal mode version of the fully quantized squeezing Hamiltonian $H_{p,s,\bar{i}}$ used in this work would be warranted to ensure the most general unitary form of energy conservation. In addition, it also seems reasonable to expect that there would be some finite `coherence time` of the quantized BH `pump' source mode such that successively emitted time-bin modes would be entangled to some (as of yet, unknown) degree.
Such energy-time or time-bin entanglement effects are known to exist
(and in fact, are currently engineered and actively studied, see \cite{Franson:1989,Brendel:1999,Barreiro:2005})
for the temporal modes that are emitted under the coherence time of the `pump' source, since such modes carry a degree of indistinguishability. Such investigations will be pursued in future work \cite{Alsing:2015}.

\ack
The author wishes to thank M.L. Fanto, C.C. Tison, S. Preble, M. Corne, W.A. Miller, S. Ray, H.A. Blair, D. Patten, C. Gerry, E. Hach, R. Birrittella and G. Ver Steeg, for stimulating discussions. The author would like to acknowledge support for this work by T. Curcic of the Air Force Office of Scientific Research (AFOSR).
The author also wishes to thank the two anonymous referees,
whose helpful comments and suggestions  strengthened the paper.
Any opinions, findings and conclusions or recommendations expressed in this material are those of the author(s) and do not necessarily reflect the views of AFRL.
%\end{acknowledgments}

%===========================================================
% Appendices
%===========================================================
%% The command \appendix is used to signify the start of the appendices. Thereafter
%% \section, \subsection, etc, will give headings appropriate for an appendix. To obtain
%% a simple heading of 'Appendix' use the code \section*{Appendix}. If it contains
%% numbered equations, figures or tables the command \appendix should precede it and
%% \setcounter{section}{1} must follow it.
%===========================================================
\appendix
\section{Analytic solution for the full Hamiltonian $\mathcal{H} = H_{p,s,\bar{i}} + H_{p,\bar{s},i}$}\label{appendix:fullH}
%\section{Analytic solution for the full Hamiltonian $\mathcal{H} = H_{p,s,\bar{i}} + H_{p,\bar{s},i}$}\label{appendix:fullH}
\setcounter{section}{1}
%===========================================================
%==========================================================
% Begin: material from 'Full Hamiltonian' v5 for appendix
%==========================================================
For the full Hamiltonian
\be{fullHamil2:app}
\mathcal{H} = H_{p,s,\bar{i}} + H_{p,\bar{s},i} = r \,(a_p^\dagger \, K^{(s\bar{i})}_-  + a_p \, K^{(s\bar{i})}_+)
  +  r \,(a_p^\dagger \, K^{(\bar{s} i)}_-  + a_p \, K^{(\bar{s} i)}_+)
\ee
considered in Section \ref{fullH}
the logical states of all modes involved are given by
\be{state:nmL}
\ket{n,m}_L = \ket{\np0-n-m}_p\ket{\ns0+n}_{s}\ket{n}_{\bar{i}}\ket{\nsbar0+m}_{\bar{s}}\ket{m}_{i},
\ee
where $\ns0$ and $\nsbar0$ are the initial number of particles/anti-particles in the $s$ and $\bar{s}$ modes in region $I$.
The output state is given by
\be{psi:out:full}
\fl \ket{\psi}_{out} = \sum_{n=0}\,\sum_{m=0} c_{n,m}(t) \, \ket{n,m}_L, \quad 0\le n+m \le \np0, \quad \ket{\psi}_{in} = \ket{0,0}_L,
\ee
where $c_{n,m}(t) = {}_L\bra{n,m}\,e^{-i\,\mathcal{H}\,t}\ket{\psi}_{in}$.
Note that the sums in \Eq{psi:out:full} can also be written as the ordered sums $\sum_{n=0}^{\np0}\,\sum_{m=0}^{\np0-n}$ or
$\sum_{m=0}^{\np0}\,\sum_{n=0}^{\np0-m}$.

Following the procedure of the Section \ref{Hpsibar_latetime_derivation} we can derive the exact differential-difference equation for $c_{n,m}(t')$ with $t'=r\,t$
\bea{cnm:eqn}
\fl \frac{d c_{n,m}}{d t'} &=&
\sqrt{\np0-n-m}
\left(  \sqrt{(n+1)\,(2\kappa+n)}       \, c_{n+1,m}(t')
      + \sqrt{(m+1)\,(2\bar{\kappa}+m)} \, c_{n,m+1}(t')
\right) \no
\fl &+&
\sqrt{\np0-n-m+1}
\left(  \sqrt{n\,(2\kappa+n-1)}        \, c_{n-1,m}(t')
      + \sqrt{m+\,(2\bar{\kappa}+m-1)} \, c_{n,m-1}(t')
\right), \qquad
\eea
where $2\kappa-1 = \ns0$ and $2\bar{\kappa}-1 = \nsbar0$.
%the number of initial particles $s$ and anti-particles $\bar{s}$ initially in region $I$.

For short-times, defined by the condition $\np0 \gg \ns0, \nsbar0, n, m, 1$ we can again factor out $\sqrt{\np0}$, define
$\tau = r\,\sqrt{\np0}\,t$ and obtain the factorized (separable) amplitudes
%\begin{subequations}
\bea{cnm:shorttime}
\fl c_{n,m}(\tau) &=& c^{<\,(s,\bar{i})}_n(\tau) \,\, c^{<\,(\bar{s},i)}_m(\tau), \\
\label{cncm:shorttime}
\fl c^{<\,(s,\bar{i})}_n(\tau) &=& \frac{(-i\,\tanh\tau)^n}{(\cosh\tau)^{n_{s0}+1}} \, \sqrt{\binom{n_{s0} + n}{n}},
\quad
c^{<\,(\bar{s},i)}_m(\tau) = \frac{(-i\,\tanh\tau)^m}{(\cosh\tau)^{n_{\bar{s}0}+1}} \, \sqrt{\binom{n_{\bar{s}0} + m}{m}},
%\quad \tau = \sqrt{n_{p0}}\,r\, t,
\eea
%\end{subequations}
with $c^<(\tau)$ having the same form as \Eq{cn:shorttime}. This yields the separable state in the mode pairs $(s,\bar{i})$ and $(\bar{s},i)$
%\begin{subequations}
\bea{psioutnm:shorttime}
\fl \ket{\psi_{<}(\tau)}_{out} &=& \sum_{n,m} \, c_{n,m}(\tau) \, \ket{n,m}_L,\no
\label{psioutnm:shorttime:sep}
\fl &\approx&  \ket{n_{p0}}_p\otimes\,\sum_{n=0}^{\infty} \, c^<_n(\tau) \,\ket{n_{s0}+n}_s\ket{n}_{\bar{i}} \,
\sum_{m=0}^{\infty} \, c^<_m(\tau) \,\ket{n_{\bar{s}0}+m}_{\bar{s}}\ket{m}_{i}, \\
\fl &\equiv& \ket{n_{p0}}_p\otimes\,\ket{\psi_{<}(\tau)}_{s,\bar{i}} \,\otimes\, \ket{\psi_{<}(\tau)}_{\bar{s},i},
\eea
%\end{subequations}
with probability distribution
%$p_<(n,\tau) = |c^<_{n,m}(\tau)|^2 = |c^{<\,(s,\bar{i})}_n(\tau)|^2 \,\, |c^{<\,(\bar{s},i)}_m(\tau)|^2$.
$p_<(n,m,\tau) = |c^<_{n,m}(\tau)|^2 = |p_<(n,\tau)|^2 \,\, |p_<(m,\tau)|^2$, with each factor on the
right hand side having the form of \Eq{pn:shortime}.

To develop a pde for long-times, where $\np0, n, m, \gg \ns0, \nsbar0, 1$, we must define two $G$ functions in analogy to \Eq{G:eqn}
\bea{Gnm:eqn}
G_{2\kappa}(n,m)       &=& g(\np0-n-m) \, g(n) \, g(2\kappa + n -1), \no
G_{2\bar{\kappa}}(n,m) &=& g(\np0-n-m) \, g(m) \, g(2\bar{\kappa} + m -1),
\eea
with $g(n)$ still given by \Eq{g:eqn}.
We further define
\be{Cnm:defn}
\tilde{C}_{n,m}(t) = \sqrt{G_{2\kappa}(n,m)\,G_{2\bar{\kappa}}(n,m)} \,\tilde{c}_{n,m}(t),
\qquad \tilde{c}_{n,m}(t) = (-i)^{n+m} \, c_{n,m}(t),
\ee
to derive the exact equation for $\tilde{C}_{n,m}(t')$
\bea{Cnmtilde:eqn}
\fl 0 = \frac{d\tilde{C}_{n,m}(t')}{dt'} &+&
G_{2\kappa}(n,m)\,
\left[ \sqrt{\frac{G_{2\bar{\kappa}}(n,m)}{G_{2\bar{\kappa}}(n+1,m)}} \, \tilde{C}_{n+1,m}(t') \,- \,
       \sqrt{\frac{G_{2\bar{\kappa}}(n,m)}{G_{2\bar{\kappa}}(n-1,m)}} \, \tilde{C}_{n-1,m}(t')
\right], \no
\fl &+&
G_{2\bar{\kappa}}(n,m)\,
\left[ \sqrt{\frac{G_{2\kappa}(n,m)}{G_{2\kappa}(n+1,m)}} \, \tilde{C}_{n,m+1}(t') \,- \,
       \sqrt{\frac{G_{2\kappa}(n,m)}{G_{2\kappa}(n,m-1)}} \, \tilde{C}_{n,m-1}(t')
\right]. \qquad
\eea
An examination of the terms in the square roots reveals that
$$
\frac{G_{2\kappa}(n,m)}{G_{2\kappa}(n,m\pm 1)} = \frac{g(\np0-n-m)}{g(\np0-n-(m\pm 1))} \approx 1,
$$
(and similarly for $\kappa\rightarrow\bar{\kappa}$)
which only utilizes $\np0\gg 1$ and thus is well within our long-time approximation. This allow us to write the approximate pde as
\be{Cnmtilde:pde:1}
\fl 0\approx\frac{\partial \tilde{C}(n,m,t')}{\partial t'}
+ 2 G_{2\kappa}(n,m)\,\frac{\partial \tilde{C}(n,m,t')}{\partial n}
+ 2 G_{2\bar{\kappa}}(n,m)\,\frac{\partial \tilde{C}(n,m,t')}{\partial m}
\ee
We now define
\bea{nm:defns}
n = \np0\,\cos^2\phi\,\sin^2\theta, \no
m = \np0\,\sin^2\phi\,\sin^2\theta,
\eea
so that $n+m = \np0\,\sin^2\theta$ in analogy with the previous section.
Using
\Bea
\frac{\partial\theta}{\partial n}&=&\frac{\partial\theta}{\partial m} = \frac{1}{2\sqrt{(n+m)\,(\np0-(n+m))}}, \no
\frac{\partial\phi}{\partial n} &=& -\frac{\partial\phi}{\partial m} = \frac{1}{2(n+m)\sqrt{m/n}},
\Eea
we derive
\bea{Cnmtilde:pde:2}
\fl \frac{\partial \tilde{C}(\theta,\phi,t')}{\partial t'}
&+& \Big[\cos\phi\, v_{\ns0}(\theta,\phi) + \sin\phi\, v_{\nsbar0}(\theta,\phi) \Big]\,
\frac{\partial \tilde{C}(\theta,\phi,t')}{\partial\theta}, \no
\fl &+& \left[\frac{-\sin\phi}{\tan\theta} \, v_{\ns0}(\theta,\phi) + \frac{\cos\phi}{\tan\theta}\, v_{\nsbar0}(\theta,\phi) \right]\,
\frac{\partial \tilde{C}(\theta,\phi,t')}{\partial\phi} = 0,
\eea
where we have defined
\bea{vs}
v_{\ns0}(\theta,\phi)    &=& \sqrt{2\kappa + n}       =  \sqrt{1+\ns0+\np0\cos^2\phi\,\sin^2\theta}, \no
v_{\nsbar0}(\theta,\phi) &=& \sqrt{2\bar{\kappa} + m} =  \sqrt{1+\nsbar0+\np0\sin^2\phi\,\sin^2\theta}.
\eea

In general, \Eq{Cnmtilde:pde:2} is a quite complicated pde in $\theta$, $\phi$ and $t'$.
We develop a reasonable special case of \Eq{vs} by considering the condition
\bea{specialcase}
1 + \ns0 &=& (2+\ns0+\nsbar0)\,\cos^2\phi, \quad  \Rightarrow \quad \frac{1+\nsbar0}{1+\ns0} = \tan^2\phi \equiv \frac{m}{n}, \no
1 + \nsbar0 &=& (2+\ns0+\nsbar0)\,\sin^2\phi.
\eea
For example, for the case of pure spontaneous emission $\ns0=\nsbar0=0$ the above special case condition requires $m=n$, or that the number of emitted  signal particles $n$ in region $I$ is equal to the number of emitted signal anti-particles $m$ in region $I$, something we expect from the symmetry of the particle production. In general, \Eq{specialcase} requires that the ratio $n/m$ of emitted particle/anti-particle in region $I$  is equal to the initial ratio of the particles/anti-particles (plus unity) in the radiation field. This simplifies \Eq{vs} to
$v_{\ns0}(\theta,\phi) = \cos\phi\,v(\theta)$ and $v_{\nsbar0}(\theta,\phi)=\sin\phi\,v(\theta)$ where
\be{vtheta}
v(\theta) = \sqrt{2 + \ns0 + \nsbar0 + \np0\sin^2\theta} = \sqrt{2\kappa+2\bar{\kappa}+n+m},
\ee
and yields the pde
\bea{Cnmtilde:pde:specialcase}
\frac{\partial \tilde{C}(\theta,\phi,t')}{\partial t'} &+&
v(\theta)\frac{\partial \tilde{C}(\theta,\phi,t')}{\partial\theta} = 0, \no
\frac{\partial \tilde{C}(u, t')}{\partial t'} &+&
\frac{\partial \tilde{C}(u, t')}{\partial u} = 0,
\eea
with
\be{u:eqn:full}
 u(\theta) = \int_0^\theta \, \frac{d\theta'}{v(\theta')}.
\ee
A further detailed analysis reveals that the next order correction to \Eq{Cnmtilde:pde:specialcase} is
${\mathcal{O}}\big(\partial_\phi\tilde{C}(\theta,\phi,t')/\np0\big)$ which we drop by invoking $\np0\gg 1$.

We therefore find from \Eq{u:eqn:full} and \Eq{vs}
%\begin{subequations}
\bea{nmbar:longtime}
%\bar{n}_>(\tau) + \bar{m}_>(\tau) = \np0 \, cn^2\left(\tau - T \,| \, k_e \right),
\bar{n}_>(\tau) &=& \frac{\ns0+1}{\ns0+\nsbar0+2}\,\np0\,cn^2\left(\tau - T_q \,| \, k_e \right), \\
\bar{m}_>(\tau) &=& \frac{\nsbar0+1}{\ns0+\nsbar0+2}\,\np0\,cn^2\left(\tau - T_q \,| \, k_e \right),
\eea
%\end{subequations}
where, with a redefinition of $\tau$ within our long-time approximation, we have
\be{taunm:kenm}
\fl \tau = \sqrt{\np0 + \ns0 + \nsbar0 + 2}, \quad k_e = \frac{\np0}{\ns0 + \nsbar0 + 2},
\quad cn^2\left(\tau - T \,| \, k_e \right) \myover{{\rightarrow} {k_e\to 1}} \sech^2\left(\tau - T\right),
\ee
and
\bea{Tqnm}
\fl T_q =  \textsf{a}(k_e) + \frac{1}{2} \ln\left(\frac{\np0+\ns0+\nsbar0+2}{\ns0+\nsbar0+2}\right)
= \textsf{a}(k_e) - \frac{1}{2} \ln(1-k_e),
\,\, \myover{{\ln 4} {k_e\rightarrow 0}} \le \textsf{a}(k_e) \le \myover{{\pi/2} {k_e\rightarrow 1}}.\,\qquad
\eea
This yields the long-time entangled state
\bea{psioutnm:longtime}
\ket{\psi_{t_>}(\tau)}_{out} &=& \sum_{n,m} \, c^>_{n,m}(\tau) \, \ket{n,m}_L,\no
% &=& \sum_{n=0}^{\infty} \, \sum_{m=0}^{\infty}
&=&\sum_{n,m}\, c^>_{n,m}(\tau)
\ket{\np0-n-m}_p\ket{n_{s0}+n}_s\ket{n}_{\bar{i}}\ket{n_{\bar{s}0}+m}_{\bar{s}}\ket{m}_{i}.
\eea
Following the discussion for \Eq{zstar:eqn} we determine the crossover point between the short-time and long-time solutions
by equating
\be{zstarnm:eqn}
\bar{n}_<(\tau^*) + \bar{m}_<(\tau^*) = \bar{n}_>(\tau^*) + \bar{m}_>(\tau^*)
% \Leftrightarrow (\ns0+1)\,\frac{\zeta^2}{1-\zeta^2} = \np0\,\left[1 - \frac{(\zeta-\zeta_{T_q})^2}{(1-\zeta\zeta_{T_q})^2}\right],
\ee
with crossover time $z^*=\tanh^2\tau^*$.
The matching of the total mean number of particles allows us to
replace $\bar{n}_<$ in the expression for $p_<(n)=|c_n^{<(s,\bar{i})}|$ by $\bar{n}_>$ for $z>z^*$,
and similarly
replace $\bar{m}_<$ in the expression for $p_<(m)=|c_n^{<(\bar{s},i)}|$ by $\bar{m}_>$,
to construct $p_>(n,m,\tau)$ from $p_<(n,m,\tau)$, as we did in \Eq{pn:longtime}.

\Eq{zstarnm:eqn} yields the same equation for $z^*$ as previously in \Eq{zstar:eqn} but now with
$(\ns0+1) \to (\ns0+1) + (\nsbar0+1)$ in the right hand side equation, as well as in  the equation for $z^*$.
In the limit $\np0\to\infty$ we obtain the exact same value of $z^* = (e^{\pi/2}/2-1)^{-2} = 0.506407$ as before
in \Eq{zstar}.
%======================================================
% End: material from 'Full Hamiltonian' v5 for appendix
%======================================================

%======================================================
% Begin: material from orig Sec7.1 v5 for appendix
%======================================================
\section{Derivation of quantum states for sending a `0' and a `1'}\label{appendix:Sec:7.1_v5}
For the channel capacity calculation of Section \ref{Sec:7.1} we send a logical `0' with in-state $\ket{\psi}^{(0)}_{in}$ and send a logical `1' with in-state $\ket{\psi}^{(1)}_{in}$ given by
%\begin{subequations}
\bea{in_states}
\label{in_states:0}
\ket{\psi}^{(0)}_{in} = \ket{\np0}_{p}\ket{0}_{s}\ket{0}_{\bar{i}}\ket{1}_{c}\ket{0}_{\bar{s}}\ket{0}_{\bar{i}}, \\
\label{in_states:1}
\ket{\psi}^{(1)}_{in} = \ket{\np0}_{p}\ket{0}_{s}\ket{0}_{\bar{i}}\ket{0}_{c}\ket{1}_{\bar{s}}\ket{0}_{\bar{i}},
\eea
%\end{subequations}
where in \Eq{in_states:0} there is now 1 input boson in mode $c$ vs 1 in the early-time mode $a$ (both region $I$), and in \Eq{in_states:1} there is still 1 input boson in the region $I$ mode $\bar{s}$, as considered
previously in section \ref{AvS:review}.

Consider first the case of sending a `1'. The squeezing/beam splitter Hamiltonian
creates the output state  $\ket{\psi}^{(1)}_{out}$  given by
\bea{out_state:1}
\fl \ket{\psi}^{(1)}_{out} &=& \sqrt{1-z}\, \sum_{n=0}^\infty \, \sum_{k=0}^\infty\,\sqrt{z^n} f^{(1)}_k(n)\,\ket{\np0-n-m}_p\ket{k}_s\ket{n}_{\bar{i}}\ket{n-k}_c \no
\fl &\otimes& \sqrt{(1-z)^2/z}\, \sum_{m=0}^\infty \, \sqrt{m\,z^m} \, \ket{m}_{\bar{s}}\ket{m-1}_{i}, \,\,
f^{(1)}_k(n) = \sqrt{\binom{n}{k}}\cos^k\theta\,(-i\sin\theta)^{n-k},\,\qquad
\eea
which leads to the separable density matrix $\rho^{(s,\bar{s})}(1) = Tr_{p,i,\bar{i}}[\ket{\psi}^{(1)}_{out}\bra{\psi}]$
%$\equiv \rho^{(s)}(1)\otimes\rho^{(\bar{s})}(1)$
%\begin{subequations}
\bea{rho:out:1}
\fl \rho^{(s,\bar{s})}(1) &=&   \sum_{k=0}^\infty p^{(s)}_k(1) \ket{k}_s\bra{k} \otimes \sum_{m=0}^\infty p^{(\bar{s})}_m(1) \, \ket{m}_{\bar{s}}\bra{m} \equiv \rho^{(s)}(1)\otimes \rho^{(\bar{s})}(1),\\
\label{psk1:psbarm1}
\fl p^{(s)}_k(1) &=& \frac{1-z}{1-z\sin^2\theta} \, \left( \frac{z\cos^2\theta}{1-z \sin^2\theta} \right)^k, \quad p^{(\bar{s})}_m(1) = m \frac{(1-z)}{z} \,\, \big( (1-z)\,z^m \big).
\eea
%\end{subequations}
In \Eq{rho:out:1} the amplitude $f^{(1)}_k(n)$ arises from $H^{s,c}_{bs}$ acting on modes $a$ and $c$ with intermediate state (after squeezing) of $\ket{n}_a\ket{0}_c$ for each value of $n$, and preserving the total  number of particles $n$ in modes $(s,c)$.
After the applying the beam splitter transformation, the output signal $(s)$ density matrix has the form $\rho^{(s)}(1) = \sum_{n=0}^\infty \sum_{k=0}^{n} (1-z) z^n |f^{(1)}_k(n)|^2 \ket{k}_s\bra{k}$ which has the mixed form
$\rho^{(s)}(1) = \sum_{n=0}^\infty \rho_{n}^{(s)}(1)$ for each beam splitter rotated state of fixed number of particles $n$. To determine the probabilities the series is resummed as
$\rho^{(s)}(1) = \sum_{k=0}^\infty \big[\sum_{n=0}^\infty (1-z) \, z^n \, |f^{(1)}_k(n-k)|^2 \,\big] \ket{k}_s\bra{k}$ where the inner sum on $n$ can be computed in closed form
(utilizing the identity $\sum_{n=0}^\infty \binom{m+n}{n}\,z^n = (1-z)^{-(m+1)}$)
and results in the probabilities $p^{(s)}_k(1)$ shown
in \Eq{psk1:psbarm1}.
%\equiv \sum_{k=0}^\infty p^{(s)}_k(1) \ket{k}_s\bra{k}$

For the case of sending a `0', the calculations proceeds as in the previous case, but now with intermediate state \Eq{in_states:0}, and with  initial $(s,c)$ state $\ket{n}_s\ket{1}_c$ after squeezing. The beam splitter now preserves the total number of particles $n+1$ initially in the modes $(s,c)$. The probabilities  $p^{(s)}_k(0)$ and $p^{(\bar{s})}_m(0)$ are directly related to the previous probabilities  $p^{(s)}_k(1)$ and $p^{(\bar{s})}_m(1)$. We obtain
\bea{out_state:0}
\ket{\psi}^{(0)}_{out} &=& \sqrt{1-z}\, \sum_{n=0}^\infty \, \sum_{k=0}^\infty\,\sqrt{z^n} f^{(0)}_k(n)\,\ket{\np0-n-m}_p\ket{k}_s\ket{n}_{\bar{i}}\ket{n+1-k}_c \no
&\otimes& \sqrt{(1-z)}\, \sum_{m=0}^\infty \, \sqrt{z^m} \, \ket{m}_{\bar{s}}\ket{m}_{i}, \\
f^{(0)}_k(n) &=& \sqrt{n+1-k}\,\cos\theta f^{(1)}_k(n) - i \sqrt{k}\,\sin\theta f^{(1)}_{k-1}(n), \,
\eea
which leads to the separable density matrix $\rho^{(s,\bar{s})}(0) = Tr_{p,i,\bar{i}}[\ket{\psi}^{(0)}_{out}\bra{\psi}^{(0)}]$ %$\equiv \rho^{(s)}(0)\otimes\rho^{(\bar{s})}(0)$
%\begin{subequations}
\bea{rho:out:0}
\rho^{(s,\bar{s})}(0) &=&   \sum_{k=0}^\infty p^{(s)}_k(0) \ket{k}_s\bra{k} \otimes \sum_{m=0}^\infty p^{(\bar{s})}_m(0) \, \ket{m}_{\bar{s}}\bra{m} \equiv \rho^{(s)}(0)\otimes \rho^{(\bar{s})}(0),\\
\label{psk0:psbarm0}
p^{(s)}_k(0) &=& \, \left( \cos^2\theta + k\,\tan^2\theta\frac{(1-z)^2}{z} \right) p^{(s)}_k(1),
%\frac{1-z}{1-z\sin^2\theta} \, \left( \frac{z\cos^2\theta}{1-z \sin^2\theta} \right)^k,
\quad p^{(\bar{s})}_m(0) =  (1-z)\,z^m, \qquad    \\
p^{(\bar{s})}_m(0) &=& \frac{(1-z)}{z}\quad p^{(\bar{s})}_m(1),
\eea
%\end{subequations}
After the applying the beam splitter transformation, the output signal $(s)$ density matrix has the form $\rho^{(s)}(0) = \sum_{n=0}^\infty \sum_{k=0}^{n+1} (1-z) z^n |f^{(0)}_k(n)|^2 \ket{k}_s\bra{k}$ which has the mixed form
$\rho^{(s)}(0) = \sum_{n=0}^\infty \rho_{n}^{(s)}(0)$ for each beam splitter rotated state of fixed number of particles $n+1$. To determine the probabilities the series is resummed as
$\rho^{(s)}(0) = \sum_{k=0}^\infty \big[\sum_{n=0}^\infty (1-z) \, z^n \, |f^{(0)}_k(n+1-k)|^2 \,\big] \ket{k}_s\bra{k}$ where the inner sum on $n$ can be again computed in closed form
and results in the probabilities $p^{(s)}_k(0)$ shown in \Eq{psk0:psbarm0}.
\section*{References}
\bibliographystyle{iopart-num}
\bibliography{HawkingRad_as_DepletedPump_PDC}

\providecommand{\newblock}{}
\begin{thebibliography}{10}
\expandafter\ifx\csname url\endcsname\relax
  \def\url#1{{\tt #1}}\fi
\expandafter\ifx\csname urlprefix\endcsname\relax\def\urlprefix{URL }\fi
\providecommand{\eprint}[2][]{\url{#2}}

\bibitem{Adami:2014}
Adami C and Steeg G~V 2014 {\em Class. Quantum Grav.\/} {\bf 31} 075015

\bibitem{Bradler_Adami:2014}
Bradler K and Adami C 2014 {\em J. High Energy Physics\/} {\bf May:2014}
  article:95

\bibitem{Hawking:1975}
Hawking S~W 1975 {\em Commun. Math. Phys.\/} {\bf 43} 199

\bibitem{Unruh:1976}
Unruh W~G 1976 {\em Phys. Rev. D\/} {\bf 14} 870

\bibitem{Gerlach:1976}
Gerlach U~H 1976 {\em Phys. Rev. D\/} {\bf 14} 1479

\bibitem{Giddings:1992}
Giddings S~B and Nelson W~M 1992 {\em Phys. Rev. D\/} {\bf 46} 2486
  (arxiv:hep--th/9204072)

\bibitem{Mathur:2009}
Mathur S~D 2009 {\em Class. Quantum Grav.\/} {\bf 26} 220401 (arxiv:0909.1038)

\bibitem{Walls:1994}
Walls D~F and Milburn G~J 1994 {\em Quantum Optics\/} (Springer-Verlag, N.Y.)

\bibitem{Nielsen:2000}
Nielsen M~A and Chuang I~L 2000 {\em Quantum computation and quantum
  information\/} (Cambridge University Press, Cambridge)

\bibitem{Nation:2010}
Nation P~D and Blencowe M~P 2010 {\em New J. Phys.\/} {\bf 12} 095013
  (arxiv:1004.0522)

\bibitem{Lloyd:2014}
Lloyd S and Preskill J 2014   arxiv:1308.4209

\bibitem{Susskind:1993}
Susskind L, Thorlacius L and Uglum J 1993 {\em Phys. Rev. D\/} {\bf 48} 3743
  (arXiv:hep--th/9306069)

\bibitem{Susskind:1994}
Susskind L and Thorlacius L 1994 {\em Phys. Rev. D\/} {\bf 49} 966
  (arXiv:hep--th/9308100)

\bibitem{Almheiri:2013}
Almeheiri A, Marolf D, Polchinski J and Sully J 2013   arxiv:1207.3123

\bibitem{Braunstein:2009}
Braunstein S~L 2009   arxiv:0907.1190

\bibitem{Braunstein:2013}
Braunstein S~L, Pirandola S and Zyczkowski K 2013 {\em Phys. Rev. Lett.\/} {\bf
  110} 101301)

\bibitem{Maldacena:2013}
Maldacena J and Susskind L 2013   arxiv:1306.0533

\bibitem{Walls:1970}
Walls D~F 1970 {\em Phys. Rev. A\/} {\bf 1} 446

\bibitem{Bonifacio:1970}
Bonifacio R and Preparata G 1970 {\em Phys. Rev. A\/} {\bf 2} 336

\bibitem{Gerry:2004}
Gerry C and Knight P~L 2004 {\em Introductory Quantum Optics\/} (Cambridge
  Univeristy Press,Cambridge)

\bibitem{Agarwal:2013}
Agarwal G~S 2013 {\em Quantum Optics\/} (Cambridge University Press, Cambridge)

\bibitem{Kwiat:1994}
Kwiat P~G and et al 1995 {\em Phys. Rev. Lett.\/} {\bf 75} 4335

\bibitem{Kwiat:1999}
Kwiat P~G and et al 1999 {\em Phys. Rev. A\/} {\bf 60} 773

\bibitem{O'Brien:2007}
O'Brien J and et al 2007 {\em Science\/} {\bf 318} 1567

\bibitem{Pan:2012}
Pan J~W, Chen Z~B, Lu C~Y, Weinfurter H, Zeilinger A and Zukowski M 2012 {\em
  Rev. Mod. Phys.\/} {\bf 84} 777

\bibitem{Fanto:2011}
Fanto M~L and et al 2011 {\em SPIE\/} {\bf 8057} 805705

\bibitem{Peters:2012}
Peters C~J and et al 2012 {\em SPIE\/} {\bf 8400} 84000Z

\bibitem{Sorkin:1987}
Sorkin R~D 1986 {\em Class. Quantum Grav.\/} {\bf 4} L149

\bibitem{Page:1993a}
Page D~N 1993 {\em Phys.Rev.Lett.\/} {\bf 71} 1291 (arxiv:gr--qc:9305007v2)

\bibitem{Page:1993b}
Page D~N 1993 {\em Phys.Rev.Lett.\/} {\bf 71} 3743 (arxiv:gr--qc:9306083v2)

\bibitem{Alsing:2012}
Alsing P~M and Fuentes I 2012 {\em Class. Quantum Grav.\/} {\bf 29} 224001
  (arxiv:1210.2223)

\bibitem{Perelomov:1986}
Perelomov A~M 1986 {\em eneralized coherent states and their applicaitons\/}
  (Springer-Verlag, N.Y.)

\bibitem{Abramowitz:1972}
Abramowitz M and Stegun I~A 1972 {\em Handbook of mathematical funcitons\/}
  (Dover. Publ., N.Y.)

\bibitem{Schumacher:1997}
Schumacher B and Westmoreland M~D 1997 {\em Phys. Rev. A\/} {\bf 56} 131

\bibitem{Bandilla:2000}
Bandilla A, Drobn´y G and Jex I 2000 {\em J. Opt. B: Quantum Semiclass. Opt.\/}
  {\bf 2} 265

\bibitem{Loudon:1983}
Loudon R 1983 {\em The quantum theory of light (2nd Ed.)\/} (Oxford Univ.
  Press, N.Y.)

\bibitem{Holstein:1940}
Holstein T and Primakoff H 1940 {\em Phys. Rev.\/} {\bf 58} 1098

\bibitem{Sanders:2000}
Rowe R, de~Guise H and Sanders B 2001 {\em J. Math. Phys.\/} {\bf 42} 2315

\bibitem{Perelomov:1972}
Perelomov A~M 1972 {\em Comm. Math. Phys.\/} {\bf 26} 222

\bibitem{Vidal:2002}
Vidal G and Werner R~F 2002 {\em Phys. Rev. A\/} {\bf 65} 032324

\bibitem{Plenio:2005}
Plenio M~B 2005 {\em Phys. Rev. Lett.\/} {\bf 95} 090503

\bibitem{Hayden:2013}
Hayden P, Headrick M,  and Maloney A 2013 {\em Phys. Rev. D\/} {\bf 87} 046003
  (arxiv:1107.2940v2)

\bibitem{Popescu:2006}
Popescu S, Short A and Winter A 2006 {\em Nature Phys.\/} {\bf 2} 754
  (arxiv:quant--ph/0511225)

\bibitem{Terhal:2003}
Terhal B~M 2003   arxiv:quant--ph/0307120

\bibitem{Koashi_Winter:2004}
Koashi M and Winter A 2004 {\em Phys. Rev. A\/} {\bf 69} 022309
  (arxiv:quant--ph/0310037)

\bibitem{Alsing:2015}
Alsing P~M, C~Gerry R~B and Corne M~A 2015   (in preparation)

\bibitem{Mottola:2013}
Mottola E and Vaulin R 2013 {\em Physics Today\/} {\bf November} 9--10

\bibitem{Yurke:1987}
Yurke B and Potasek M 1987 {\em Phys.Rev. A\/} {\bf 36} 3464

\bibitem{Alsing:2004}
Alsing P~M, McMahon D and Milburn G~J 2004 {\em J. Opt. B: Quantum Semiclass.
  Opt.\/} {\bf 6} S834

\bibitem{Skaar:2004}
Skaar J, Escartin J~C~G and Landro H 2004 {\em Am. J. Phys.\/} {\bf 72} 1385

\bibitem{Franson:1989}
Franson J~D 1989 {\em Phys. Rev. Lett.\/} {\bf 62} 2205

\bibitem{Brendel:1999}
Brendel J, Gisin N, Tittel W and Zbinden H 1999 {\em Phys. Rev. Lett.\/} {\bf
  82} 2594 (arxiv:quant--ph/9809034)

\bibitem{Barreiro:2005}
J~T~Barreiro Nathan K~Langford N~A~P and Kwiat P~G 2005 {\em Phys. Rev.
  Lett.\/} {\bf 95} 260501 (arxiv:quant--ph/0507128)

\end{thebibliography}
%======================================================================================
%===========================================================
\end{document}